\documentclass[prd,preprint,superscriptaddress,amsmath,amssymb,nofootinbib]{revtex4}

\pdfoutput=1

\usepackage{slashed,bbold}
\usepackage{graphicx,color}
\usepackage[toc,title,page]{appendix}

\begin{document}

{\begin{flushright}{KIAS-P19041}
\end{flushright}}

\title{  Radiatively scotogenic type-II seesaw and a relevant phenomenological analysis}

\author{Chuan-Hung Chen}
\email{physchen@mail.ncku.edu.tw}
\affiliation{Department of Physics, National Cheng-Kung University, Tainan 70101, Taiwan}

\author{Takaaki Nomura}
\email{nomura@kias.re.kr}
\affiliation{School of Physics, KIAS, Seoul 02455, Korea}

\date{\today}

\begin{abstract}
When a small vacuum expectation value of Higgs triplet ($v_\Delta$) in the  type-II seesaw model is required to explain  neutrino oscillation data, a fine-tuning issue occurs on the mass-dimension lepton-number-violation (LNV) scalar coupling.  Using the scotogenic approach, we investigate how a small LNV term is arisen through a radiative correction when an $Z_2$-odd vector-like lepton ($X$) and an $Z_2$-odd right-handed Majorana lepton ($N$) are introduced to the type-II seesaw model. Due to the dark matter (DM) direct detection constraints,  the available DM candidate is the right-handed Majorana particle, whose mass depends on and is close to the $m_X$ parameter. Combing the constraints from the DM measurements, the $h\to \gamma\gamma$ decay, and the oblique $T$-parameter, it is found that the preferred range of $v_\Delta$ is approximately in the region of  $10^{-5}-10^{-4}$ GeV; the mass difference between the doubly and the singly charged Higgs is less than 50 GeV, and the influence on the $h\to Z\gamma$ decay is not significant. Using the constrained parameters, we analyze the decays of each Higgs triplet scalar in detail, including the possible three-body decays when the kinematic condition is allowed. It is found that with the exception of doubly charged Higgs, scalar mixing effects play an important role in the Higgs triplet two-body decays when the scalar masses are near-degenerate. In the non-degenerate mass region, the branching ratios of the Higgs triplet decays are dominated by the three-body decays.

\end{abstract}

\maketitle

\section{Introduction}

An extension of the standard model (SM) is necessary due to the observed massive neutrinos. If  the origin of neutrino masses arises from a similar Brout-Englert-Higgs mechanism in the SM~\cite{Englert:1964et,Higgs:1964pj,Guralnik:1964eu}, where the $W^\pm$ and $Z$ gauge bosons, the quarks, and the charged leptons obtain their masses through a Higgs doublet $(H)$, it is natural to introduce a Higgs triplet ($\Delta$) to the SM as a  neutrino mass source. Hereafter, we call the Higgs triplet model  the type-II seesaw model~\cite{Schechter:1980gr,Magg:1980ut,Cheng:1980qt,Lazarides:1980nt,Mohapatra:1980yp}.  Since only the left-handed leptons couple to the Higgs triplet, neutrinos are  the Majorana particles.

In addition to the Yukawa couplings, the neutrino masses are associated with the vacuum expectation value (VEV) of the Higgs triplet. In the minimal type-II seesaw model, it is known that the $\Delta$ VEV indeed is dictated by the lepton-number softly breaking term $\mu_\Delta H^T i\tau_2 \Delta^\dag H$, which appears in the scalar potential.   Thus, a fine-tuning issue on $\mu_\Delta$ is caused when the condition of $\mu_D \ll O(m_W)$  is required to explain the neutrino mass~\cite{Chun:2003ej,Franceschini:2013aha,Cai:2017jrq}. 

 From the astrophysical observation, dark matter (DM) is introduced to explain more than $80\%$ of non-baryonic matter. If DM is a kind of weakly interacting massive particle (WIMP), a radiatively scotogenic mechanism for generating the neutrino masses can be applied~\cite{Ma:2006km,Fraser:2015mhb}, where the particles in the dark sector are the mediators in the loop Feynman diagrams. Various applications of scotogentic models can be found in~\cite{Ma:2014cfa,Molinaro:2014lfa,Vicente:2014wga,Merle:2015gea,Culjak:2015qja,Merle:2015ica,Yu:2016lof,Ahriche:2016cio,Ferreira:2016sbb,Rocha-Moran:2016enp,Chowdhury:2016mtl,Hessler:2016kwm,Diaz:2016udz,Borah:2017dfn,Abada:2018zra,Hagedorn:2018spx,Hugle:2018qbw,Rojas:2018wym,Borah:2018rca,CentellesChulia:2019gic,Ma:2019yfo,Kang:2019sab,Chen:2019nud,Brdar:2013iea,Baumholzer:2018sfb}. 
 
 In order to naturally obtain a small $\mu_D$ parameter in the type-II seesaw model,  in this study, we consider that  $\mu_\Delta H^T i\tau_2 \Delta^\dag H$ is suppressed at the tree level due to the lepton-number symmetry; then, the necessary $\mu_\Delta$ term  is radiatively induced through the scotogenic mechanism~\cite{Kanemura:2012rj,Nomura:2016dnf,Nomura:2017emk}. Since the minimal type-II seesaw model does not include any particles that belong to the invisible side, we inevitably have to add  new dark representations to the type-II seesaw model.  Because the Higgs triplet cannot couple to singlet fermions, the minimum representation that directly couples to the Higgs triplet is the $SU(2)_L$ doublet fermion ($X$).  Due to $H$ and $X$ being the $SU(2)_L$ doublets, in order to form a gauge invariant interaction, we can add one more singlet fermion $(N)$ into the model such that the $H$, $X$, and $N$  coupling can generate the $\mu_\Delta$ term through the one-loop level.    

If the new representation set is assumed to be a minimal choice, due to the gauge anomaly free condition, the new doublet fermion can be a vector-like lepton doublet, and the singlet fermion can be a right-handed Majorana lepton without carrying any SM gauge quantum numbers.  In addition, to have a stable DM candidate, we impose a $Z_2$-symmetry to the vector-like lepton doublet and right-handed singlet; that is, $X$ and $N$ belong to the dark representations.  Thus, the loop-induced  $\mu_\Delta$ term indeed arises from the lepton-number soft breaking effects in the invisible sector. 

The main characteristics in the simple extension of the type-II seesaw model can be summarized as follows: (a) The Dirac-type neutral component of $X$, denoted by $X^0$, becomes a Majorana-type  lepton when the mixing  with $N$  from the $XHN$ coupling occurs after electroweak symmetry breaking (EWSB); (b) the spin-independent (SI) and the spin-dependent (SD) DM-nucleon scatterings arise from the mediation of the $Z$ boson and the SM Higgs, respectively; (c) although the $X^0$- and $N$-DM candidates can produce the observed DM relic density, the $X^0$ candidate is excluded by the constraints of the DM direct detection experiments; therefore, the DM candidate in this study is dominated by the Majorana particle $N$; (d) the loop-induced VEV of $\Delta$ can be in the range of $10^{-5}-10^{-4}$ GeV, whereas the Higgs triplet Yukawa couplings constrained by the neutrino oscillation data are in the range of $10^{-8}-10^{-7}$, and (e) the doubly charged Higgs ($H^{\pm\pm}$) favors decaying to the same sign $W$-boson and  lepton pairs when $H^{\pm\pm}$ is as heavy as  $m_{H^{\pm\pm}}\sim 400$ and  800 GeV, respectively. In addition, we analyze the constraints from the Higgs diphoton decay and the oblique $T$ parameter~\cite{Peskin:1991sw}; as a result, $|m_{H^{\pm\pm}}-m_{H^\pm}|\lesssim 50$ GeV is allowed and the new physics influence on the $h\to Z\gamma$ decay is not significant. 

In addition to the DM candidate and the origin of the neutrino masses,  similar to the conventional type-II seesaw model, it is of interest to explore and probe the new scalars of the Higgs triplet at the LHC, especially the search for $H^{\pm\pm}$. With an integrated luminosity of 12.9 fb$^{-1}$ at $\sqrt{s}=13$ TeV, CMS reports that the bounds on $m_{H^{\pm\pm}}$ through the $ \ell^\pm \ell^\pm$ ($\ell=e,\mu)$, $\ell^\pm \tau^\pm$, and $\tau^\pm \tau^\pm$ channels are between 800 and 820 GeV, between 643 and 714 GeV, and  535 GeV, respectively, where $BR(H^{++}\to \ell^+ \ell'^+)=100\%$ ($\ell'=e,\mu,\tau$) for each lepton pair is used~\cite{CMS:2017pet}. Using 36 fb$^{-1}$ of the integrated luminosity at $\sqrt{s}=13$ TeV and the same sign dilepton channels, ATLAS obtains the $m_{H^{\pm\pm}}$  lower bound from 770 to 870 GeV  with $BR(H^{++}\to \ell^+ \ell^+)=100\%$. Moreover,  the $m_{H^{\pm\pm}}$ lower bound via the $H^{++}\to W^+ W^+$ channel measured by ATLAS is given to be between 200 and 220 GeV~\cite{Aaboud:2018qcu,Ucchielli:2018koe}.  

Based on the lower bound measurements of $m_{H^{\pm\pm}}$, since the preferred $X$ mass in this study is close to 1 TeV, $H^{\pm\pm}$ decaying to the same sign charged heavy $X^\pm$ lepton pair is kinematically suppressed. Thus, the possible decay channels of the Higgs triplet are similar to the those of the conventional type-II seesaw model. Nevertheless, since the $\mu_\Delta$ parameter is dynamically  generated in the model and mainly depends on the $XHN$ coupling, which is determined by the observed DM relic density and the DM direct detection experiments, the allowed $\Delta$ VEV is limited in the narrow region of $10^{-5}-10^{-4}$ GeV, so, the Higgs triplet decay patterns are strongly correlated with the scalar couplings $\lambda_1 H^\dag H Tr(\Delta^\dag \Delta)$ and $\lambda_4 H^\dag \Delta \Delta^\dag H$, where the $\lambda_4$ sign determines the mass ordering of the Higgs triplet scalars. Because the doubly charged Higgs search in the LHC has been broadly studied in the literature~\cite{Akeroyd:2005gt,delAguila:2008cj,Melfo:2011nx,Aoki:2011pz,Akeroyd:2011zza,Arhrib:2011vc,Akeroyd:2012nd,Chun:2012zu,Chun:2013vma,Chabab:2014ara,Han:2015hba,Guo:2016dzl,Mitra:2016wpr,Ghosh:2017pxl,Dev:2018sel,Dev:2018kpa,Du:2018eaw,Antusch:2018svb,Bhattacharya:2018fus,Barman:2019tuo,Primulando:2019evb, Chiang:2012dk}, we  thus  focus the analysis on   the decays of each Higgs triplet scalar in detail.

The paper is organized as follows: In Sec. II, we discuss the extension of the SM, including the derivations of heavy $Z_2$-odd particle mixing and their gauge couplings. In addition to the  loop-induced $\mu_\Delta$ term, we show all scalar mass spectra and the associated scalar mixings, the Higgs-triplet Yukawa couplings, and neutrino mass in Sec. III.  In Sec. IV, we study the possible constraints, such as neutrino data, DM relic density and DM direct detections, the oblique $T$ parameter, and $h\to \gamma\gamma$.  We discuss the influence on $h\to Z\gamma$ and show the decays of each Higgs triplet  in Sec. V. A conclusion is given in Sec. VI.

\section{The Model }

In addition to the SM particles, we add one Higgs triplet $\Delta$, one vector-like lepton doublet $X_{R, L}$, and one $SU(2)$ singlet heavy neutrino into the SM, where their representations in $SU(2)_L\times U(1)_Y$ are given in Table~\ref{tab:rep_charge}. In order to avoid the Dirac neutrino  mass term, we require that $X$ and $N$ are $Z_2$-odd states and that the others are $Z_2$-even; therefore, the lightest neutral particles of $X$ and $N$ could be the DM candidate. In addition, in order to dynamically generate the finite dimension-3 lepton-number violating term in the scalar potential, we 
assign that  $X_{L(R)}$, $N$ and $\Delta$  carry the  lepton numbers as $0(1)$, $0$ and $2$, respectively, where the lepton number symmetry is softly broken by the $X$ Dirac mass term.
The detailed charge assignments of the introduced particles are shown in Table~\ref{tab:rep_charge}. 

%


\begin{table}[htp]
\caption{Representations and charge assignments of the introduced particles. }
\begin{center}
\begin{tabular}{c|c|c|c} \hline \hline

Particle ~& ~ $SU(2)_L\times U(1)_Y$ ~& $Z_2$  & Lepton $\#$  \\ \hline 
$X_L$~ &~ $(2,\, -1)$ ~& ~$-1$ ~&~  $0$ \\ \hline 
$X_R$~ &~ $(2,\, -1)$ ~& ~$-1$ ~&~  $1$ \\ \hline
$N$~ & ~$(1,\, 0)$ ~&  ~$-1$ ~&~ $0$ \\ \hline
$\Delta$~ &~ $(3,\, 2)$ ~&~ $+1$ ~&~ $-2$ \\ \hline \hline

\end{tabular}
\end{center}
\label{tab:rep_charge}
\end{table}%

Based on the chosen representations and charge assignments, the gauge invariant Yukawa couplings can be written as:
\begin{align}
-{\cal L}_Y & = \bar L {\bf y}^{\ell} H \ell_R + L^T C i\tau_2 \, \Delta \, {\bf y}^\ell_{\Delta} \,  L + y_R X^T_R C i\tau_2 \, \Delta  \, X_R \nonumber \\
&+ y_X \bar X_L \tilde {H}  N + \frac{m_N}{2}  N^T C  N + m_X \bar X_L X_R + H.c.\,,  \label{eq:Yu}
\end{align}
where the flavor indices are suppressed; $C=i\gamma^2 \gamma^0$ is charge conjugation matrix; $H$ is the SM Higgs doublet, $\tilde{H}=i \tau_2 H^*$, $\tau_2$ is the Pauli matrix, and $L^T=(\nu , \ell)$ is the SM lepton doublet. 
 It can be seen that the lepton number symmetry is explicitly broken by the $m_X$ dimension-3 terms.
The Higgs doublet, vector-like lepton doublet, and Higgs triplet are respectively expressed as:
 \begin{align}
 H & =   \left(\begin{array}{c}
    G^+ \\ 
   \Phi^0\\ 
  \end{array} \right) \,, ~ X =   \left(\begin{array}{c}
   X^0 \\ 
  X^- \\ 
  \end{array} \right)  \,, \nonumber \\
 \Delta & =   \left(\begin{array}{cc}
    \delta^+/\sqrt{2} & \delta^{++}  \\ 
   \Delta^0 & -\delta^+/\sqrt{2}\\ 
  \end{array} \right) \,,
 \end{align} 
 with $\Phi^0 = (v_h + Re(\Phi^0) + i Im(\Phi^0))/\sqrt{2}$ and $\Delta^0 = (v_\Delta + Re(\Delta^0) + i Im(\Delta^0))/\sqrt{2}$, in which $v_h$ and $v_\Delta$ are the VEVs of the $\Phi^0$ and $\Delta^0$ fields, respectively. The VEVs and scalar masses are  determined by the scalar potential. 
 
 \subsection{   Heavy Majorana masses }

Because of the $X_L H N$  and $X_R \Delta X_R$ couplings, it is found that the Dirac-type $X^0$ not only mixes with Majorana particle $N$ but also  has a Majorana mass, which is related to  $v_\Delta X^T_R C X_R$ when $\Delta^0$ obtains a VEV. Thus, using the basis of $(X_R, X^C_L, N)$,  the Majorana-type heavy  fermion mass matrix  is written as:
  \begin{equation}
  M_{M}=
\left(
\begin{array}{ccc}
 m_0  & m_X  &  0 \\
 m_X  &  0  &   y_X v_h/\sqrt{2} \\
  0 &  y_X v_h/\sqrt{2}  &  m_N  
\end{array}
\right)\,, \label{eq:M_M}
  \end{equation}
  with $m_0 =\sqrt{2} y_{R} v_\Delta$. Since $v_\Delta$ is induced from one-loop in this study, it is expected that $m_0 \ll m_{N, X}$.  It is found that   the $M_{M}$ eigenvalues can be approximately expressed as follows: For $m_N > m_X$, 
 \begin{equation}
m_{N_1} \approx m_X - e_X\,,   -m_{N_2} \approx (m_X + e_{\delta} )\,, m_{N_3}\approx m_N+ e_N\,,  \label{eq:heavy_M1}
 \end{equation}
where we use $N_i$ as the  Majorana particle eigenstates, and $e_{N,X}$ and $e_{\delta}$ are obtained as:
 \begin{align}
  %
 %
 e_N & = \frac{y^2_X v^2_h}{2m_N}\,, \nonumber \\
  e_X & = m_X+ \frac{e_N}{2} - \sqrt{(m_X + e_N/2)^2 - m_X e_N}\,, \nonumber \\
 e_\delta & = e_N -e_X\,.
   \label{eq:es1}
 \end{align}
For $m_N < m_X$, they are:
 \begin{equation}
 m_{N_1} \approx m_X + e_X\,,   -m_{N_2} \approx (m_X + e_{\delta} )\,, m_{N_3}\approx m_N- e_N\,,  \label{eq:heavy_M2}
 \end{equation}
 where the corresponding $e_{N,X}$ and $e_\delta$ are given as:
  \begin{align}
 %
 e_N & = \frac{2 m^2_X}{m_N + m_X} \left( -\left(1-\frac{m^2_N}{m^2_X} \right) + \sqrt{\left( 1-\frac{m^2_N}{m^2_X}\right)^2 + \left(1+\frac{m_N}{m_X}\right) 
\frac{y^2_X v^2_h}{m^2_X}} \right)\,, \nonumber \\
 e_X & = \frac{1}{2} \left( 1+ \frac{m_N}{m_X}\right) e_N \,, ~
e_\delta = \frac{1}{2} \left( 1- \frac{m_N}{m_X}\right) e_N \,.
 \label{eq:es2}
  \end{align}
 Based on the obtained eigenvalues, the $3\times 3$ orthogonal matrix elements ($O_{ij}$), which transform the  $(X_R, X^C_L, N)$ state to the $(N_1, N_2, N_3)$ state, can be formulated as:
  \begin{align}
  O_{11} & = {\cal N}^{-1}_{1} \frac{m_X}{m_{N_1} -m_0}\,, ~  O_{12}  = \frac{1}{{\cal N}_{1}} \,, ~ O_{13} = - {\cal N}^{-1}_{1} \frac{y_X v}{\sqrt{2} (m_N - m_{N_1})}\,, \nonumber \\
 O_{21} & = -{\cal N}^{-1}_{2} \frac{m_X} { m_0-m_{N_2}}\,, ~O_{22} =\frac{1}{{\cal N}_{2}}\,, ~ O_{23} =-{\cal N}^{-1}_{2} \frac{y_X v}{\sqrt{2}(m_N-m_{N_2})}\,, \nonumber \\
 O_{31} & = {\cal N}^{-1}_{3} \frac{m_X}{m_{N_3} -m_0} \,, ~ O_{32} = \frac{1}{{\cal N}_{3}}\,, ~ O_{33} =-{\cal  N}^{-1}_{3} \frac{y_X v_h }{\sqrt{2} (m_N -m_{N_3} )}\,,\label{eq:O_mixing}
  \end{align}
where ${\cal N}^2_{i} = \sum_k O^2_{ik}$ are the normalization factors. 

\subsection{Gauge couplings of $Z_2$-odd particles}

If we define the Majorana states $\chi_i$ as $\chi_i=N_i + N^C_i=\chi^C_i$, which satisfy  $P_R \chi_i =N_i$ and $P_L \chi_i=N^C_i$,  the charged current interactions of the heavy fermions  can be expressed as: 
 \begin{align}
 {\cal L}^{CC} 
 %
& = - \frac{g}{\sqrt{2}} O_{i1} \bar \chi_i  \gamma^\mu P_R X^{-}_{R} W^+_\mu  - \frac{g}{\sqrt{2}} O_{i2} \bar\chi_i \gamma^\mu P_L X^-_L W^+_\mu + H.c.\,,
 \end{align}
 where the mixing matrix elements $O_{ij}$ for the neutral $Z_2$-odd particles are included. 
  The neutral current interactions of the  $Z$-gauge boson and the photon with the $Z_2$-odd particles can be obtained as:
 \begin{align}
 {\cal L}^{NC} & = - \frac{g c^Z_{ij}}{2c_W} \bar \chi_i \gamma^\mu \frac{\gamma_5}{2} \chi_j Z_\mu + \frac{g c_{2W}}{2c_W} \overline{X^-} \gamma^\mu X^{-} Z_\mu \nonumber \\
 &  - eQ_X \overline{X^{-}} \gamma^\mu X^{-} A_\mu\,,  \label{eq:Z_Z2_odd}
 \end{align}
where $c_W=\cos\theta_W$ and $c_{2W}=\cos2\theta_W$ with Weinberg angle $\theta_W$; $X^{-}$ includes $X^-_{R}$ and $X^{-}_{L}$, $Q_{X}=-1$ is the $X^{-}$ electric charge, and $c^Z_{ij}$ show the FCNC effects and are defined as:
 \begin{equation}
 c^Z_{ij} = \left( O {\rm diag}(1,\, -1,\, 0) O^T \right)_{ij}=O_{i1}O_{j1} - O_{i2} O_{j2}\,.
 \end{equation}  
 From Eq.~(\ref{eq:Z_Z2_odd}), it can be seen that  the $Z$-boson coupling to the $Z_2$-odd particle is through axial-vector currents; therefore,  it will lead to the SD DM-nucleon elastic scattering. 
 
When $N_1(\chi_1)$  is  the DM candidate, in order to satisfy the DM direct detection constraints, we must  require $c^Z_{11}$ to be small enough.  From Eq.~(\ref{eq:O_mixing}), if we drop the $m_0$ and $y_X v_h/\sqrt{2}$ effects, it can be seen that $c^Z_{11}=0$. However, the case leads to $m_{N_1} =m_{N_2}$ and $c^Z_{12}=1$, where the DM-nucleon scattering occurs through $\chi_{1R} \chi_{2R} Z$ coupling (or $X^0 X^0 Z$ coupling). Hence, in addition to  the $c^Z_{11}$ magnitude, we have to take proper $m_0$ and $y_X v_h/\sqrt{2}$ in such a way that the mass splitting  between $N_1$ and $N_{2(3)}$ is large enough, so that  the DM scattering off the nucleon through $N_1N_{2,3} Z$ coupling can be kinematically suppressed.  If we take $m_N> m_X$,  the mass splitting between $N_1$ and $N_2$ can be found to be $\Delta m_{12} = e_X + e_\delta \approx e_N $, and  the $c^Z_{11}$ coefficient can be expressed as:
 \begin{equation}
 c^Z_{11} \approx \frac{2 }{{\cal N}^2_{1}}  \frac{e_X + m_0}{m_X}\,.
 \end{equation} 
If $N_3(\chi_3)$ is the DM candidate, because $c^Z_{33}$ is small, we will show that the SD DM-nucleon scattering cross section is under the current  PICO-60~\cite{Amole:2017dex}  and Xenon1T~\cite{Aprile:2019dbj} upper limits. 

\section{ Scalar potential and Yukawa sector}

\begin{figure}[phtb]
\begin{center}
\includegraphics[scale=0.65]{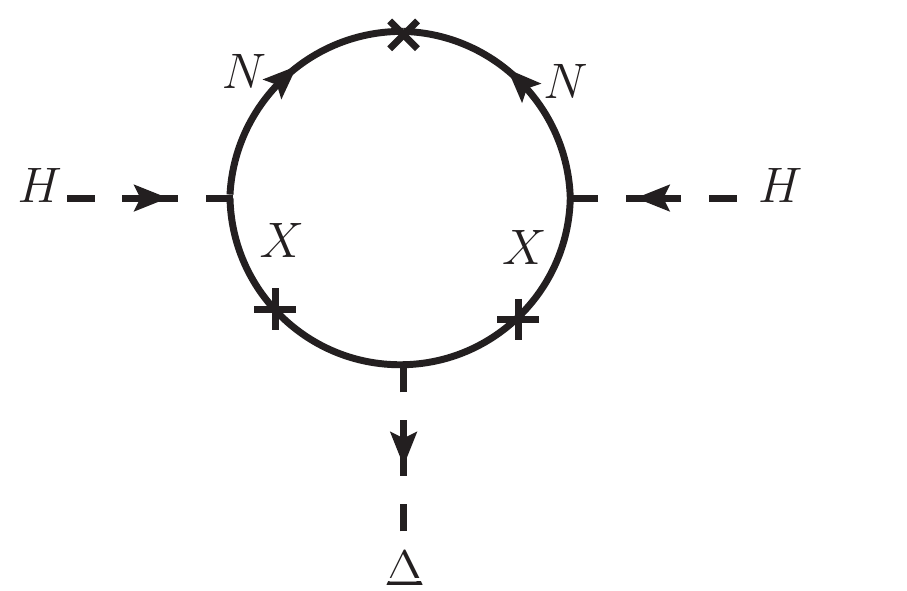}
 \caption{One-loop Feynman diagram for producing the $H^T i\tau_2 \Delta^\dag H$ term, where the cross symbols denote the mass insertions of the $N$ and $X$ leptons. }
\label{fig:mu_loop}
\end{center}
\end{figure}

\begin{figure}[phtb]
\begin{center}
\includegraphics[scale=0.6]{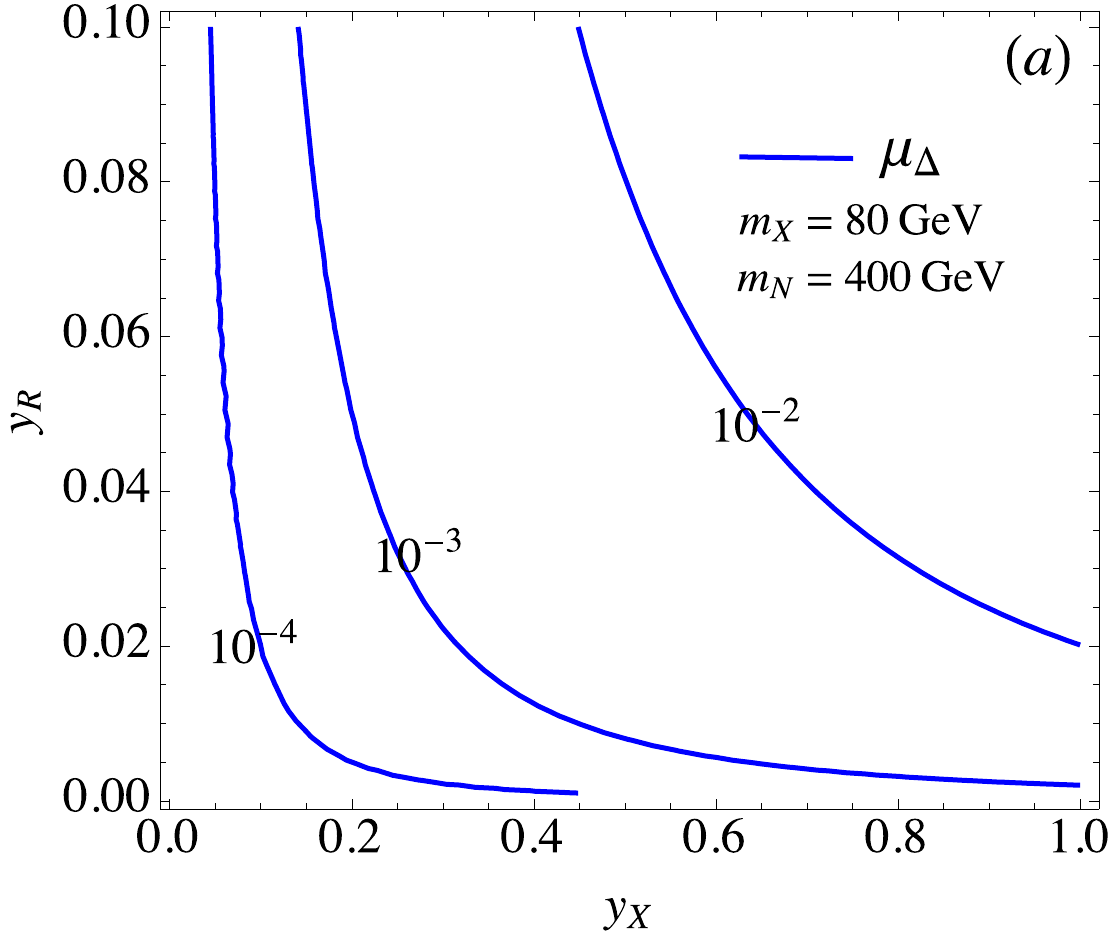}
\includegraphics[scale=0.6]{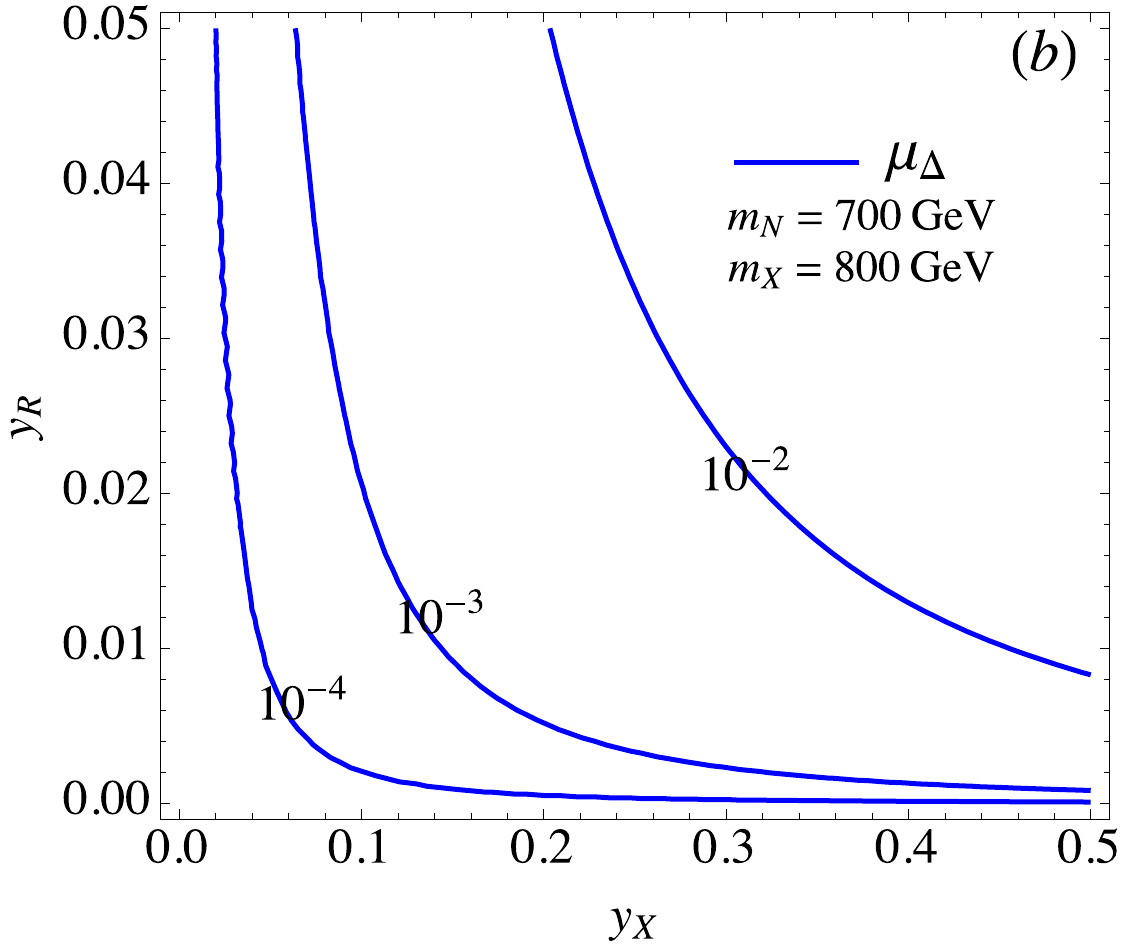}
 \caption{Contours of $\mu_\Delta$  as a function of $y_X$ and $y_R$ for (a)  $(m_X, m_N)=(80, 400)$ GeV and (b) $(m_X,m_N)=(800, 700)$ GeV. }
\label{fig:mu_D_ab}
\end{center}
\end{figure}

 According to  the convention in~\cite{Bonilla:2015eha,Primulando:2019evb},  we write the gauge invariant scalar potential as:
\begin{align}
V(H,\Delta) & = -\mu^2 H^\dagger H + \frac{\lambda}{4}(H^\dagger H)^2 + M^2_\Delta \, Tr(\Delta^\dag \Delta) + \lambda_1( H^\dag H) Tr (\Delta^\dag \Delta)\nonumber \\
& +  \lambda_2 \left( Tr(\Delta^\dag \Delta)\right)^2  + \lambda_3 Tr(\Delta^\dag \Delta)^2 + \lambda_4 H^\dag \Delta \Delta^\dag H \,,
\label{eq:SV}
\end{align}
where we take $\mu^2, \lambda >0$ for the purpose of spontaneously breaking the electroweak gauge symmetry. It can be seen that due to the lepton-number conservation, the dimension-3 $H^T i \tau_2 \Delta^\dag H$   term is suppressed at the tree level. Without this term, the Higgs triplet cannot obtain a VEV and the SM neutrinos are still massless. In order to generate the finite dimension-3 term, we require that the right-handed $Z_2$-odd lepton doublet only couples to the Higgs triplet by assigning  proper lepton numbers to $X_R$ and $X_L$, which are shown in Table.~\ref{tab:rep_charge. }
 Thus, the finite  $H^T i \tau_2 \Delta^\dag H$ term can be dynamically generated through a fermion loop,  in which the $m_X$ lepton number violating effect is involved. The associated Feynman diagram is shown in Fig.~\ref{fig:mu_loop}, where the cross symbols denote the  mass insertions of the $N$ and $X$ leptons. Thus,  the resulting dimension-3 term can be expressed as:
\begin{equation}
V(H,\Delta)_{\rm dim-3}  =\mu_\Delta H^T i\tau_2 \Delta^\dag H+ H.c.\,, \label{eq:V_dim3}
\end{equation}
where the $\mu_\Delta$ coefficient is obtained as:
\begin{align}
\mu_\Delta & = \frac{y^2_X y_R m_N}{8 \pi^2} I_\Delta \left( \frac{m^2_X}{m^2_N}\right)\,,  \label{eq:mu_Del}\\
I_\Delta(x) & =  - \frac{x}{1-x}- \frac{x\ln x}{(1-x)^2 }\,. \nonumber 
\end{align}
For clarity, we show the contours of $\mu_\Delta$ as a function of $y_X$ and $y_R$ in  Fig.~\ref{fig:mu_D_ab}(a), where $m_X=80$ GeV and $m_N=400$ GeV are used. Clearly, we can easily obtain $\mu_\Delta < 10^{-2}$ GeV without extremely fine-tuning the $y_R$ and $y_X$ parameters. For comparison, we make a contour plot with $m_X=800$ GeV and $m_N=700$ GeV in Fig.~\ref{fig:mu_D_ab}(b). We will show that the former and latter plots correspond to  the cases for which $\chi_1$ and $\chi_3$ are the DM candidates, respectively. 

Combining Eqs.~(\ref{eq:SV}) and (\ref{eq:V_dim3}), the minimum of  the scalar potential can be obtained through $\partial V/\partial v_h =0$ and $\partial V/\partial v_\Delta=0$, and the minimum conditions can be written as:
\begin{align}
& - \mu^2 + \frac{\lambda}{4} v^2_h + \frac{\lambda_1 + \lambda_4}{2} v^2_\Delta = \sqrt{2} \mu_\Delta v_\Delta \,, \nonumber \\
& \left( M^2_\Delta + \frac{\lambda_1 + \lambda_4}{2}v^2_h + ( \lambda_2 + \lambda_3) v^2_\Delta \right) v_\Delta = \frac{\mu_\Delta v^2_h}{\sqrt{2}}\,. \label{eq:mini_V}
\end{align}
Because we focus on the case of $\mu_\Delta <  10^{-2}$ GeV, i.e., $v_\Delta\ll 1$ GeV,  when we neglect the small $\mu_\Delta v_\Delta$ and $v^2_\Delta$ effects, the VEVs of $\Phi^0$ and $\Delta^0$ can be respectively obtained as $v_h\approx \sqrt{4 \mu^2/\lambda}$ and 
\begin{align}
v_\Delta \approx  \frac{\mu_\Delta v^2_h}{\sqrt{2}[M^2_\Delta +v^2_h(\lambda_1 + \lambda_4)/2] }\,. \label{eq:V_v_D}
\end{align}
To obtain $v_\Delta >0$, we  require $\mu_\Delta >0$, which is equivalent to  $y_R>0$. Because of  $v_\Delta \ll$1 GeV, the influence on the electroweak $\rho$-parameter  can be neglected. We note that in addition to  $\mu_\Delta$ and $M_\Delta$, $v_\Delta$ also depends on the $\lambda_{1,4}$ parameters. We will discuss the correlation between $v_\Delta$ and $\lambda_{1,4}$ when the constraints on the $\lambda_{1,4}$ parameters are studied. 

 The vacuum stability of  scalar potential has been studied in the literature~\cite{Arhrib:2011uy,Kannike:2012pe,Bonilla:2015eha}. Following the results in~\cite{Bonilla:2015eha}, the conditions for the scalar potential  bounded from below in our notations can be written as:
 \begin{align}
& \lambda >0\,, ~ \lambda_2 + \lambda_3 >0 \,,~ 2 \lambda_2 + \lambda_3 >0 \,, \nonumber \\
& \lambda_1 + \sqrt{\lambda(\lambda_2 + \lambda_3)}>0\,, ~ \lambda_1 + \lambda_4 + \sqrt{\lambda(\lambda_2 + \lambda_3)}>0\,, 
 \end{align}
 and
 \begin{align}
 & |\lambda_4| \sqrt{\lambda_2 + \lambda_3} - \lambda_3 \sqrt{2} > 0  ~{\rm or} \nonumber \\
  &  2\lambda_1 + \lambda_4 + \sqrt{\left(2 \lambda \lambda_3 -\lambda^2_4 \right)\left( 2 \lambda_2/\lambda_3 +1 \right)} >0\,.
 \end{align}
 For the sake of satisfying perturbativity, we take $\lambda, |\lambda_i| \leq 4 \pi$ before we find the stricter constraints.

\subsection{Scalar mass spectra and  scalar couplings}

In addition to the SM-like Higgs boson, the type-II seesaw model has two doubly and two singly charged Higgs, and one CP-even and one CP-odd scalar. The scalar mass spectra and the scalar-scalar couplings can be obtained from the scalar potential. Since the doubly charged Higgs does not mix with the other scalars, its mass can be easily obtained  as:
\begin{align}
m^2_{H^{\pm\pm}} &= M^2_\Delta + \frac{ \lambda_2}{2} v^2_h + (\lambda_2 + \lambda_3) v^2_\Delta \nonumber \\
  &=  \frac{\mu_\Delta v^2_h}{\sqrt{2} v_\Delta } - \frac{\lambda_4}{2} v^2_h\,,
\end{align}
where the minimal conditions in Eq.~(\ref{eq:mini_V}) have been applied in the second line.  The mass-square matrices for $(G^-, \Delta^{-})$, $(G^0,\, Im\Delta^0)$, and $(Re\Phi^0, Re\Delta^0)$ can be respectively derived as:
\begin{equation}
( G^-, \Delta^{-})
\left(
\begin{array}{cc}
 \sqrt{2} v_\Delta \left( -\frac{\lambda_4 v_\Delta}{2\sqrt{2}} + \mu_\Delta \right) &   -v_h\left( -\frac{\lambda_4 v_\Delta}{2\sqrt{2}} +\mu_\Delta\right)    \\
  -v_h \left(-\frac{\lambda_4 v_\Delta}{2\sqrt{2}} +\mu_\Delta \right)  &     \frac{v^2_h}{\sqrt{2} v_\Delta} \left( -\frac{\lambda_4 v_\Delta}{2\sqrt{2}} + \mu_\Delta \right)
\end{array}
\right) \left(
\begin{array}{c}
  G^+      \\
  \Delta^+    
\end{array}
\right)\,, \label{eq:G+D+}
\end{equation}
\begin{equation}
\frac{1}{2} ( G^0, Im\Delta^{0})
\left(
\begin{array}{cc}
 2\sqrt{2} \mu_\Delta v_\Delta &   -\sqrt{2} \mu_\Delta v_h   \\
    -\sqrt{2} \mu_\Delta v_h  &    \mu_\Delta v^2_h/(\sqrt{2} v_\Delta)
    \end{array}
\right) \left(
\begin{array}{c}
  G^0      \\
  Im\Delta^0    
\end{array}
\right)\,, \label{eq:G0ImD0}
\end{equation}
\begin{equation}
\frac{1}{2} ( Re\Phi^0, Re\Delta^0 )
\left(
\begin{array}{cc}
\lambda v^2_h /2 &   (\lambda_1 + \lambda_4) v_h v_\Delta -\sqrt{2} v_h \mu_\Delta   \\
    (\lambda_1 + \lambda_4) v_h v_\Delta -\sqrt{2} v_h \mu_\Delta  &    \frac{\mu_\Delta v^2_h}{\sqrt{2} v_\Delta} + 2 v^2_\Delta (\lambda_2 + \lambda_3)
    \end{array}
\right) \left(
\begin{array}{c}
 Re\Phi^0   \\
Re\Delta^0  
\end{array}
\right)\,. \label{eq:hReD0}
\end{equation}
It can be easily verified that the determinants of the mass-square matrices in Eqs.~(\ref{eq:G+D+}) and (\ref{eq:G0ImD0}) vanish; that is, there exists a massless boson, which corresponds to the Goldstone boson, in each matrix.  The detailed eigenvalues of the mass-square matrices and the associated mixing angles are shown in Appendix~\ref{ap:mixing}. 

Because the off-diagonal elements in Eq.~(\ref{eq:hReD0}) are much smaller than $v^2_h \mu_\Delta/(\sqrt{2} v_\Delta)$, the mixing effect between $Re\Phi^0$ and $Re\Delta^0$ can be approximately neglected if we only concentrate on the scalar spectrum. Thus, from the mass-square matrices,  the mass squares for the physical bosons, such as  the charged scalar $H^\pm$, the CP-odd pseudoscalar $A^0$, and the two CP-even $H^0$ and $h$, can be written as:
 \begin{align}
 m^2_{H^\pm} & = \left(  \frac{v^2_h}{\sqrt{2} v_\Delta}+\sqrt{2} v_\Delta \right) \left( -\frac{\lambda_4 v_\Delta}{2\sqrt{2}} + \mu_\Delta\right)\,, \nonumber \\
 m^2_{A^0} &=  \mu_\Delta \left( \frac{v^2_h }{\sqrt{2} v_\Delta} + 2 \sqrt{2} v_\Delta \right) \,, \nonumber \\
 m^2_{H^0} & \approx m^2_{A^0} - 2 \sqrt{2} v_\Delta \mu_\Delta+ 2 v^2_\Delta (\lambda_2 + \lambda_3)\,,
 \end{align}
and $m^2_h\approx \lambda v^2_h/2$, respectively, where $h$ is the SM-like Higgs boson. If we ignore the small $v_\Delta$ and $\mu_\Delta$ effects, it can be found that:
 \begin{align}
 m_{H^0}\approx m_{A^0} & \approx \frac{v^2_h \mu_\Delta }{\sqrt{2} v_\Delta} \, , \nonumber \\
 m^2_{H^{\pm\pm}}-m^2_{H^\pm} & \approx - \frac{\lambda_4 v^2_h}{4}\,, \nonumber \\ 
 m^2_{H^\pm}-m^2_{H^0(A^0)} & \approx -\frac{\lambda_4 v^2_h}{4}, \label{eq:m_scalar}
  \end{align}
 where the mass splittings in the Higgs triplet components can be constrained by the electroweak oblique parameters~\cite{Peskin:1991sw}. From Eq.~(\ref{eq:m_scalar}), we have the mass ordering $m_{H^0(A^0)}> m_{H^\pm}> m_{H^{\pm\pm}}$ when $\lambda_4 >0$; however, the order is reversed when $\lambda_4 < 0$.  

In order to study the Higgs precision measurement  constraint,  we write the Higgs trilinear  couplings to the triplet scalars  as:
 \begin{align}
 - {\cal L}_V& \supset \lambda_1 v_h hH^{--}H^{++} + \left(\lambda_1 + \frac{\lambda_4}{2} \right) v_h h H^{-} H^{+} \nonumber \\
 & + \frac{1}{2} \left(\lambda_1 + \lambda_4 \right) v_h h\left(H^{0} H^0 + A^{0} A^{0}\right) + \frac{1}{2} \left((\lambda_1 + \lambda_4)v_\Delta -\sqrt{2} \mu_\Delta \right) hhH^0\,. \label{eq:hTT}
 \end{align}
The Higgs triplet couplings to the gauge bosons can be obtained from the kinetic terms, written as:
 \begin{equation}
 {\cal L}_{\rm kin}  = Tr[ (D_\mu \Delta)^\dag (D^\mu \Delta)] \,, \label{eq:kin}
 \end{equation}
where the covariant derivative of the Higgs triplet is given as:
  \begin{equation}
  D_\mu \Delta = \partial_\mu \Delta  +  i\, g  \left[  \frac{\boldsymbol{\tau}}{2} \cdot {\bf W}_\mu, \, \Delta \right] + i\, g' B_\mu \Delta\,.  \label{eq:cov}
  \end{equation}
The detailed trilinear couplings to gauge bosons can be found in  Appendix~\ref{ap:gauge}. 

\subsection {Yukawa couplings and neutrino masses}

Using the heavy Majorana flavor mixing matrix in Eq.~(\ref{eq:O_mixing}), the scalar Yukawa couplings to the heavy $Z_2$-odd fermions can be straightforwardly obtained  as: 
\begin{align}
 -{\cal L}^{\rm odd}_Y & \supset \frac{1}{2}   \left(\sqrt{2} y_{R} O_{i1} O_{j1}\right) \left( \bar\chi_i  \chi_j  H^0 + i \, \bar\chi_i \gamma_5 \chi_j A^0 \right) + \frac{1}{2} \frac{y_X c^h_{ij}}{\sqrt{2}} \bar\chi_i \chi_j  \, h\nonumber \\
 &-\left[ \sqrt{2} y_R O_{i1} \bar\chi_i  X^-_R  H^+  + \frac{1}{2} (2y_R) X^{-T}_{R} C X^-_R H^{++}   + H.c. \right]\,, \label{eq:h_chi_chi}
 \end{align}
with $c^h_{ij} =O_{i2}O_{j3} + O_{i3} O_{j2}$. 

In addition to the SM lepton coupling to the Higgs doublet, the SM left-handed leptons also couple to the Higgs triplet. When we derive the lepton couplings to the Higgs triplet in physical states, we have to simultaneously  consider the ${\bf y}^\ell$ and ${\bf y}^\ell_\Delta$ terms in Eq.~(\ref{eq:Yu}). In terms of the components of the Higgs doublet and triplet, the relevant  Yukawa couplings of $Z_2$-even leptons are written as:
\begin{align}
- {\cal L}^{\rm even}_Y & \supset  \bar \ell_{L} {\bf y}^\ell\ell_{R} \frac{v+h}{\sqrt{2}}  + \nu^T_{L} C {\bf y}^\ell_\Delta \nu_L \frac{v_\Delta + H^0 + i A^0}{\sqrt{2}}\nonumber \\
&  -\sqrt{2} \nu^T {\bf y}^\ell_\Delta \ell_L H^+ - \ell^T_L C {\bf y}^\ell_\Delta \ell_L H^{++} + H.c.\,,  \label{eq:SMlepton}
\end{align}
where we have neglected the small  $\mu_\Delta$ and $v_\Delta$ effects.  To diagonalize the charged lepton and Majorana neutrino mass matrices, we introduce the unitary matrices for which the transformations are defined as: $\nu_L \to U^\nu_\ell \nu_L$ and $\ell_{L(R)} \to U^{\ell}_{L(R)} \ell_{L(R)}$. If we define ${\bf h}^\ell\equiv U^{\ell*}_L {\bf y}^\ell_\Delta U^{\ell\dag}_L$ and the Pontecorvo-Maki-Nakagawa-Sakata (PMNS) matrix as $U^\dag_{\rm PMNS} = U^\nu_L U^{\ell\dag}_{L}$, Eq.~(\ref{eq:SMlepton}) with respect to the lepton physical states can be written as:
 \begin{align}
 -{\cal L}^{\rm even}_Y & \supset  \bar \ell_L {\bf m}^{\rm dia}_\ell  \ell_R + \bar \ell_L \left(\frac{{\bf m}^{\rm dia}_\ell }{v}\right) \ell_R h + \frac{1}{2} \nu^T_L C {\bf m}^{\rm dia}_\nu \nu_L  +  \frac{1}{2} \nu^T_L C \left(\frac{ {\bf m}^{\rm dia}_{\nu}}{v_\Delta} \right)  \nu_L \left( H^0 + i A^0 \right) \nonumber \\
& - \sqrt{2} \nu^T C U^T_{\rm PMNS} {\bf h}^\ell \ell_L H^+ - \frac{1}{2} \ell^T_L C (2 {\bf h}^\ell) \ell_L H^{++} + H.c.\,, \label{eq:Yukawa}
 \end{align}
where the diagonal mass matrices are given as:
 \begin{align}
 {\bf m}^{\rm dia}_\ell&={\rm diag}(m_e, m_\mu, m_\tau) = U^\ell_L \frac{{\bf y}^\ell v_h}{\sqrt{2}} U^{\ell\dag}_R\,, \nonumber \\
   {\bf m}^{\rm dia}_\nu &={\rm diag}(m_1, m_2,m_3)= U^T_{\rm PMNS} (\sqrt{2} v_\Delta {\bf h}^\ell)  U_{\rm PMNS}\,. \label{eq:lepton_mass}
 \end{align}
In order to explain the neutrino data, it is necessary to have $v_\Delta {\bf h}^\ell \sim 10^{-2}$ eV. It will be shown that the partial decay widths of the Higgs triplet scalars decaying to leptons are sensitive to $v_\Delta$, which is dictated by the parameters, such as $M_\Delta$, $\lambda_1$, $\lambda_4$, and $\mu_\Delta$. 

\section{The Constraints}

In this section, we discuss the constraints, such as the neutrino mass data, the observed DM relic density, the DM direct detections, the T-parameter, and the Higgs to diphoton precision measurement. It will be found that the $\chi_1$-DM candidate will be excluded by the upper limits of the DM-nucleon scattering cross sections. Since the cross section upper limit of the SD DM-neutron scattering in Xenon1T~\cite{Aprile:2019dbj} is smaller than that of the SD DM-proton scattering  in PICO-60~\cite{Amole:2017dex}, we take the Xenon1T data as the upper limit of the SD DM-nucleon scattering cross section and use it to bound the parameters.

\subsection{ Constraint from the neutrino data}

 From Eq.~(\ref{eq:lepton_mass}), the matrix elements of ${\bf h}^{\ell}$ can be written as:
  \begin{equation}
  {\bf h}^\ell_{ij} = \frac{1}{\sqrt{2} v_\Delta} \left( U^*_{\rm PMNS} \right)_{ik} m_{\nu k} \left(U^*_{\rm PMNS} \right)_{jk} \,,
  \end{equation}
where the sum in  $k$ for all active light neutrinos is indicated. It can be seen that the ${\bf h}^{\ell}_{ij}$ magnitudes strongly depend on the $v_\Delta$ value. 
Using the PMNS matrix parametrized as~\cite{PDG}:
 \begin{align}
\label{MNS}
U_{\rm PMNS} & = \begin{pmatrix} 
c_{12} c_{13} & s_{12} c_{13} & s_{13} e^{-i \delta} \\
-s_{12} c_{23} - c_{12} s_{23} s_{13} e^{i \delta} & c_{12} c_{23} -s_{12} s_{23} s_{13} e^{i \delta} & s_{23} c_{13} \\
s_{12} s_{23} - c_{12} c_{23} s_{13} e^{i \delta} & -c_{12} s_{23} - s_{12} c_{23} s_{13} e^{i \delta} & c_{23} c_{13} 
\end{pmatrix} \nonumber \\
& \times \text{diag}(1, e^{i \alpha_{21}/2}, e^{i\alpha_{31}/2}) \equiv U_\nu \times \text{diag}(1, e^{i \alpha_{21}/2}, e^{i\alpha_{31}/2}) \,,
\end{align}
where  $s_{ij} \equiv \sin \theta_{ij}$, $c_{ij} \equiv \cos \theta_{ij}$; $\delta$ is the Dirac CP violating phase, and $\alpha_{21, 31}$ are  Majorana CP violating phases, and the experimental data through the neutrino oscillation measurements can be given as~\cite{PDG}:
 \begin{align}
\Delta m^2_{21} & = (7.53\pm0.18)\times 10^{-5} ~{\rm eV^2}\,, ~ \sin^2\theta_{12}=0.307 \pm 0.013 \,, \nonumber \\
\Delta m^2_{32} &= ( 2.51 \pm 0.05, \, -2.56 \pm 0.04)\times 10^{-3} ~{\rm eV^2} ~ (\rm {NO,\, IO})\,, \nonumber \\
\sin^2\theta_{23} &= ( 0.597^{+0.024}_{-0.030},\, 0.592^{+0.023}_{-0.030} )  ~ (\rm {NO,\, IO})\,, \nonumber \\
\sin^2\theta_{13} & = (2.12 \pm 0.08)\times 10^{-2}\,,
 \end{align}
where $\Delta m^2_{ij}\equiv m^2_i - m^2_j$, and $\Delta m^2_{32}>0$ and $\Delta m^2_{32} < 0$ denote the normal ordering (NO) and inverted ordering (IO), respectively.  The uncertain sign in $m^2_{32}$ originates from the undetermined neutrino mass ordering. Since the neutrino oscillation experiments cannot detect the Majorana CP phases, for simplicity, we take $\alpha_{31, 32}=0$ in the following numerical estimates. 

According to the recent results obtained by a global fit analysis,  the central values of $\theta_{ij}$, $\delta$, and $\Delta m^2_{ij}$  are given as~\cite{deSalas:2017kay}:
\begin{align}
 {\rm NO} :\, &  \theta_{12} = 34.5^{\circ}\,, \,\theta_{23}=47.7^{\circ}\,, \,\theta_{13} =8.45^{\circ}\,, \, \delta=218^{\circ}\,,\nonumber \\
&  \Delta m^2_{21} =7.55 \times 10^{-5}\, {\rm eV^2}\,, \, \Delta m^2_{31} = 2.50\times 10^{-3}\, {\rm eV^2}\,, \nonumber \\
 {\rm IO}:\, &  \theta_{12} = 34.5^{\circ}\,,\,\theta_{23}=47.9^{\circ}\,,\,\theta_{13} =8.53^{\circ}\,, \, \delta=281^{\circ}\,, \nonumber \\
 &  \Delta m^2_{21} =7.55 \times 10^{-5}\, {\rm eV^2}\,, \, \Delta m^2_{31} =- 2.42\times 10^{-3}\, {\rm eV^2}\,, 
  \end{align}
  where $m_{1(3)}=0$ for NO (IO) is taken. Using these results, the corresponding ${\bf h}^\ell_{ij}$ Yukawa matrix element values are shown in Table~\ref{tab:V_hell_ij}, where the values are in units of $10^{-3} {\rm eV}/(2 v_\Delta)$. When $v_\Delta$ is fixed, the  ${\bf h}^\ell_{ij}$ values then can be determined. With $v_\Delta\sim 10^{-4}$ GeV, it can be seen that the ${\bf h}^\ell_{ij}$ magnitudes can be in the range of  $ \sim (0.1,\, 1)\times 10^{-7}$. Due to the small Yukawa couplings, it can be expected that the lepton-flavor violating effects will be small. 
  
  \begin{table}[htp]
\caption{ The ${\bf h}^\ell_{ij}$ Yukawa matrix element values (in units of $10^{-3} {\rm eV}/2 v_\Delta)$, where the central values obtained by a global fit analysis in~\cite{deSalas:2017kay} are applied. }
\begin{center}
\begin{tabular}{c|cccccc} \hline \hline 
 & ${\bf h}^\ell_{11}$  &  ${\bf h}^\ell_{12}$  &  ${\bf h}^\ell_{13}$  &   ${\bf h}^\ell_{22}$   &  ${\bf h}^\ell_{23}$  &  ${\bf h}^\ell_{33}$  \\  \hline
NO ($10^{-3} {\rm eV}/2v_\Delta)$ & $3.17 e^{i 0.34}$ & $3.73 e^{-i 1.93}$ & $7.33 e^{-i 2.69}$ & $29.91 e^{-i 0.013}$ & $21.38 $ & $24.93e^{i0.014}$ \\ \hline 
IO ($10^{-3} {\rm eV}/2v_\Delta)$  &  $47.60$ & $5.26 e^{-i 1.72}$ & $4.84 e^{-i 1.81}$ & $21.44 e^{i 0.008}$ & $24.84 e^{i3.13}$ & $26.51 e^{i0.009}$ \\ \hline \hline
\end{tabular}
\end{center}
\label{tab:V_hell_ij}
\end{table}%

\subsection{ Constraints from the DM relic density and the DM direct detections}

In this model, the DM candidate could be an $\chi_1$ or $\chi_3$ Majorana fermion.  Regardless of which one is the DM candidate, it is necessary to examine that whether the involved couplings can produce the current correct DM relic abundance ($\Omega_{\rm DM} h^2$), which is observed as in~\cite{Ade:2015xua}:
 \begin{equation}
 \Omega^{\rm obs}_{\rm DM} h^2 = 0.1199 \pm 0.0022\,.
 \end{equation}
Since the DM relic density is inversely proportional to the product of the DM annihilation cross section and its velocity, i.e. $<\sigma v>$,  in addition to  the thermal effects in the early time of the universe, we have to consider the DM annihilation and  co-annihilation to the SM particles in the final states. In order to deal with the thermal effects and to calculate the $Z_2$-odd particle annihilation processes, we employ micrOmegas~\cite{Belanger:2008sj} with a choice of a unitary gauge.
 For clarity,  we separately discuss the  situations of $\chi_1$- and $\chi_3$-DM in the following analysis.  
  Although DM couples to the Higgs triplet, since we take the associated $y_R$ parameter to be $\lesssim O(10^{-2})$, the effects indeed are small. Thus, we neglect the Higgs triplet contributions to the DM relic density.

\begin{figure}[phtb]
\includegraphics[scale=0.6]{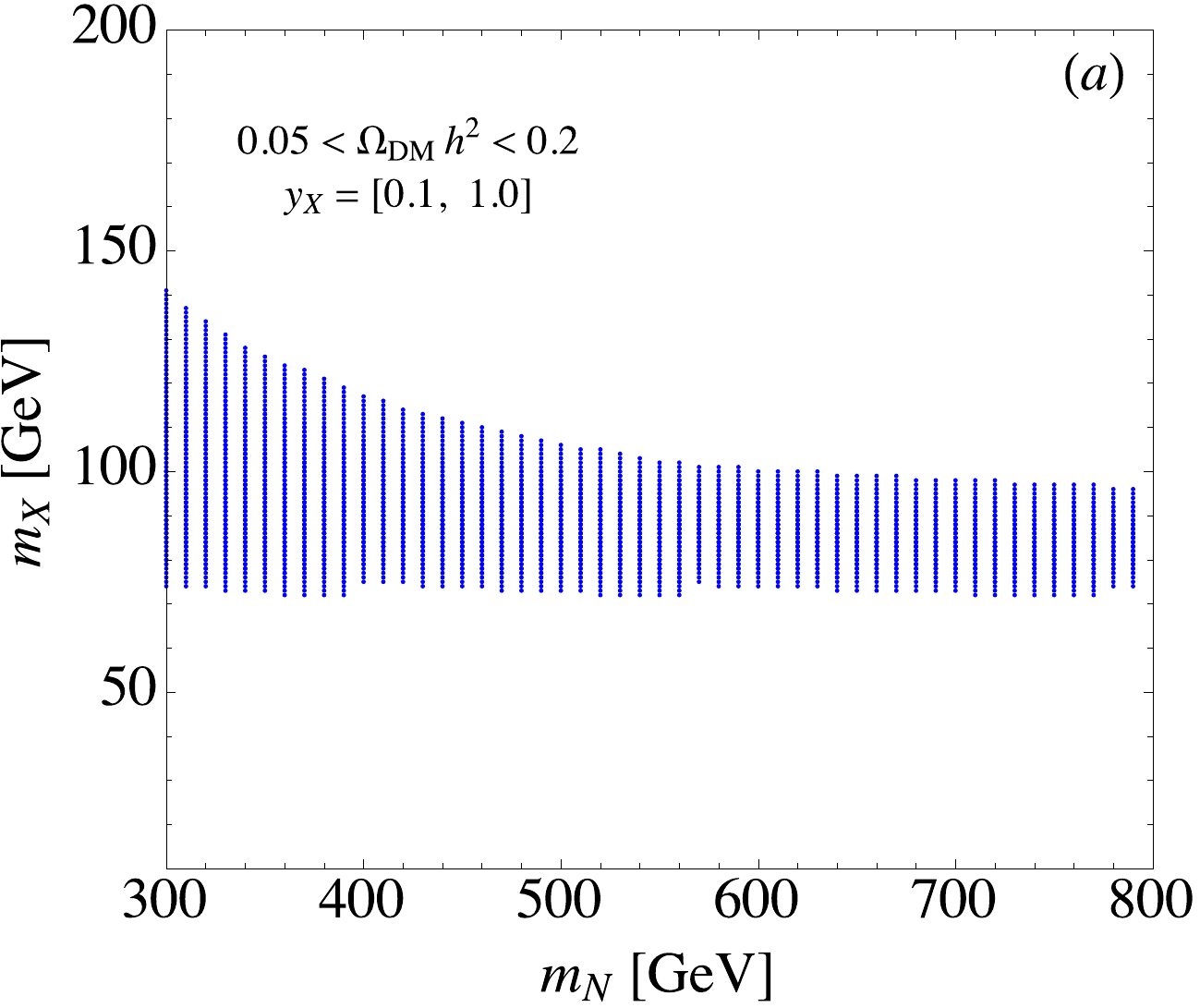}
\includegraphics[scale=0.6]{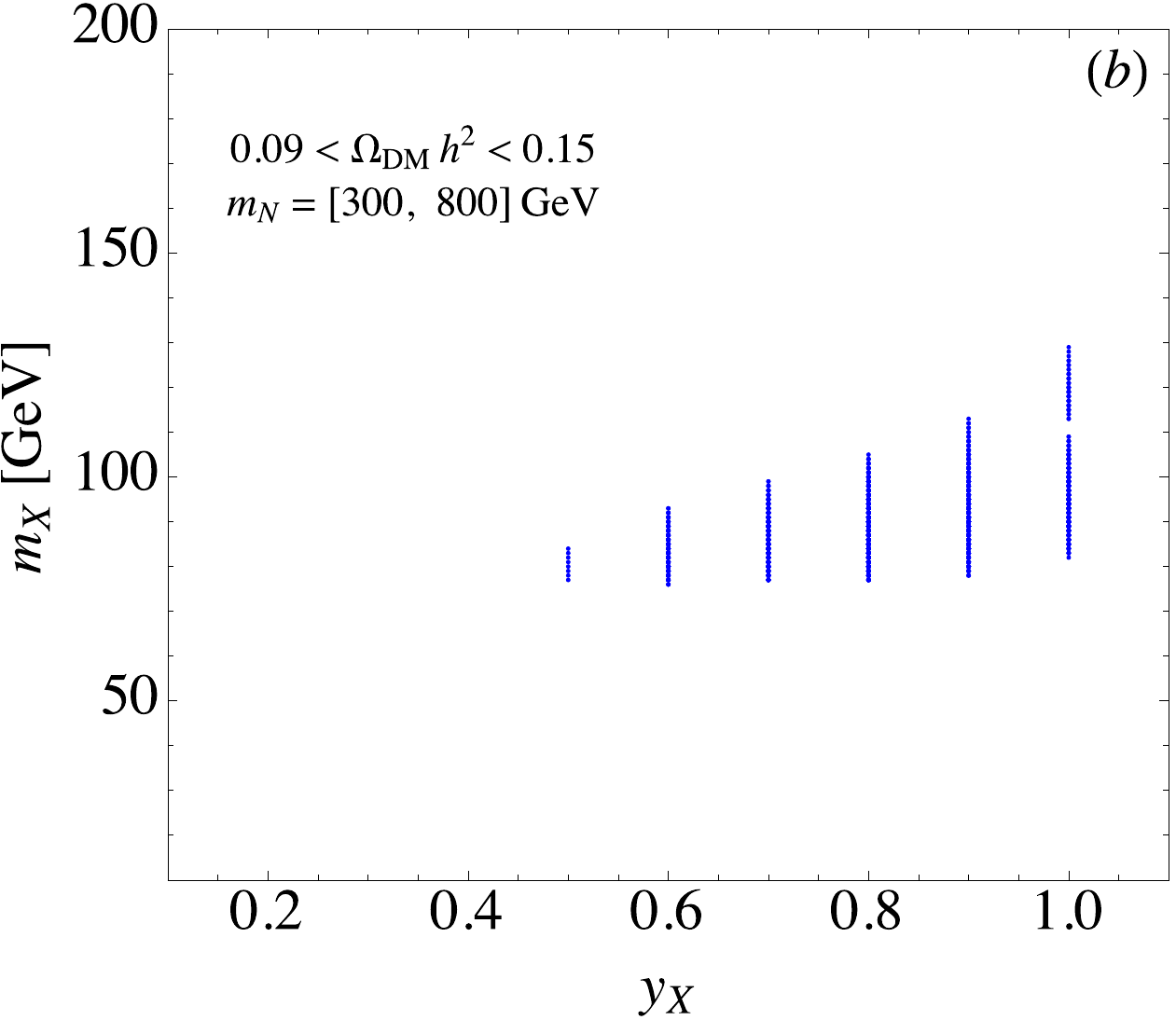}
 \caption{ Allowed parameter space, which can produce the DM relic density in the region of $0.09<\Omega_{\rm DM}h^2 <0.15$.   }
\label{fig:allowed_X}
\end{figure}

\begin{figure}[phtb]
\includegraphics[scale=0.65]{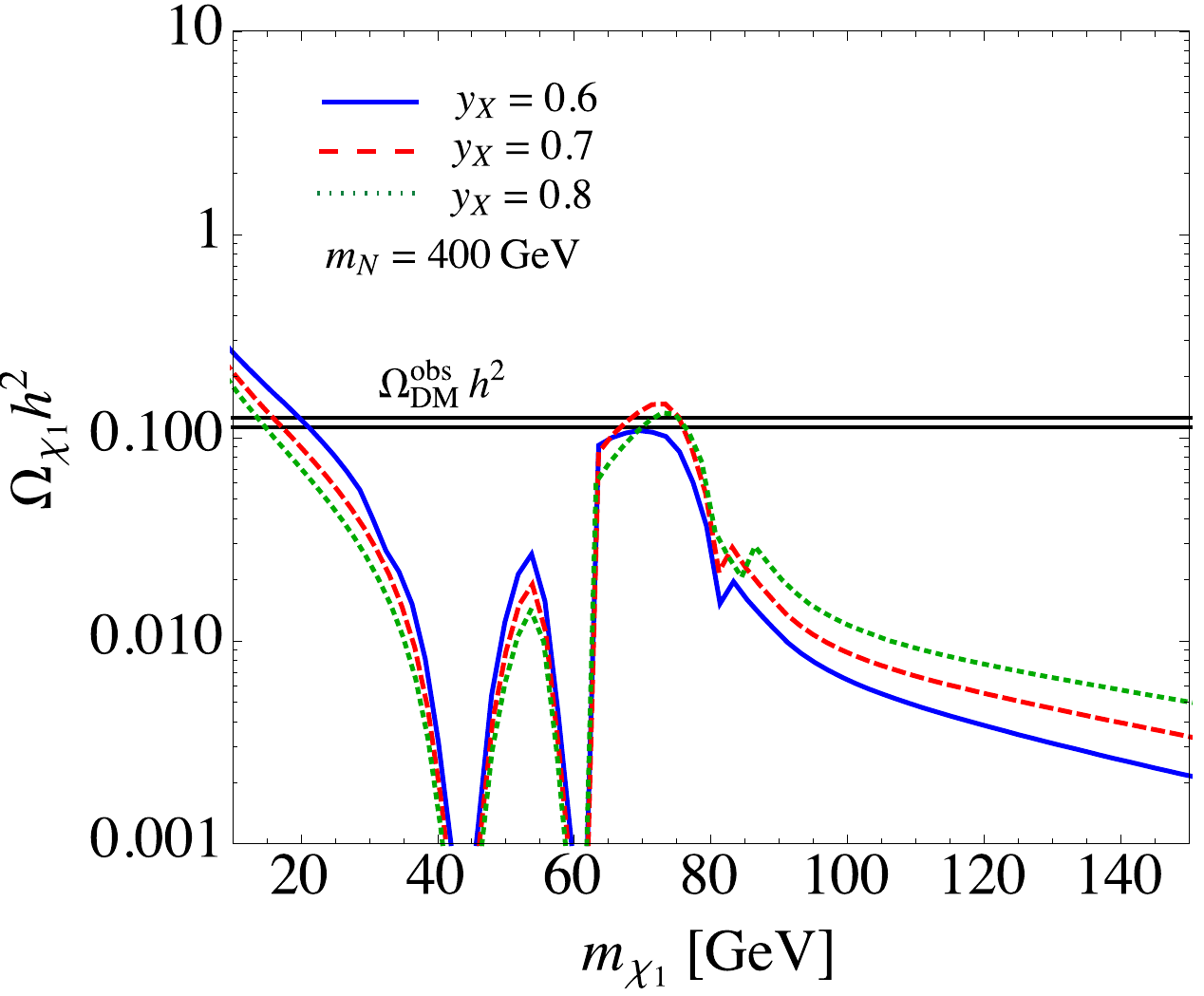}
 \caption{ $\chi_1$-DM relic abundance as a function of $m_{\chi_1}$ for $y_X=0.6$ (solid), $y_X=0.7$ (dashed), and $y_X=0.8$ (dotted), where $m_N=400$ GeV is fixed, and the horizontal lines denote $\Omega^{\rm obs}_{\rm DM} h^2$ with $3\sigma$ errors.  }
\label{fig:Omega_X}
\end{figure}

When the DM candidate is the $\chi_1$ Majorana particle, because its origin is the $SU(2)$ lepton doublet,  and it has a large coupling to the SM gauge bosons,  we require that the DM mass satisfies $m_{\chi_1}> 45$ GeV due to  the invisible $Z$ decay constraint. To avoid obtaining too large of a DM annihilation rate,   the massive gauge boson pair production should be suppressed; that is, $\chi_1$ cannot be too heavy. In order to understand the correlation between $\Omega_{\rm DM} h^2$ and  the $m_{N,X}$ and $y_X$ parameters,  the scanned parameter regions are chosen as:
 \begin{equation}
 m_N=[300,800] \, {\rm GeV}\,, ~ m_{X}=[10, 150]\, {\rm GeV}\,, ~ y_{X}=[0.1,1.0]\,,
 \end{equation} 
 where  we require that  the resulting $\Omega_{\rm DM}$ satisfies $0.09< \Omega_{\rm DM}h^2 < 0.15$. We note that, in order to get more sampling points for illustration, the region of $\Omega_{\rm DM} h^2$ is taken slightly  wider than the observed $\Omega_{\rm DM} h^2$.  We show the allowed parameter space as a function of $m_N$ and $m_X$ and as a function of $y_X$ and $m_X$ in Fig.~\ref{fig:allowed_X} (a) and (b), respectively. It can be seen that only $m_X \sim 90$ GeV and $y_X > 0.5$ can fit the condition of $0.09< \Omega_{\rm DM}h^2 < 0.15$.  Based on the results, we show  $\Omega_{\rm DM} h^2$ as a function of $m_{\chi_1}$ in Fig.~\ref{fig:Omega_X}, where $m_N=400$ GeV is used, and the solid, dashed, and dotted lines denote the results of $y_X=0.6$, $0.7$, and $0.8$, respectively. Two dips denote $m_{\chi_1} \sim m_Z/2$ and $m_{\chi_1}\sim m_h/2$. It can be found that $m_{\chi_1} \sim 70$ GeV with $y_{X}\sim 0.7$ can fit the observed $\Omega_{\rm DM}h^2$ and can escape the constraint from the invisible $Z$ decay. Hence, without considering the DM direct detection constraints, the neutral component of the $Z_2$-odd lepton doublet could be the DM candidate in this model.

\begin{figure}[phtb]
\includegraphics[scale=0.6]{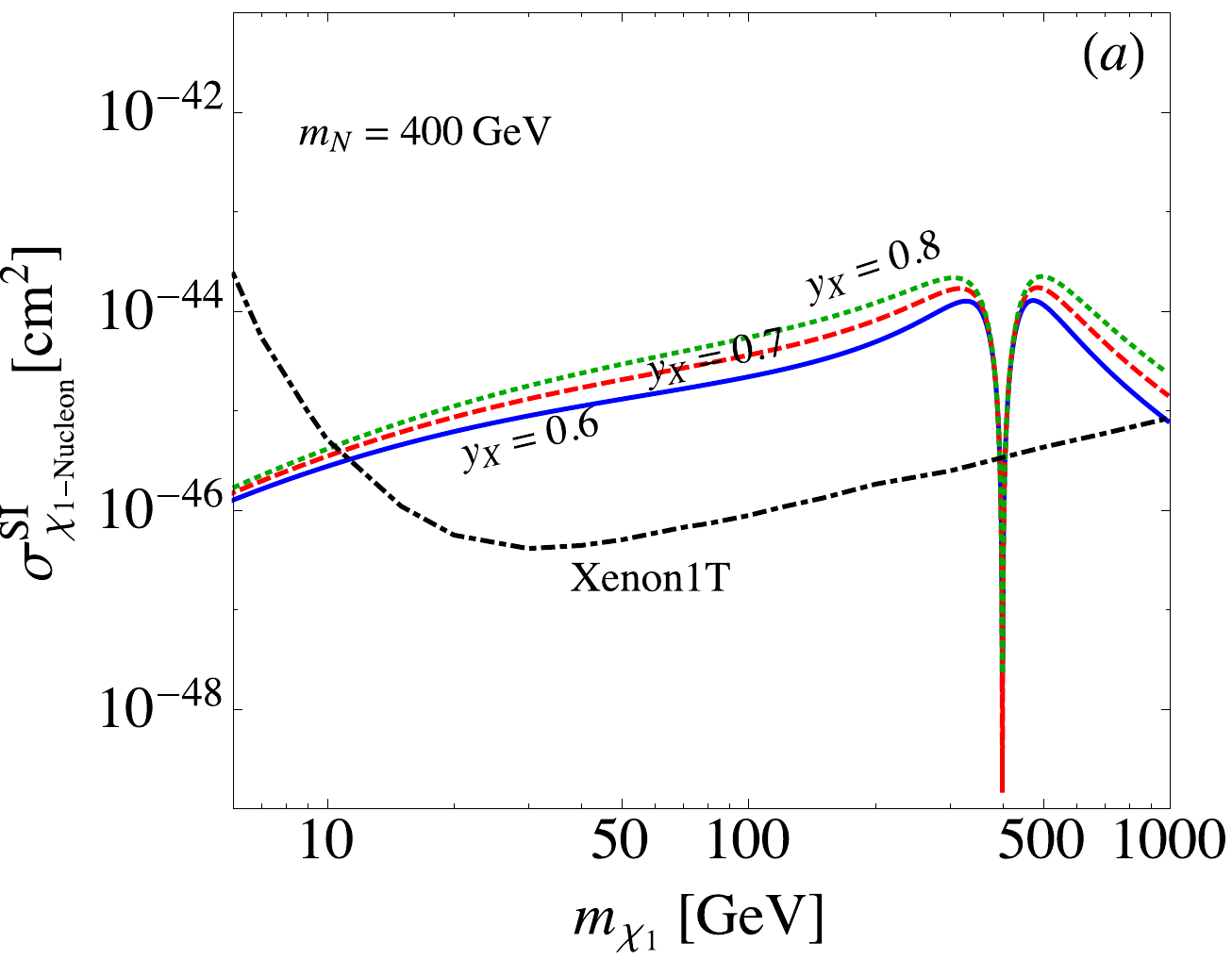}
\includegraphics[scale=0.6]{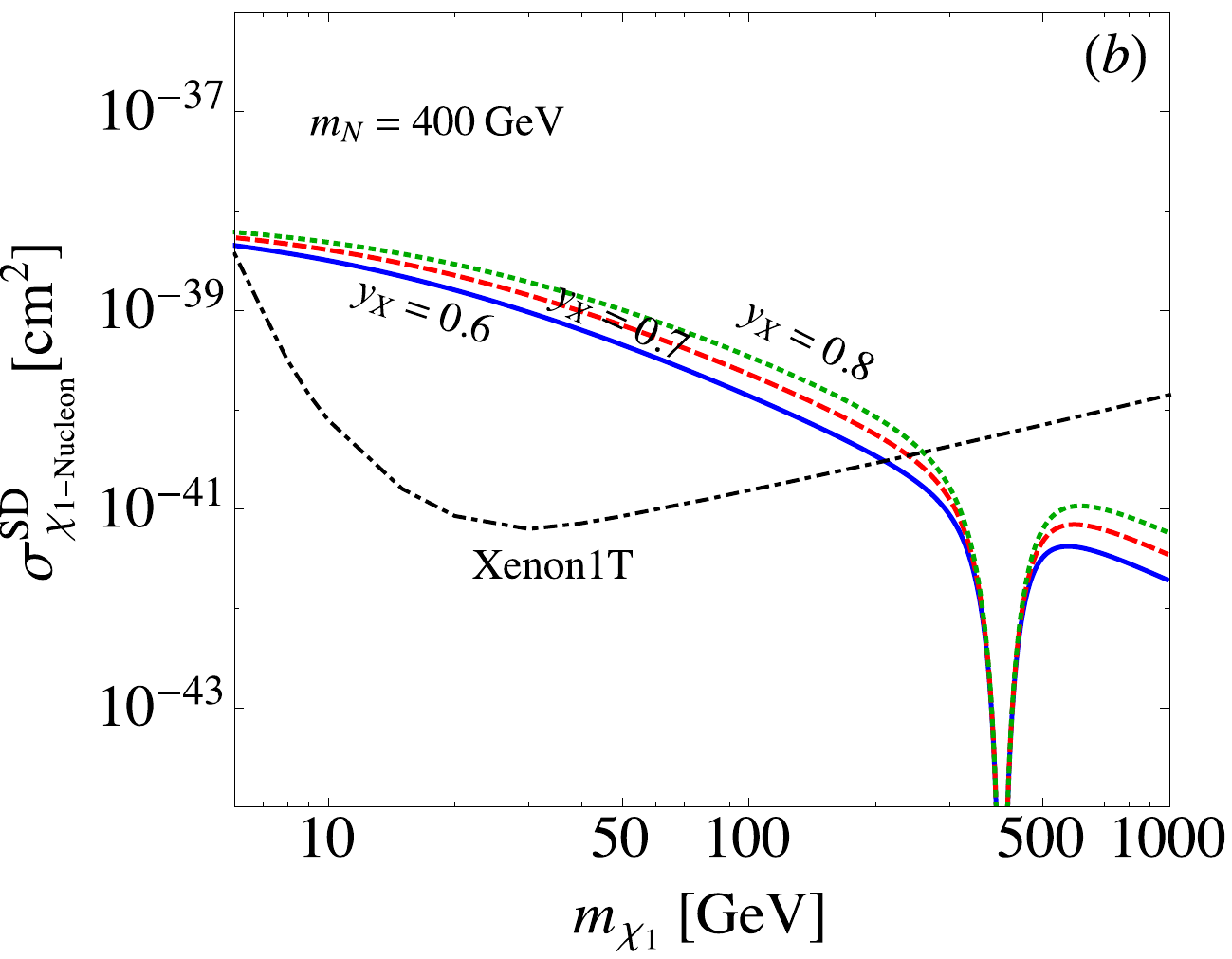}
 \caption{ (a) $h$-mediated spin-independent and  (b) $Z$-mediated spin-dependent DM-nucleon scattering cross sections as a function of $m_{\chi_1}$, where the solid, dashed, and dotted lines denote $y_X=0.6$, $y_X=0.7$, and $y_X=0.8$, respectively, and  $m_N=400$ GeV is used. The dot-dashed lines in (a) and (b) are the Xenon1T results shown in~\cite{Aprile:2018dbl,Aprile:2019dbj}.}
\label{fig:SI_SD_X}
\end{figure}

In addition to the DM relic density, we have to examine  whether the same parameter space, which can fit $\Omega^{\rm obs}_{\rm DM}h^2$, is excluded by the DM direct detection experiments.  In the model, it is found that the SI DM scattering off a nucleon is  dictated by the Higgs mediation, whereas the SD scattering is through the $Z$-mediated effects.  According to the interactions in Eq.~(\ref{eq:Z_Z2_odd}) and Eq.~(\ref{eq:h_chi_chi}), the relevant four-Fermi effective interactions for $\chi_1$  and the SM particles can be expressed as:
\begin{align}
{\cal L}_{\chi N} & \supset \frac{y_X c^h_{11}}{\sqrt{2}v m^2_h} (\bar \chi_1  \chi_1) \sum_q m_q \bar q q  \nonumber \\
& -\frac{g c^Z_{11} }{2c_W m^2_Z} \bar \chi_1 \gamma^\mu \gamma_5  \chi_1 \, \sum_q \bar q  \gamma_\mu \left( g^q_V + g^q_A \gamma_5 \right) q \,, \\
g^u_V &= \frac{g}{2c_W}\left( \frac{1}{2} - \frac{4}{3}s^2_W\right)\,, ~g^u_A=\frac{1}{2}\,,\nonumber \\
g^d_V &= \frac{g}{2c_W}\left( -\frac{1}{2} + \frac{2}{3}s^2_W\right)\,, ~g^d_{A}=- \frac{1}{2}\,. \nonumber 
\end{align}
Accordingly,  the $h$-mediated SI DM-nucleon scattering cross section can be written as~\cite{Arcadi:2019lka}:
 \begin{align}
  \sigma^{SI}_{h} & = \frac{y^2_X (c^h_{11})^2 }{8\pi} \frac{m^2_n \mu^2_{\chi_1 n} f^2_N}{ v^2 m^4_h}\,,  \label{eq:spin-inde_X}
  \end{align}
  where  $f_N\approx 0.3$, and  $\mu_{\chi_1 n}=m_{\chi_1} m_n/(m_{\chi_1} + m_n)$ is the DM-nucleon reduced mass.  The $Z$-mediated DM-nucleon scattering cross-section can be expressed  as~\cite{Alves:2015pea}
  \begin{align}
  \sigma^{SD}_{Z} & \approx \frac{3\mu^2_{\chi_1n}}{\pi }\left( \frac{g c^Z_{11}}{2 c_W m^2_Z}\right)^2 \left[ g^u_A \Delta^n_u + g^d_A\left( \Delta^n_d + \Delta^n_s \right) \right]\,,  \label{eq:spin-de_X}
    \end{align}
 where the quark spin fractions of the nucleon are taken as $\Delta^n_u=0.84$, $\Delta^n_{d}=-0.43$, and $\Delta^n_s=-0.08$~\cite{Belanger:2008sj}.  Using Eq.~(\ref{eq:spin-inde_X}) and Eq.~(\ref{eq:spin-de_X}), we show $\sigma^{\rm SI}_{h}$ and $\sigma^{\rm SD}_{Z}$ as a function of $m_{\chi_1}$ in Fig.~\ref{fig:SI_SD_X}(a) and (b), respectively. A comparison with the results in Fig.~\ref{fig:Omega_X} clearly shows that the allowed parameter regions, which can fit the observed $\Omega_{\rm DM}h^2$, are excluded by the current Xenon1T  SI and SD measurements~\cite{Aprile:2018dbl,Aprile:2019dbj}. Thus, it can be concluded that  $\chi_{1}$ cannot be the DM candidate due to the strict constraints from the direct detection experiments. 

\begin{figure}[phtb]
\includegraphics[scale=0.6]{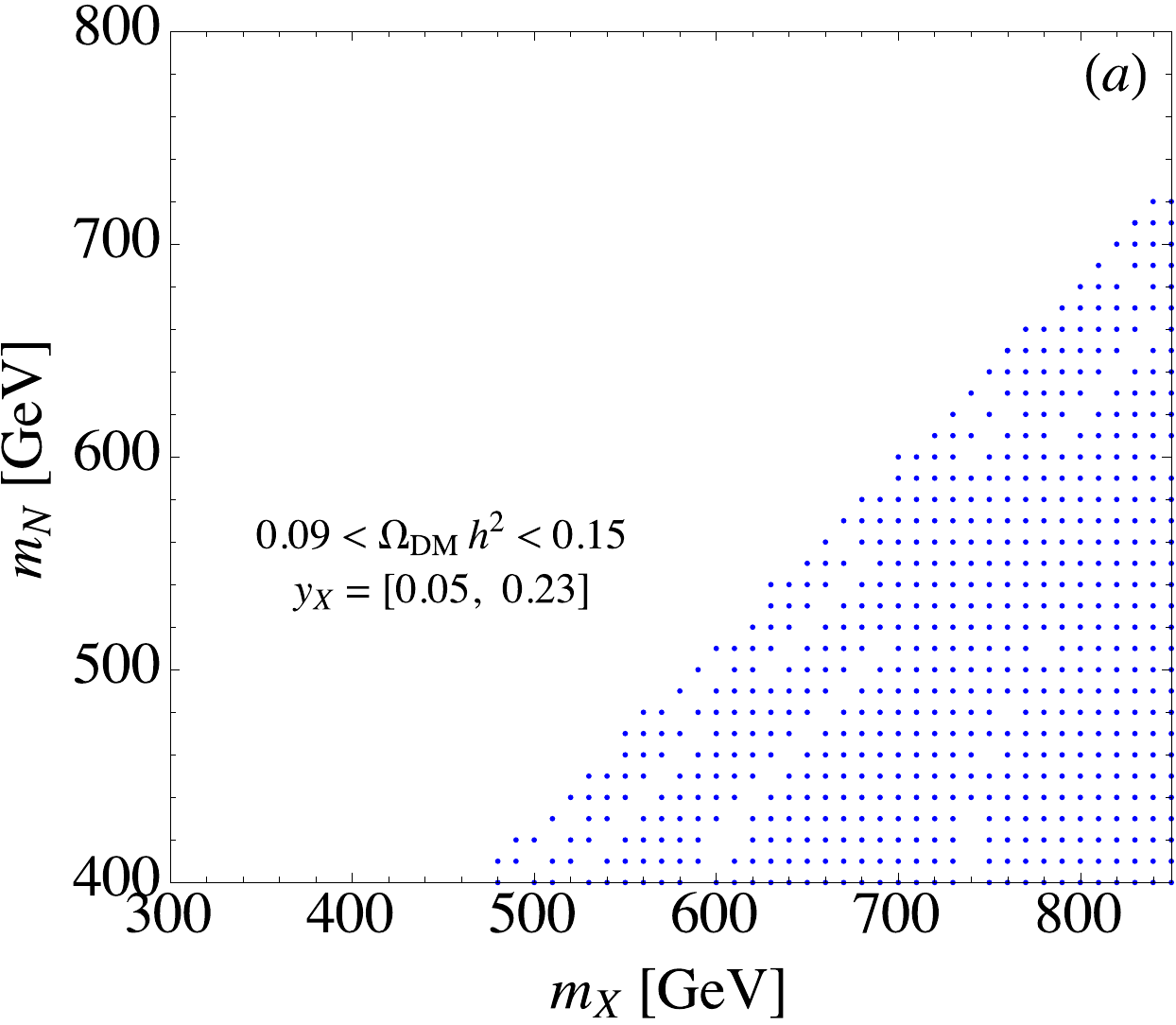}
\includegraphics[scale=0.6]{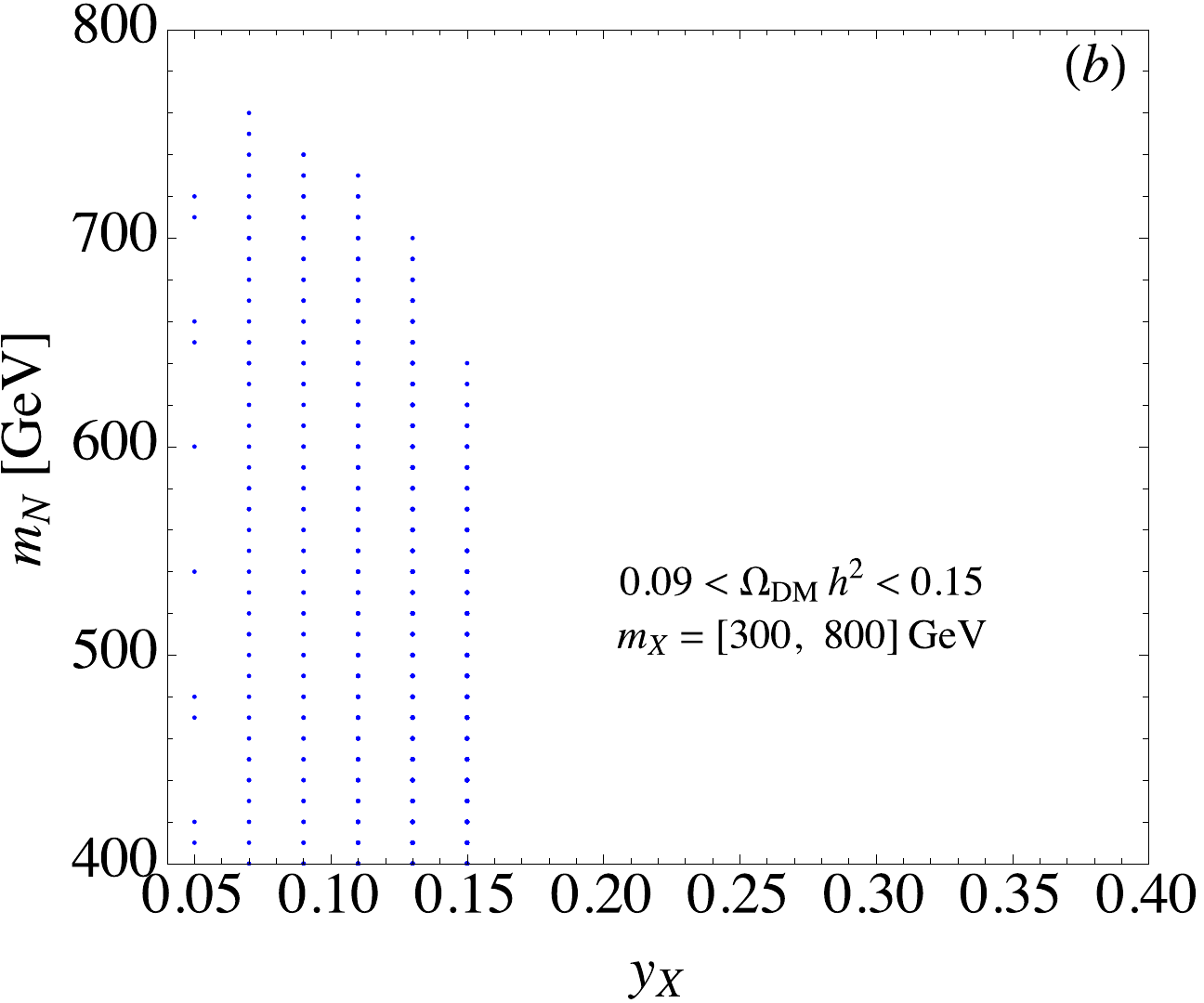}
 \caption{ Legend is the same as in Fig.~\ref{fig:allowed_X}, but for the $m_{\chi_3}$ case.   }
\label{fig:allowed_N3}
\end{figure}

\begin{figure}[phtb]
\includegraphics[scale=0.65]{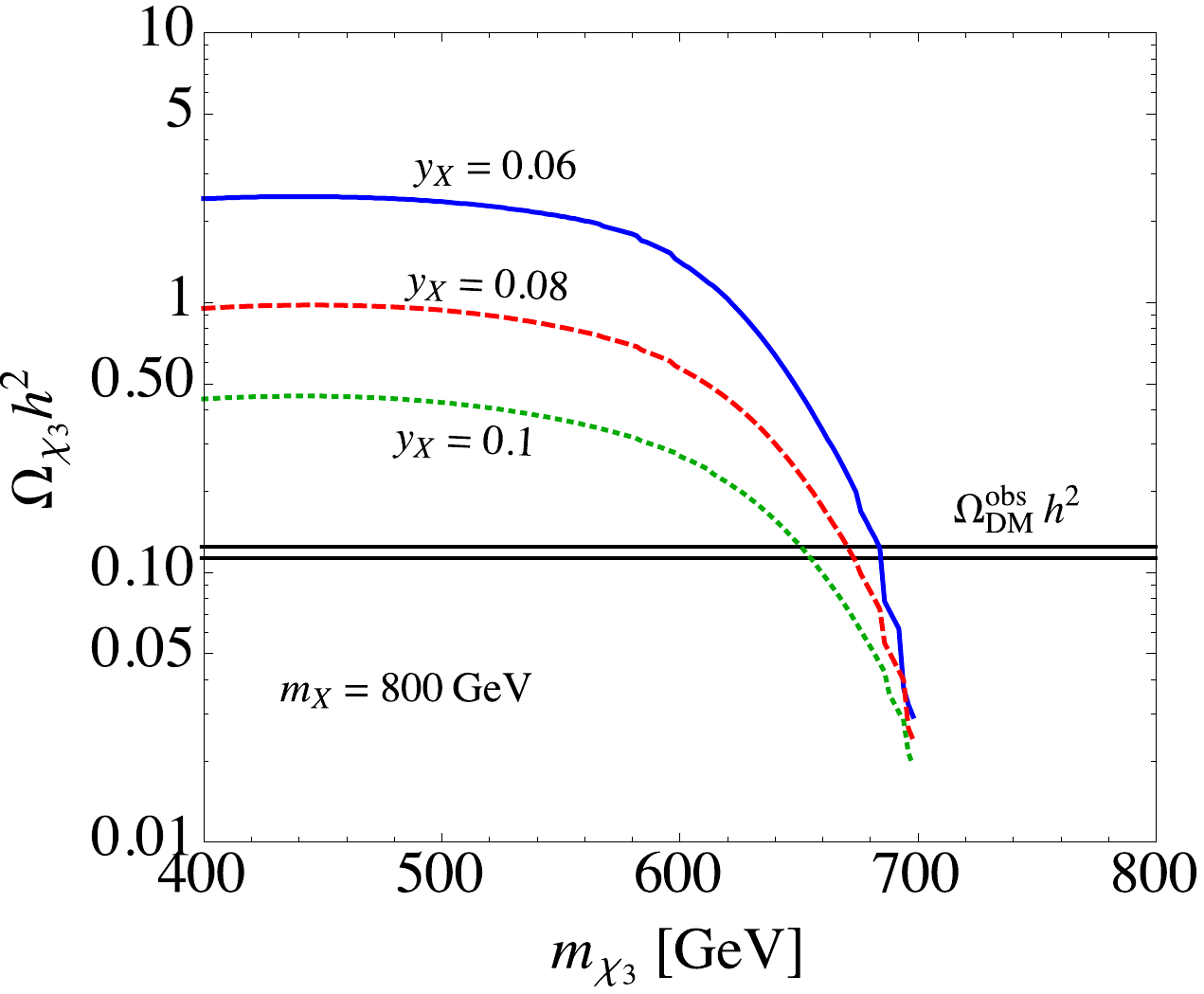}
 \caption{ $\chi_3$-DM relic abundance for $y_X=0.06$ (solid), $y_X=0.08$ (dashed), and $y_X=0.10$ (dotted), where $m_X=800$ GeV is fixed, and the horizontal lines denote $\Omega^{\rm obs}_{\rm DM} h^2$ with $3\sigma$ errors.  }
\label{fig:Omega_N}
\end{figure}

Next, we discuss $\chi_3$ as the DM candidate. Since $\chi_3$ originates from an $SU(2)$ singlet right-handed lepton, without the $y_X$ coupling, it can a heavy $Z_2$-odd sterile neutrino and doesn't couple to the SM particles. Therefore, the $\chi_3$ effects are all related to the  $y_X$ parameter and the main interactions are through the Higgs couplings, i.e. the $\chi_i \chi_3 h$ couplings shown in Eq.~(\ref{eq:h_chi_chi}).  Similar to the $\chi_1$ case, to understand the correlation between $\Omega_{\rm DM}h^2$ and  the $m_{N,X}$ and $y_X$ parameters,  we choose the scanned parameter regions to be:
 \begin{equation}
 m_N=[300,800] \, {\rm GeV}\,, ~ m_{X}=[400, 900]\, {\rm GeV}\,, ~ y_{X}=[0.05, 2.3]\,,
 \end{equation} 
 and the resulting $\Omega_{\rm DM}h^2$ is required to be in the region of $0.09< \Omega_{\rm DM}h^2 < 0.15$. As a result, the correlations between $m_N$ and $m_X$ and between $m_N$ and $y_X$ are shown in Fig.~\ref{fig:allowed_N3}(a) and (b), respectively. From the plots, it can be seen that when $\chi_3$ is the DM candidate, the DM mass prefers to be heavy,  and $y_X$ is of the order of $0.1$.  In addition, according to the result shown in Fig.~\ref{fig:allowed_N3}(a),  it can be seen  that the allowed maximum $m_N$ follows an approximate relation with $m_X$ as $m_X-m_N\sim 100$ GeV.
 Based on the results, we show $\Omega_{\rm DM}h^2$ as a function of $m_{\chi_3}$ in Fig.~\ref{fig:Omega_N}, where $m_X=800$ GeV is fixed, and the solid, dashed, and dotted lines denote the results of $y_{X}=0.06$, $0.08$, and $0.10$, respectively. It can be seen that  $m_{\chi_3} \sim (680,\, 670,\, 650)$ GeV with $y_X\sim (0.06,\, 0.08,\, 0.1)$ can fit the observed $\Omega_{\rm DM}h^2$. As mentioned earlier, the maximum of $m_N$ is close to 700 GeV when $m_X=800$ GeV is taken; therefore, the three lines end at $m_{\chi_3}\approx 700$ GeV. Due to $m_{\chi_3} > m_{Z,h}$, we can evade the constraints from the invisible $Z$ and $h$ decays. 

\begin{figure}[phtb]
\includegraphics[scale=0.6]{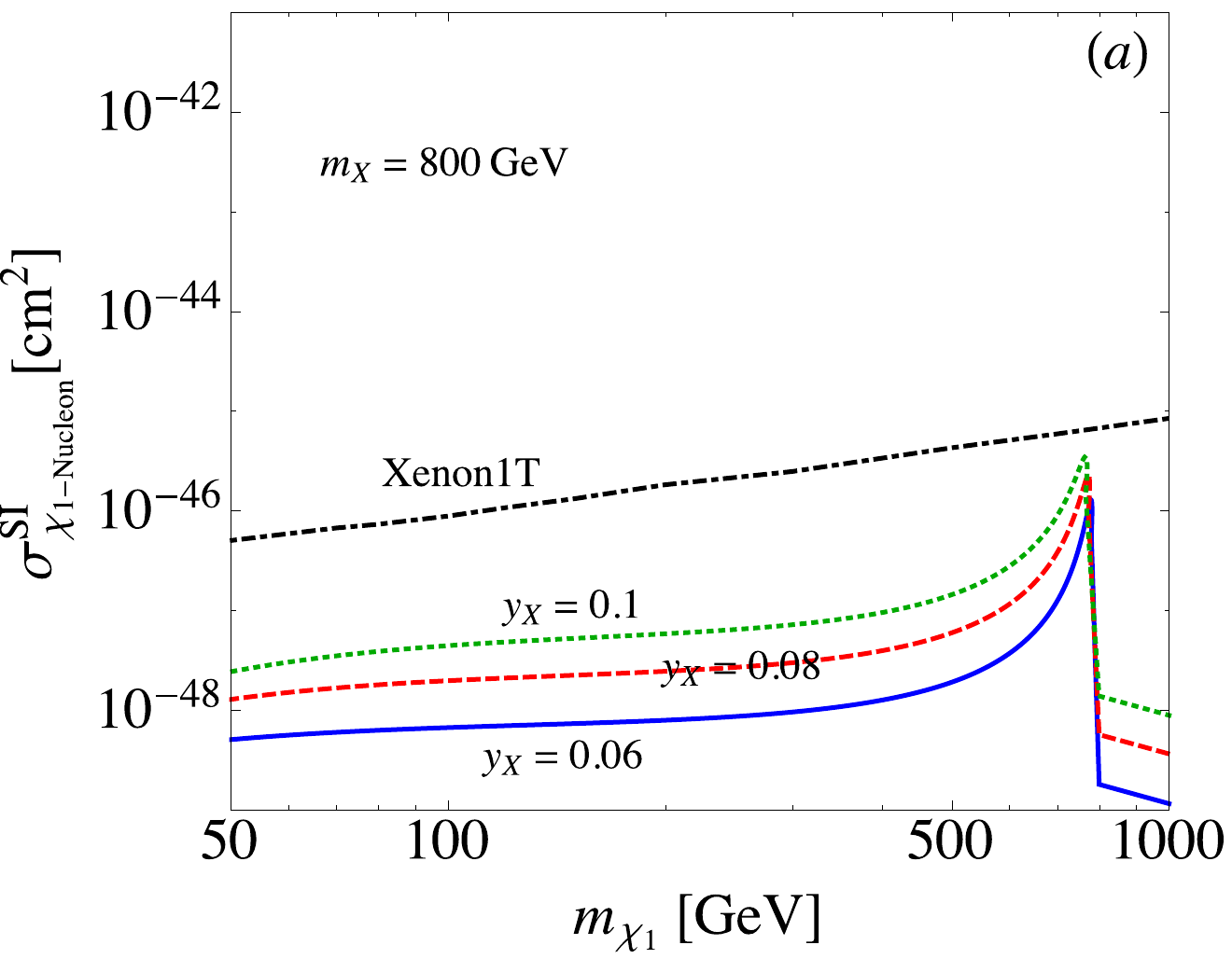}
\includegraphics[scale=0.6]{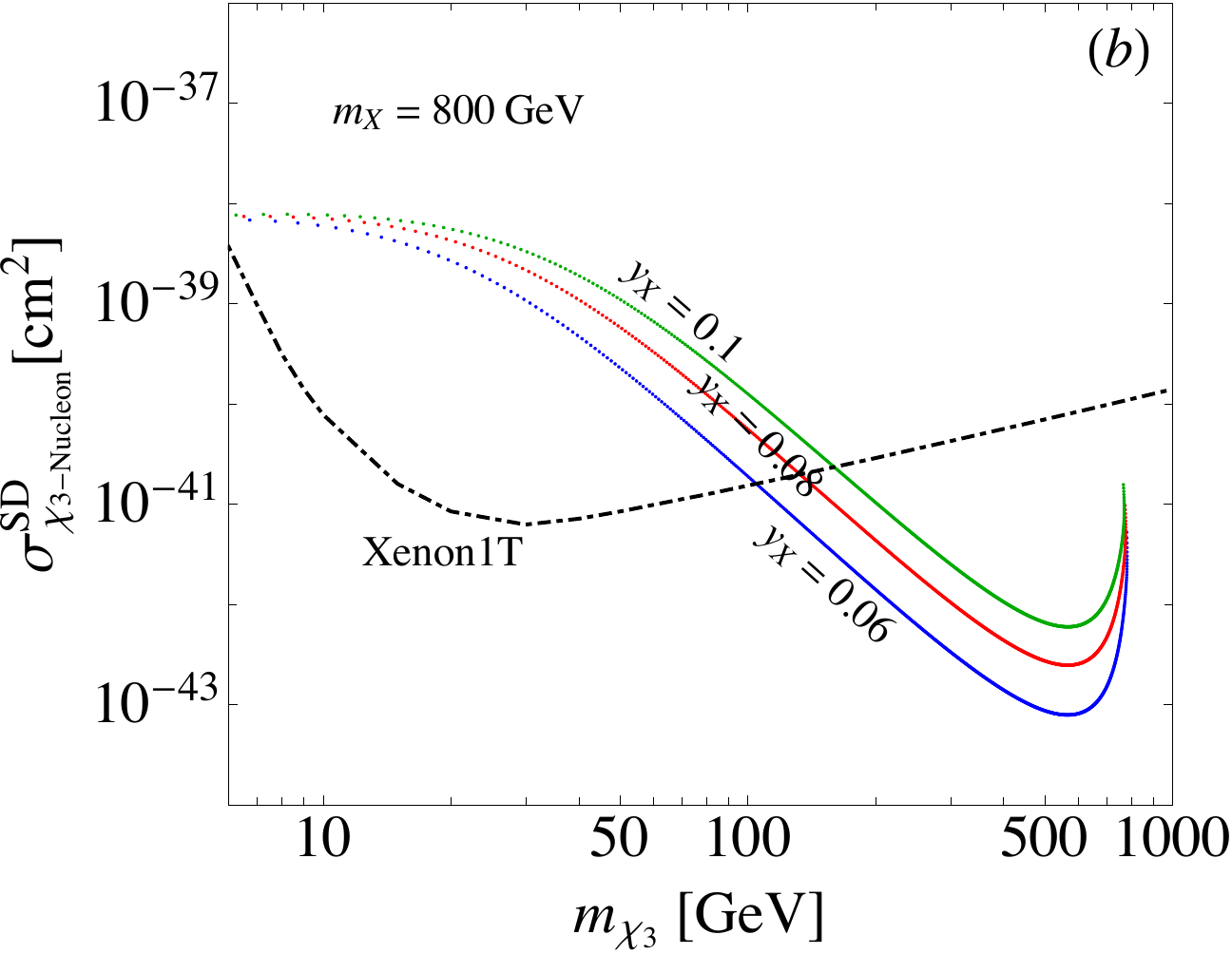}
 \caption{ (a) $h$-mediated spin-independent and  (b) $Z$-mediated spin-dependent DM-nucleon scattering cross sections, where the solid, dashed, and dotted lines denote $y_X=0.06$, $y_X=0.08$, and $y_X=0.1$, respectively, and  $m_X=800$ GeV is used. The dot-dashed lines in (a) and (b) are the Xenon1T results shown in~\cite{Aprile:2018dbl,Aprile:2019dbj}. }
\label{fig:SI_SD_N}
\end{figure}

 Similar to the $\chi_1$ case, $\chi_3$ can contribute to the SI and SD DM-nucleon scatterings through the $h$ and $Z$ mediation, respectively. To estimate the elastic scattering cross sections, we can  use the formulas in Eqs.~(\ref{eq:spin-inde_X}) and (\ref{eq:spin-de_X})  by  replacing $c^{h,Z}_{11}$ and $\mu_{\chi_1 n}$ with $c^{h,Z}_{33}$ and $\mu_{\chi_3 n}=m_{\chi_3} m_n/(m_{\chi_3} + m_{n})$. Accordingly, we show the SI and SD $\chi_3$-nucleon scattering cross sections as a function of $m_{\chi_3}$ in Fig.~\ref{fig:SI_SD_N}(a) and (b), where $m_X=800$ GeV is used, and the solid, dashed, and dotted lines denote the results of $y_X=0.06$, $0.08$, and $0.1$, respectively.  A comparison with the results shown in Fig.~\ref{fig:Omega_N} reveals clearly that  $\sigma^{SI}_{h}$ and $\sigma^{SD}_Z$ at the  $m_{\chi_3}$ value, which is determined by $\Omega^{\rm obs}_{\rm DM} h^2$, are all under the Xenon1T upper limits~\cite{Aprile:2018dbl,Aprile:2019dbj}. That is, the  DM candidate in the model is the $\chi_3$ $Z_2$-odd Majorana lepton. Note that a steep behavior in Fig.~\ref{fig:SI_SD_N}(a) occurs when $m_{\chi_3}$ approaches $m_X=800$ GeV, which is the upper limit of $m_N$.

\subsection{T-parameter and $h\to \gamma \gamma$ constraints}

 From Eq.~(\ref{eq:V_v_D}), it can be seen that when $\mu_\Delta$ is fixed, $v_\Delta$ is determined by the $M_\Delta$ and $\lambda_{1,4}$ parameters.  According to Eq.~(\ref{eq:m_scalar}),  the mass ordering of the Higgs triplet bosons and their mass splittings are dictated by the $\lambda_4$ parameter.  Moreover, the Higgs couplings to the doubly and singly charged Higgses also depend on $\lambda_{1, 4}$. Thus, it can be expected that the electroweak oblique $T$ parameter~\cite{Peskin:1991sw} and the Higgs to diphoton precision measurement may give a strict constraint on the $\lambda_{1,4}$ parameters, where their values in principle could be $|\lambda_{1,4}|\leq 4 \pi$. Following the results obtained in~\cite{Lavoura:1993nq}, the $T$-parameter, which arises from the Higgs triplet, can be formulated as~\cite{Lavoura:1993nq}:
\begin{align}
T&=\frac{1}{8\pi c^2_W s^2_W}  \left[ G\left( \frac{m^2_{H^{\pm\pm}}}{m^2_Z}, \frac{m^2_{H^\pm}}{m^2_Z} \right) + G\left( \frac{m^2_{H^\pm}}{m^2_Z}, \frac{m^2_{H^0}}{m^2_Z}  \right) \right] \,, \\
G(x,y)&=x+y -\frac{2 x y}{x-y} \ln\frac{x}{y}\,.
\end{align}
Basically, the mass splitting in the vector-like lepton doublet can also contribute to the T-parameter, where  the mass difference is dictated by $e_N$.    Using $y_X = 0.1$, $m_X=800$ GeV, and $m_{\chi_3}=700$ GeV, we obtain $e_N\approx 3.2$ GeV, where the resulting $T$ can be estimated to be $T\approx 0.8\times 10^{-3}$~\cite{Chen:2019nud}. Since the influence on $T$-parameter is not significant, we drop the vector-like lepton doublet contribution in this study. 

Next, we discuss the new physics contributions to $pp\to h \to \gamma \gamma$.  As shown in Appendix~\ref{ap:mixing}, because the $h$-$H^0$ mixing angle is suppressed,  the Higgs couplings to the SM quarks can be taken as unmodified. Thus, the $h$ production cross section in the $pp$ collisions is still from the SM contributions.  Since the $h\to \gamma \gamma$ decay arises from the charged particle loops,  in addition to the top and bottom quarks and  the $W$-boson in the SM,  the new physics effects in this model are from the doubly and singly charged Higgses. We note that although we have an $Z_2$-odd $X^-$ in the model, the $h$ coupling to $X^-$ has two suppression factors, where one is  the $h$-$H^0$ mixing effect, and the other is the small $y_R$ parameter. Thus, we neglect the $X^-$ contribution to $h\to \gamma\gamma$.  Based on the results in~\cite{Arhrib:2011vc,Gunion:1989we},  we write the SM and Higgs triplet contributions to the partial decay width of $h\to \gamma\gamma$ as:
\begin{align}
\Gamma(h\to \gamma\gamma) & = \frac{G_F \alpha^2 m^3_h}{128\sqrt{2} \pi^3 } \left| \Gamma^{\rm SM}_{\gamma\gamma} + \Gamma^{\rm \Delta}_{\gamma\gamma}\right|^2 \,,  \nonumber \\
\Gamma^{\rm \Delta}_{\gamma\gamma} & = \frac{ \lambda_1 v^2_h}{2 m^2_{H^{\pm\pm}} } Q^2_{H^{\pm\pm}} A_0\left( \frac{4 m^2_{H^{\pm\pm}}}{m^2_h}\right) +\frac{ \lambda_1 v^2_h}{2 m^2_{H^{\pm}} } Q^2_{H^{\pm}} A_0\left(\frac{4 m^2_{H^{\pm}}}{m^2_h}\right) \,,
\end{align}
where $\Gamma^{\rm SM}_{\gamma\gamma} \approx 6.50 - i 0.02$; $Q_{H^{\pm\pm}}=2$ and $Q_{H^{\pm}}=1$;  $A_0(\tau)=\tau( 1-\tau f(\tau))$, and the loop function is defined as:
\begin{equation}
f(x) = 
\left\{
\begin{array}{c}
  \left( \sin^{-1}\frac{1}{\sqrt{\tau}} \right)^2 \,,~~~~~~~~~~ (\tau\geq 1)\,, \\
  -\frac{1}{4} \left(\ln\frac{1+\sqrt{1-\tau}}{1-\sqrt{1-\tau}} -i\pi \right)^2\,, ~~~~~ (\tau < 1)\,.
\end{array}
\right. \label{eq:f_function}
\end{equation}
Thus, we can write the signal strength for $pp\to h \to \gamma\gamma$ as:
 \begin{align}
 \mu_{\gamma\gamma} = \frac{\sigma(pp\to h)}{\sigma(pp\to h)^{\rm SM}} \frac{BR(h\to \gamma \gamma)} {BR(h\to \gamma\gamma)^{\rm SM}} \approx \frac{BR(h\to \gamma \gamma)} {BR(h\to \gamma\gamma)^{\rm SM}} \,.
 \end{align}
For numerical estimates, we take the Higgs width in the SM as $\Gamma^{\rm SM} \approx 4.07$ MeV~\cite{Denner:2011mq}.  The current Higgs to diphoton measurements from ATLAS and CMS at $\sqrt{s}=13$ TeV are given as $1.06\pm 0.12$~\cite{ATLAS:2019slw} and $1.15\pm 0.15$~\cite{CMS:1900lgv}, where the corresponding integrated luminosities are $79.8$ fb$^{-1}$ and $77.4$ fb$^{-1}$, respectively. 

\begin{figure}[phtb]
\includegraphics[scale=0.6]{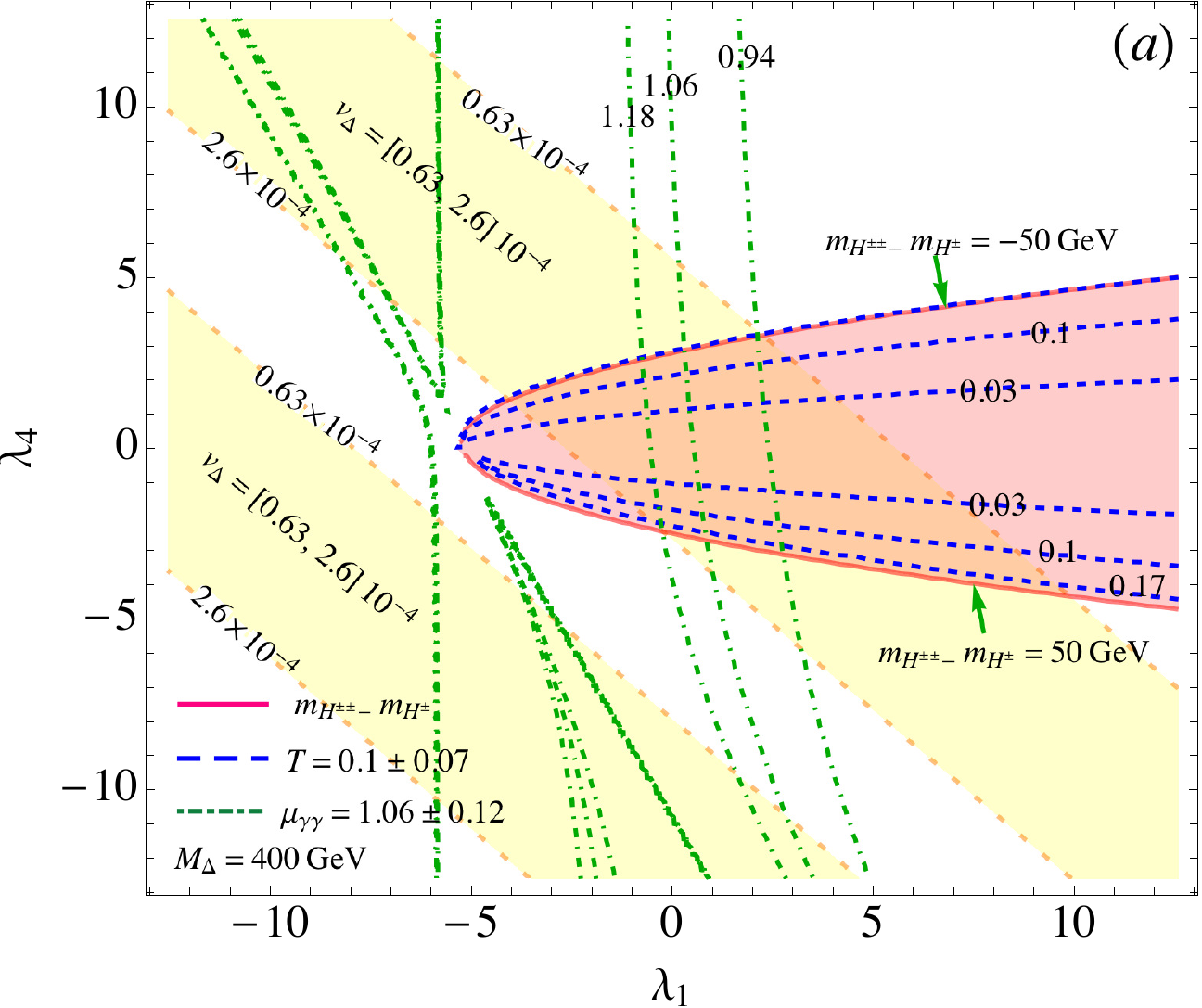}
\includegraphics[scale=0.6]{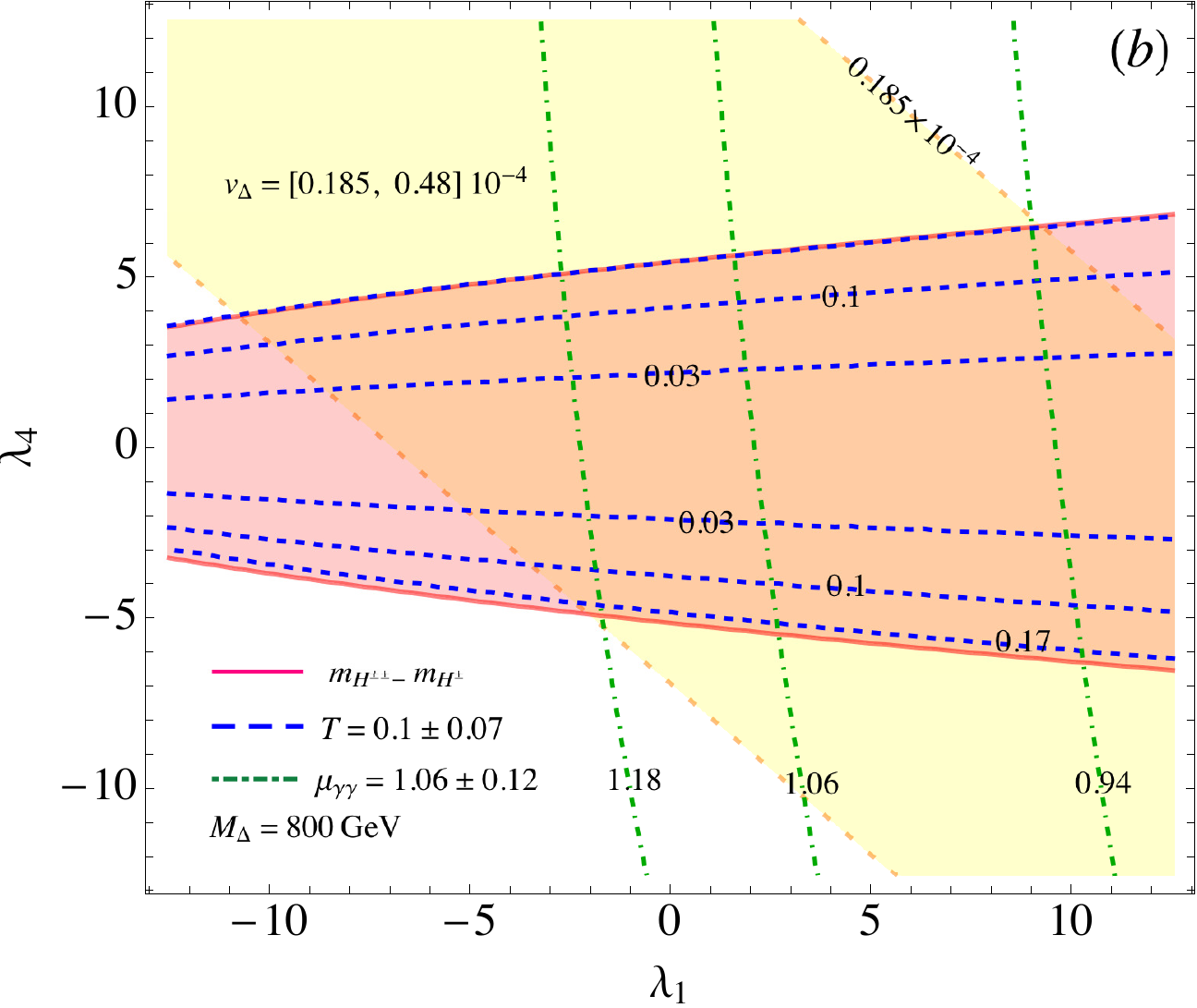}
 \caption{Constraints from the oblique $T$-parameter (dashed) and the $h\to \gamma\gamma$ (dot-dashed) precision measurement, where (a) [(b)] corresponds to the case with $M_\Delta=400 [800]$ GeV and $\mu_\Delta=4.8\times 10^{-4}$ GeV. The  area enclosed by the solid line denotes $m_{H^{\pm\pm}} - m_{H^\pm}=(-50, 50)$ GeV. The $v_\Delta$ regions are $(0.63,\, 2.6) \times 10^{-4}$ GeV and $(0.185,\, 0.48) \times 10^{-4}$ GeV for $M_\Delta=400$ GeV and $M_\Delta=800$ GeV, respectively. }
\label{fig:triplet_limit}
\end{figure}

 From Eqs.~(\ref{eq:mu_Del}) and (\ref{eq:V_v_D}), it is known that in addition to the $m_{X,N}$ and $y_{X,R}$ parameters, $v_\Delta$ also depends on the $\lambda_{1,4}$ constraints. Since the DM candidate in this model is $\chi_{3}$, and its mass is determined to be $m_{\chi_3}\sim 680$ GeV when $m_X\sim 800$ GeV is used, in order to simplify the study on the $\lambda_{1,4}$ constraints, we fix $m_{N(X)}=700(800)$ GeV, $y_X=0.1$, and $y_R=0.01$, where the corresponding $\mu_\Delta$ value is $4.8\times 10^{-4}$ GeV. 
 Using the introduced formulas for the $T$-parameter and $\mu_{\gamma\gamma}$, we show $T$-parameter, $\mu_{\gamma\gamma}$, $m_{H^{\pm\pm}}-m_{H^\pm}$, and $v_\Delta$ as a function of $\lambda_{1}$ and $\lambda_{4}$ in Fig.~\ref{fig:triplet_limit}, where the plots (a) and (b) correspond to  $M_\Delta=400$ GeV and $M_\Delta =800$ GeV, respectively. 
 
  From the resulting plots, we find: (a) Due to the $T$-parameter constraint, $|m_{H^{\pm\pm}}-m_{H^\pm}|\lesssim 50$ GeV, which is consistent with the results shown in~\cite{Chun:2012jw,Ghosh:2017pxl}; (b) using the ATLAS result of $\mu_{\gamma\gamma}=1.06 \pm 0.12$, the $\lambda_1$ parameter is bounded to be $\lambda_{1}=(-0.8, 2.63)$ and $\lambda_{1}=(-2.8, 10.2)$ for $M_\Delta =400$ GeV and $M_\Delta=800$ GeV, respectively, and (d) the allowed  $v_\Delta$ range, which fits the $T$-parameter and $\mu_{\gamma\gamma}$ constraints, is obtained as: $v_\Delta \approx (0.63,\, 2.6) [(0.185,\, 0.48)]\times 10^{-4}$ GeV for  $M_\Delta=400[800]$ GeV. It can be seen that the allowed $\lambda_1$ is mostly  in the region of $\lambda_1 >0$, and  the allowed $\lambda_1$ can reach a value of $10$ when  $M_\Delta$ approaches to $1$ TeV. In addition, the $\lambda_4$ parameter is bounded in the region of $(1.1, 3.4)$ and $(-2.78, -0.9)$ for $M_\Delta=400$ GeV and in the region of $(2.12, 6.50)$ and $(-5.9, -2.0)$ for $M_\Delta=800$ GeV. We note that the constraints cannot determine the sign of the $\lambda_4$ parameter; thus, the mass order, i.e. $m_{H^0(A^0)} \lesssim m_{H^\pm} \lesssim m_{H^{\pm\pm}}$ or $m_{H^{\pm\pm}} \lesssim m_{H^\pm} \lesssim m_{H^0(A^0)}$, is still uncertain in the model.

\section{Phenomenological analysis}

After analyzing the potential constraints, in this section, we study the relevant phenomenology in detail, such as the $h\to Z\gamma$ and $H^{\pm\pm}$, $H^\pm$, and $H^0(A^0)$ decays. 
 From the earlier analysis, since $m_X$ is taken to be $800$ GeV, the processes, in which the Higgs triplet decays to the vector-like leptons, are kinematically suppressed when we focus on the study with $m_{\Delta} < 1$ TeV; therefore, we only consider the SM particles in the final states, where the three-body decays are also included when the kinematic condition is allowed. When the final states are all leptons, for simplicity, we sum up all possible lepton flavors. In addition, since the neutrino constraints from the NO and IO are similar in most lepton Yukawa couplings, hereafter, we only use the NO constraint as the inputs. 

\subsection{Signal strength for $h\to Z\gamma$}

 We have shown that the Higgs to diphoton measurement can bound the Higgs couplings to $H^{\pm\pm}$ and $H^\pm$, which is dominated by the $\lambda_1$ parameter. Since the same couplings can also contribute to the loop-induced $h\to Z\gamma$, with the constrained parameters, we can predict the $h\to Z\gamma$ in the model. Thus, similar to the case in $h\to \gamma\gamma$, the signal strength of $h\to Z\gamma$ can be expressed as:
 \begin{align}
 \mu_{Z\gamma}= \frac{\sigma(pp\to h)}{\sigma(pp\to h)^{\rm SM}} \frac{BR(h\to Z\gamma)}{BR(h\to Z\gamma)^{\rm SM}} \approx \frac{BR(h\to Z\gamma)}{BR(h\to Z\gamma)^{\rm SM}}\,,
 \end{align}
where the $h$ production cross section is dominated by the SM effects in the model, and the current upper limit is $\mu_{Z\gamma} < 6.6$~\cite{PDG}. 

\begin{figure}[phtb]
\includegraphics[scale=0.6]{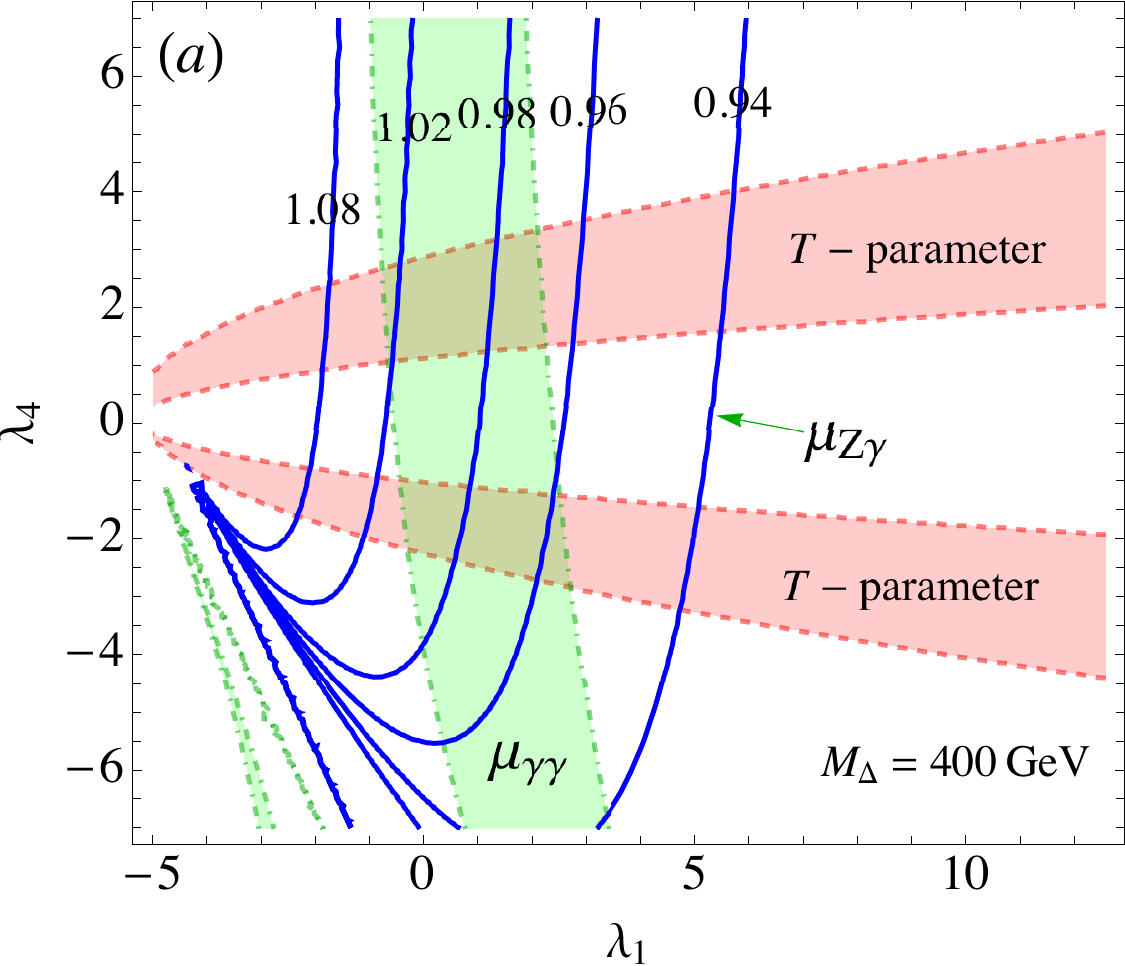}
\includegraphics[scale=0.6]{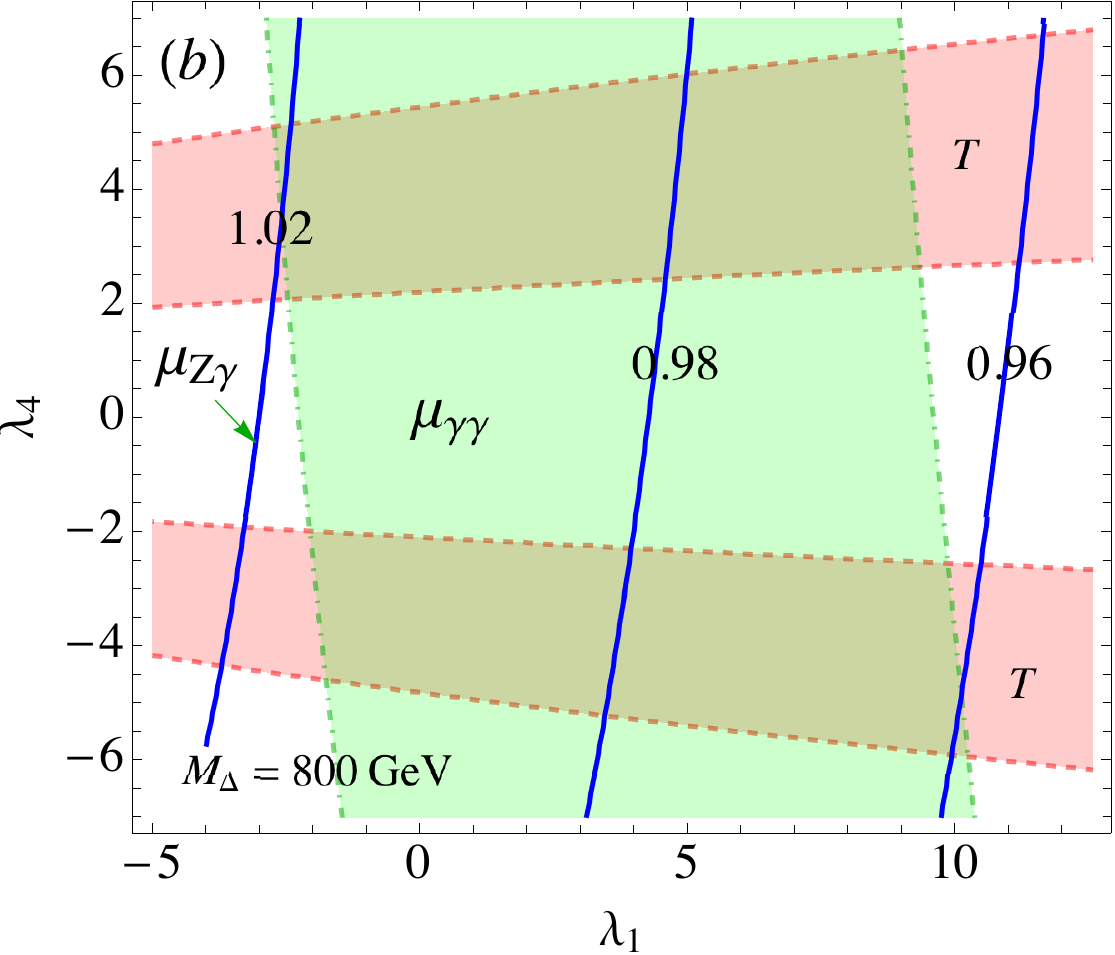}
 \caption{ Contours for signal strength of $h\to Z\gamma$ as a function of $\lambda_1$ and $\lambda_4$ for (a) $M_\Delta=400$ GeV and (b) $M_\Delta=800$ GeV, where the $T$-parameter and $\mu_{\gamma\gamma}$ constraints are also shown. }
\label{fig:hZga}
\end{figure}

Based on the results in~\cite{Chabab:2014ara,Gunion:1989we,Cahn:1978nz,Bergstrom:1985hp,Arbabifar:2012bd,Dev:2013ff}, we write the partial decay rate for $h\to Z\gamma$ as:
 \begin{equation}
 \Gamma(h\to Z\gamma)= \frac{G_F \alpha m^2_W m^3_h}{64 \pi^4}\left( 1- \frac{m^2_Z}{m^2_h}\right)^3 \left| A_{SM} + A_{\Delta}\right|^2\,,
 \end{equation}
 where the SM and Higgs triplet contributions can be expressed as~\cite{Gunion:1989we,Dev:2013ff}:
  \begin{align}
  A_{SM} &= -\frac{N_C}{c_W} \sum_f Q_f \left( 2 I^f_{3} -4 Q_f s_W^2 \right) A^h_{1/2}(\tau^f_h, \tau^f_Z) \nonumber \\
    & - c_W  A^h_1(\tau^W_h, \tau^W_Z) \,, \nonumber \\
  A_{\Delta} &= 2 sW g_{Z2H^\pm} g_{h2H^\pm} A^h_0 (\tau^{H^\pm}_h, \tau^{H^\pm}_Z) \nonumber \\
   & + 4 s_W g_{Z2H^{\pm\pm}} g_{h2H^{\pm\pm}} A^h_0(\tau^{H^{\pm\pm}}_h, \tau^{H^{\pm\pm}}_Z)  \,. \label{eq:As}
  \end{align}
Here, $N_C=3$ is the color number; $\tau^{i}_{h(Z)}=4 m^2_{i}/m^2_{h(Z)}$, $Q_f$ is the electric charge of $f$ fermion; $I^f_3$ is the third component of weak isospin of $f$ fermion, and the charged Higgs couplings to $h$ and $Z$ bosons are given as:
   \begin{align}
   g_{h2H^\pm} & = \frac{m_W}{g m^2_{H^\pm}} \left( \lambda_1 + \frac{\lambda_4}{2}\right)v_h\,,~   g_{h2H^{\pm\pm}}  = \frac{m_W}{g m^2_{H^{\pm\pm}}} \lambda_1 v_h\,, \nonumber \\
   g_{Z2H^{\pm}}&= -\tan\theta_W,\, ~ g_{Z2H^{\pm\pm}}= 2\cot2\theta_W\,.
   \end{align}
   The detailed loop functions $A^{h}_{0,1/2,1}$  can be found  in Appendix~\ref{ap:loop}.  Accordingly, we show the $\mu_{Z\gamma}$ contours  as a function of $\lambda_1$ and $\lambda_4$ in Fig.~\ref{fig:hZga}(a) and (b) for $M_\Delta=400$ GeV and $M_\Delta=800$ GeV, respectively, where the $T$-parameter and $\mu_{\gamma\gamma}$ constraints shown in Fig.~\ref{fig:triplet_limit} are included. From the plots, it can be seen that the influence from the Higgs-triplet charged particles is $\Delta \mu_{Z\gamma} =|\mu^{\rm SM}_{Z\gamma} - \mu_{Z\gamma}| \lesssim 4\%$ and is not significant. 
   
\subsection{Doubly charged Higgs decays}

The most peculiar phenomena in a type-II seesaw model should be the doubly charged-Higgs decays, where the final states in the decays are two singly charged particles. If $m_{H^\pm} > m_{H^{\pm\pm}}$, the final states are the same sign charged-lepton pair and $W$-boson pair; however, if $m_{H^\pm} < m_{H^{\pm\pm}}$,  in addition to the  leptons and the $W$-boson, we also have the three-body decays through the  decay chain $H^{++} \to H^{+} W^{*} \to H^{+}\bar f f' $, where $f(f')$ denotes the possible final states, and for simplicity, we take $f(f')$ to be massless.   Although the $H^{++}\to H^{+*} W^+$ decay  is  possible in principle, because the off-shell $H^+$ decays are associated with the small couplings, e.g. $v_\Delta$ and ${\bf h}^\ell_{ij}$, we neglect their contributions.  

\begin{figure}[phtb]
\includegraphics[scale=0.6]{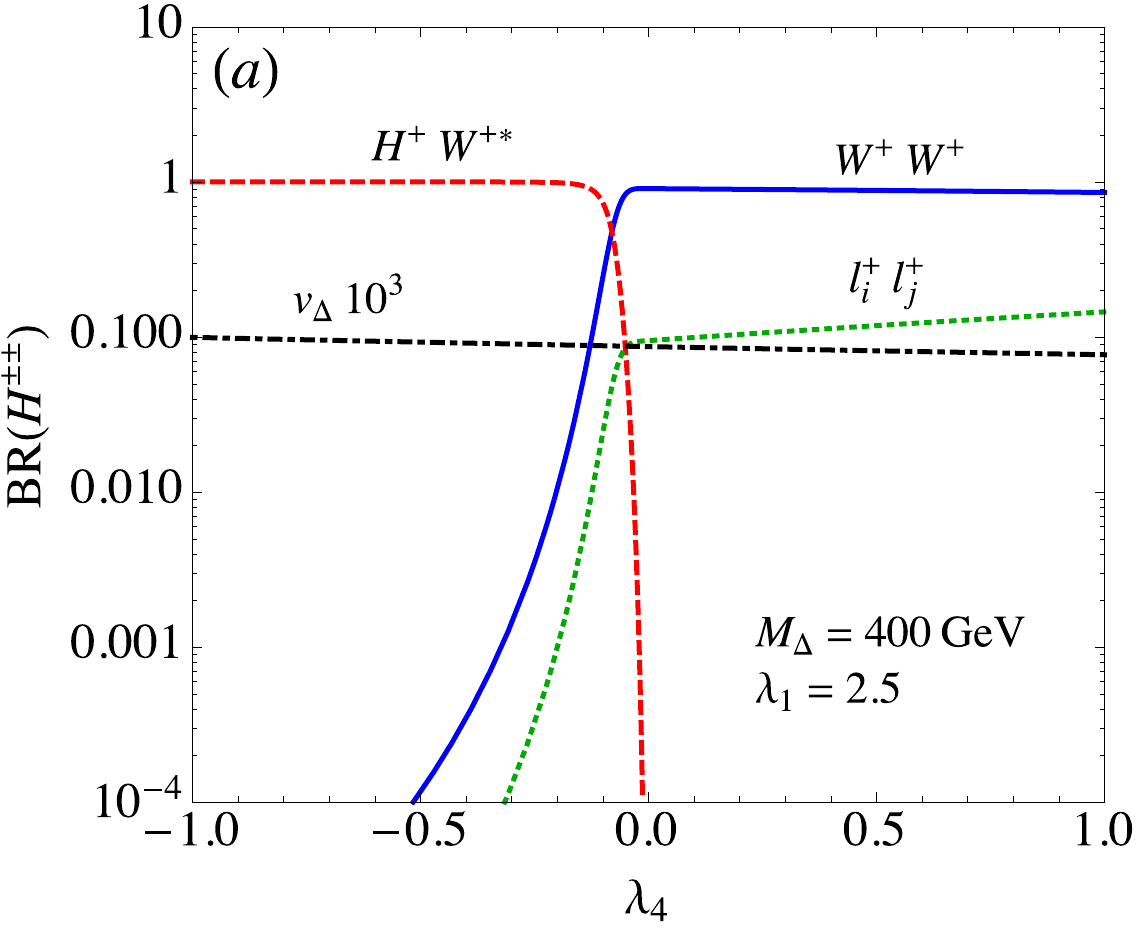}
\includegraphics[scale=0.6]{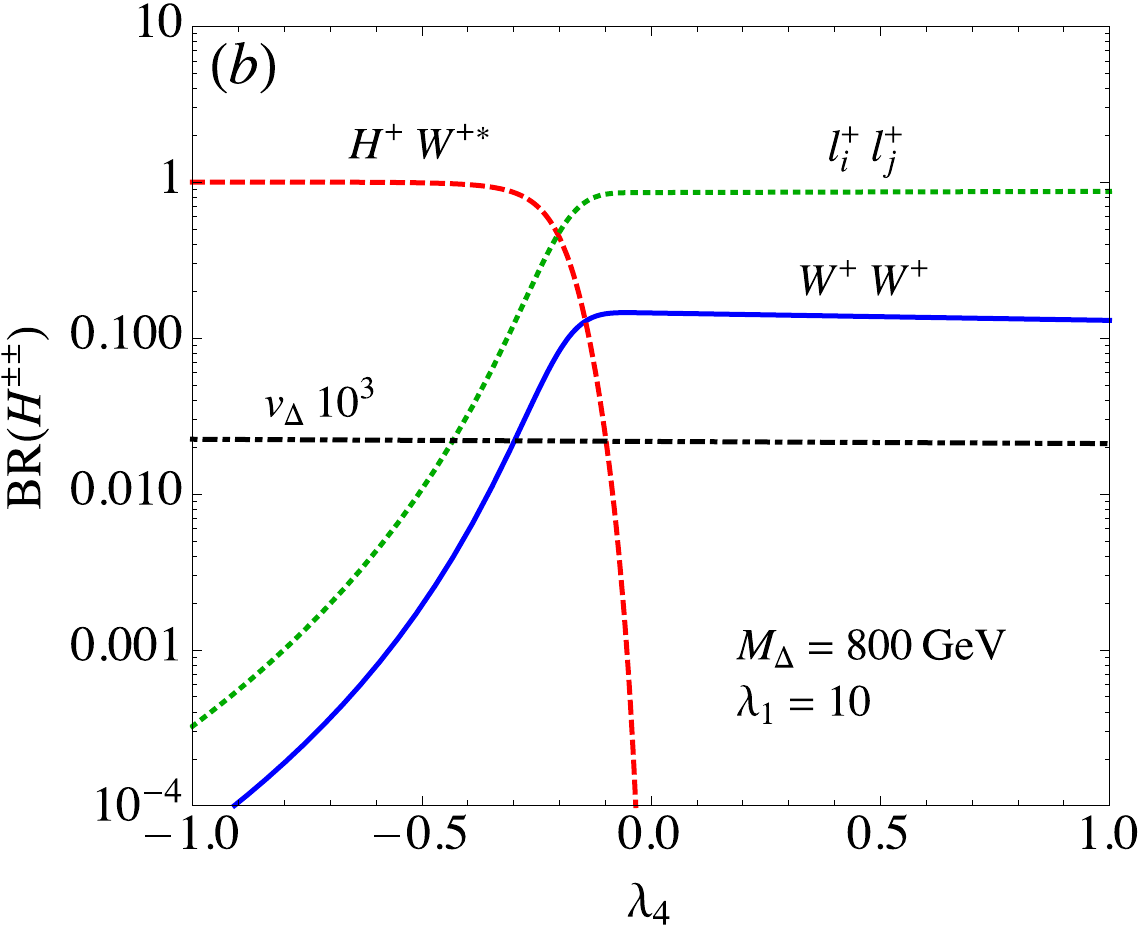}
 \caption{ BR of  the $H^{++}$ decay as a function of $\lambda_4$, where   (a) $M_\Delta=400$ GeV and $\lambda_1=2.5$ are fixed and (b) $M_\Delta=800$ GeV and $\lambda_1=10$ are used. The dot-dashed line is for $v_\Delta$. }
\label{fig:Hcc_decay}
\end{figure}

 According to the introduced gauge and Yukawa couplings, the two-body $H^{\pm\pm}$ partial decay rates  can be expressed as:
 \begin{align}
 \Gamma(H^{++}\to W^+W^+ ) & = \frac{g^4 v^2_\Delta}{16\pi m_{H^{\pm\pm}}} \left( 2 + \frac{(1-2 y_W)^2}{4 y^2_W}\right)\left(1 - 4y_W  \right)^{1/2}\,, \nonumber \\
 \Gamma(H^{++} \to  \ell^+_i \ell^+_j ) & =  \frac{S_{ij} }{4\pi} \left|{\bf h}^\ell_{ij}\right|^2 m_{H^{\pm\pm}}\,,
 \end{align}
where $y_W=m^2_W/m^2_{H^{\pm\pm}}$, $S_{ii}=1/2$, and $S_{ij}=1$ for $i\neq j$. For $\lambda_4 < 0$, $m_{H^{\pm\pm}}$ is the heaviest  Higgs triplet; then, the three-body partial decay rate for $H^{++} \to H^{+}W^{+*}$ can be expressed as:
 \begin{align}
 \Gamma(H^{++} \to H^+ W^{+*}) & =\frac{3 g^4 m_{H^{\pm\pm}}}{2^8 \pi^3} J_0(y_W, y_{H^\pm})\,, \nonumber \\
J_0(y_W,y_{H^\pm}) & =  \int^{s_{\rm max}}_{s_{\rm min}} \frac{\left( \left( 1-y_{H^\pm} + s \right)^2 - 4 s\right)^{3/2}}{(s-y_W)^{2}}\,,
 \end{align}
with $y_{H^\pm}=m^2_{H^\pm} /m^2_{H^{\pm\pm}}$, $s_{\rm min}=0$, and $s_{\rm max}=(1-\sqrt{y_{H^\pm}})^2$. The phase space integral can be simplified as:
 \begin{align}
J_0(a,b) &= \frac{1}{2 a}(1-b) \left( 9a(1+b)-2(1-b)^2 - 6 a^2 \right) \nonumber \\
 & -3 \left( 1-2 a + (b-a)^2\right)\ln\sqrt{b}  \nonumber \\
& - 3 (1-a+b) \sqrt{-\lambda(a,b)} \left( \tan^{-1}\frac{1-a-b}{\sqrt{-\lambda(a,b)}} +\tan^{-1}\frac{1+a-b}{\sqrt{-\lambda(a,b)}} \right)\,, 
 \end{align} 
with $\lambda(a,b)=1+a^2 +b^2 -2a-2b-2ab$. If we assume that the main $H^{++}$ decay modes are $W^+ W^+$, $\ell^+_i \ell^+_j$, and $H^+ W^{+*}$, the relative BRs as a function of $\lambda_4$ can be shown in Fig.~\ref{fig:Hcc_decay} (a) and (b), where $M_\Delta=400$ GeV and $\lambda_1=2.5$ are used in plot (a) and $M_\Delta=800$ GeV and $\lambda_1=10$ are used in plot (b). For clarity, we also show the corresponding $v_\Delta$ in the plots (dot-dashed). From the plots, it can be seen that  the $H^{++}\to H^+ W^{+*}$ decay is the dominant channel when $\lambda_4 < -0.1 (-0.22)$ and  $M_\Delta=400 (800)$ GeV. When $\lambda_4>0$, the dominant decay modes are $W^+W^+$ and $\ell^+_i \ell^+_j$, where   the result with  $M_\Delta=400$ GeV is $BR(H^{++}\to W^+W^+)> BR(H^{++}\to \ell^+_i \ell^+_j)$; however, the BR order  with $M_\Delta=800$ GeV is reversed due to a smaller $v_\Delta$.  We note that the relation between $m_{H^{\pm\pm}}$ and $M_\Delta (\lambda_1)$ can be written as $m_{H^{\pm\pm}}\approx \sqrt{M^2_\Delta + v^2_h \lambda_1 /2}$, which is independent of  the $\lambda_4$ parameter; therefore, the corresponding  $m_{H^{\pm\pm}}$ value can be easily obtained when $M_\Delta$ and $\lambda_1$ are fixed.

As we discussed in the introduction section, $m_{H^{\pm \pm}}$ lower bound is $770-870$ GeV when $H^{\pm \pm}$ dominantly decays into charged leptons. Thus, the scheme with $M_\Delta = 800$ GeV and $\lambda_1=10$ has $m_{H^{\pm \pm}} \approx 971$ GeV   and can be tested at the LHC. When $H^{\pm \pm}$ predominantly decays into $W^\pm W^\pm$, the lower bound of $m_{H^{\pm \pm}}$ is $\sim 220$ GeV; therefore, the scheme with $M_\Delta = 400$ GeV and $\lambda_1=2.5$, i.e. $M_{H^{\pm\pm}}\approx 485$ GeV, is safe from the constraint.

\subsection{Singly charged Higgs decays} 

In addition to the $H^+$ direct couplings to the SM particles, the singly charged Higgs can also decay through  mixing with the  SM charged-Goldstone boson $(G^+)$, where the relation between the mixing angle $\phi^+$ and the $v_\Delta$ parameter  is  shown in Appendix~\ref{ap:mixing}.  Thus, if the direct $H^+$ couplings to the SM particles are proportional to $v_\Delta$, the mixing effects with $G^+$ become important.  We find that with the exception of $\ell^+ \nu$ mode, the decay channels, such as $t \bar b$, $h W^+$, $ZW^+$, and $\gamma W^+$,  are all related to the mixing angle $\phi^+$. Hence, the partial decay rates for  the fermionic $H^+$ decays  can be expressed as:
\begin{align}
\Gamma(H^+ \to \ell^+_i \nu) & =  \frac{m_{H^\pm}}{8\pi} ({\bf h}^{\ell\dagger} {\bf h}^\ell)_{ii}\,, \nonumber \\
\Gamma(H^+ \to t \bar b) & = \frac{m_{H^\pm}}{8\pi} \frac{m^2_t}{v^2_h} s^2_{\phi^+}  \left( 1-y_t\right)^2 \,,\nonumber \\
\end{align}
with $s_{\phi^+}(c_{\phi^+})=\sin\phi^+(\cos\phi^+)$ and $y_t=m^2_t/m^2_{H^\pm}$. Since the $G^+$ coupling to a quark is proportional to the quark mass~\cite{Gunion:1989we}, we only consider the $t\bar b$ mode and the $m_b$ effect is  neglected due to $m_b\ll m_t$. 

It is found that in addition to the $G^+ hW^-$ coupling, $H^+$ can decay to the $h W^+$ final state through the mixing between $Re\Phi$ and $Re\Delta$, where the mixing effect is dictated by the mixing angle $\alpha$ shown in Eq.~(\ref{eq:alpha}). Using the gauge couplings in Eq.~(\ref{eq:Kin_TH}) and the $\phi^+$ and $\alpha$ mixing effects, the partial decay rates for the $H^+$ diboson decays can then be formulated as:
\begin{align}
\Gamma(H^+ \to h W^+ )& = \frac{g^2 m_{H^\pm}}{64 \pi} (\sqrt{2} s_\alpha+s_{\phi^+})^2 \frac{\lambda(w_W,w_h)^{3/2}}{w_W} \,, \nonumber \\
\Gamma(H^+ \to Z W^+) & = \frac{e^2 s^2_W m_{H^\pm}}{16 \pi} \left(\frac{g v_\Delta (1-3s^2_W)}{\sqrt{2} m_W s^2_W}  c_{\phi^+}+ s_{\phi^+}  \right)^2 \nonumber  \\
& \times w_Z \sqrt{\lambda(w_W,w_Z)}\left(3+\frac{\lambda(w_W,w_Z)}{4 w_W w_Z} \right) \,, \nonumber \\
\Gamma(H^+ \to \gamma W^+) & =  \frac{3 e^2 m_{H^\pm}}{16 \pi} \left(-\frac{3 g v_\Delta }{\sqrt{2} m_W} c_{\phi^+} + s_{\phi^+} \right)w_W (1-w_W) \,,
\end{align}
with $w_i = m^2_i/m^2_{H^\pm}$. It is known that the $\lambda_4$ parameter determines the order of the Higgs triplet masses. Therefore, it is expected that $H^+$ can decay to $H^{++}$ and $H^0(A^0)$ through the three-body decay when $\lambda_4 >0$ and $\lambda_4 < 0$, respectively. Similar to the $H^{++} \to H^+ W^{+*}$ decay, we write the partial decay rates for $H^+ \to (H^{++} W^{-*}, H^0(A^0) W^{+*}$)  as: 
\begin{align}
\Gamma(H^+ \to H^{++} W^{-*}) &= \frac{3 g^4 m_{H^{\pm}}}{2^8 \pi^3} J_0(w_W, w_{H^{\pm\pm}})\,,~~~~ {\rm \lambda_4 > 0}\,,\nonumber \\
\Gamma(H^+ \to S W^{+*}) & = \frac{3 g^4 m_{H^{\pm}}}{2^9 \pi^3} J_0(w_W, w_{S})\,,~~~~~~~~ {\rm \lambda_4 < 0}\,,
\end{align}
with $S=H^0(A^0)$. 

\begin{figure}[phtb]
\includegraphics[scale=0.6]{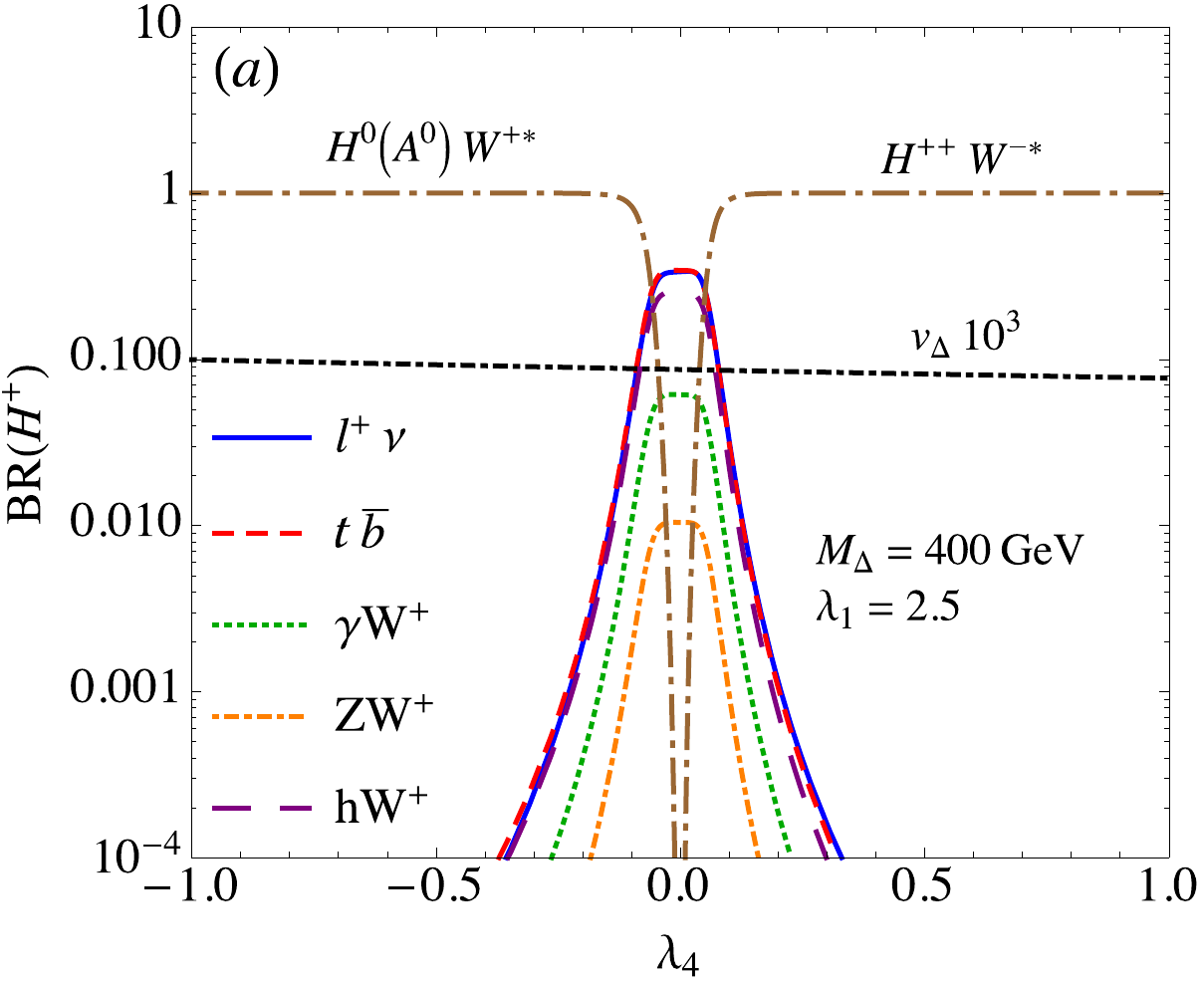}
\includegraphics[scale=0.6]{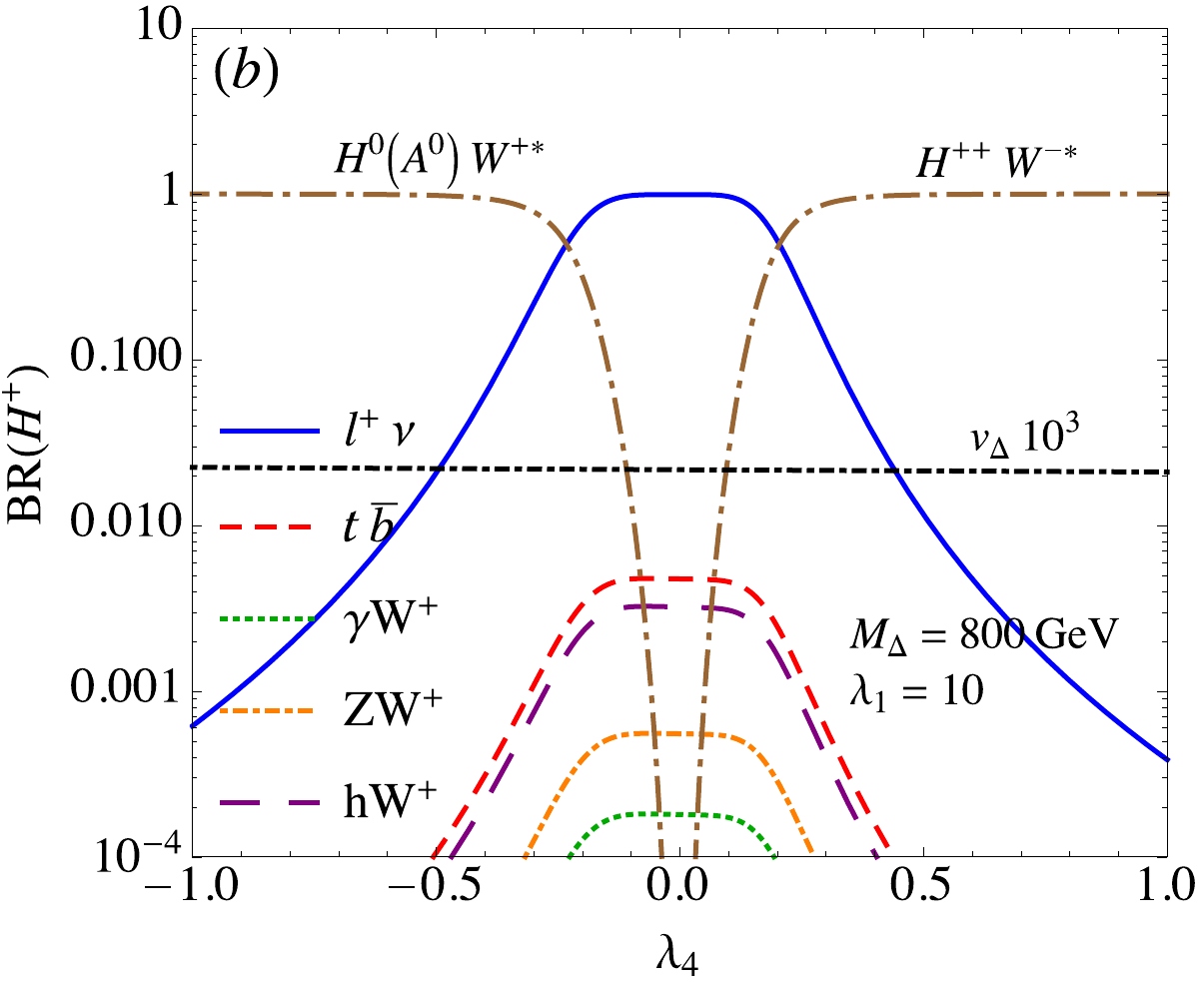}
 \caption{ This legend is the same as that shown in Fig.~\ref{fig:Hcc_decay}  with the exception of  the $H^+$ decays. }
\label{fig:Hc_decay}
\end{figure}

\begin{figure}[phtb]
\includegraphics[scale=0.6]{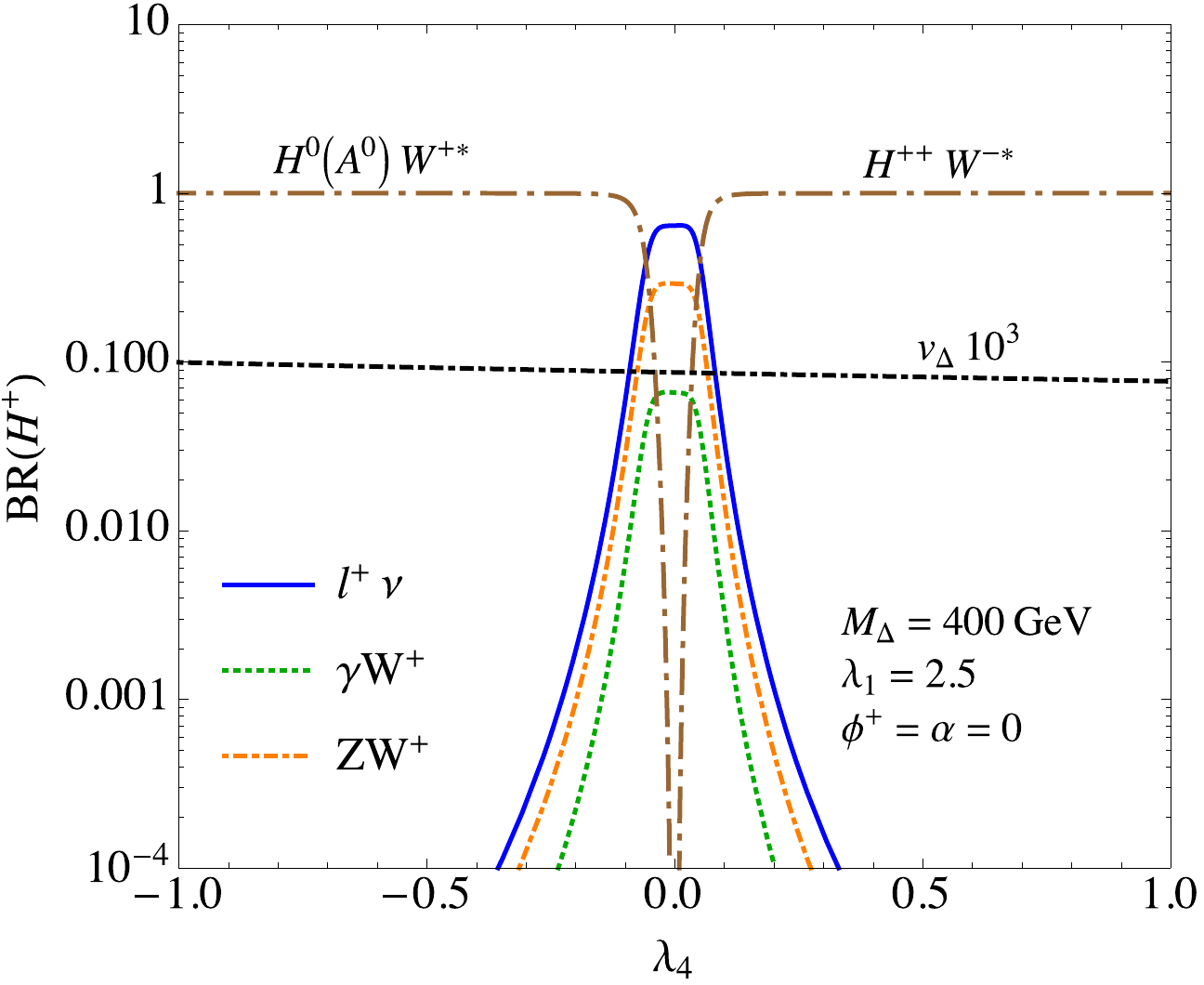}
 \caption{ BRs for $H^+$ decaying to $\ell^+ \nu$, $\gamma W^+$,  $ZW^+$, $H^{++}W^{-*}$, and $SW^{+*}$, where $\phi^+=\alpha=0$, $M_{\Delta}=400$ GeV, and $\lambda_1=2.5$ are used. }
\label{fig:Hc_no-mixing}
\end{figure}

Based on the  partial decay rate formulations, we show the BR for each decay mode as a function of $\lambda_4$  in Fig.~\ref{fig:Hc_decay}(a) and (b), where the plots (a) and (b) correspond to ($M_\Delta=400$ GeV,  $\lambda_1=2.5$) and ($M_\Delta=800$ GeV, $\lambda_1=10$), respectively, and we have summed all possible charged lepton flavors in the $\ell^+ \nu$ mode. From the plots, it can be clearly seen that when $|\lambda_4|>0.1 (0.3)$ for $M_\Delta=400 (800)$ GeV, the three-body decay channels are the main decays, where the associated mass differences in scalars are $|m_{H^+}-m_{H^{\pm\pm},S}| >1.55 (2.32)$ GeV. That is, in the model, the two-body $H^+$ decays can have the significant signals in the scheme with $m_{H^{\pm\pm}}\approx m_{H^\pm}\approx m_{S}$. In such a degenerate scheme, it is found that for $m_\Delta=400$ GeV, the BRs of the two-body decays follow $BR(\ell \nu)\approx BR(t\bar{b}) \gg  BR(hW^+)> BR(\gamma W^+) >BR(ZW^+)$, and for $M_\Delta=800$ GeV, the situation becomes $BR(\ell \nu)\gg BR(t \bar{b}) > BR(hW^+)\gg BR(ZW^+) > BR(\gamma W^+)$. For illustration, we show the numerical values with $\lambda_4=0$ in Table~\ref{tab:vHc}.  In addition, in order to understand  the scalar mixing influence on the BRs, we show the BRs with $\phi^+=\alpha=0$ in Fig.~\ref{fig:Hc_no-mixing}, where $M_\Delta=400$ GeV and $\lambda_1=2.5$ are used. It can be seen that without the $\phi^+$ and $\alpha$ mixing effects, the contributions to the $t\bar{b}$ and $hW^+$ modes vanish, and the BR order follows $BR(H^+\to \ell^+ \nu)>BR(H^+\to ZW^+)> BR(H^+\to \gamma W^+)$.

\begin{table}[htp]
\caption{BRs of the $H^+$ decays with $\lambda_4=0$, where $\lambda_1=2.5$ for $M_\Delta=400$ GeV and $\lambda_1=10$ for $M_\Delta=800$ GeV are used. }
\begin{tabular}{c|ccccc} \hline \hline

Mode ~ & $\ell^+ \nu$  &  $t \bar{b}$  &  $hW^+$  &  $\gamma W^+$  & $Z W^+$ \\ \hline 
$(M_\Delta=400\ {\rm GeV}, BR)$ ~ &  0.34  & 0.34  & 0.25 & 0.06 & 0.01 \\ \hline 
$(M_\Delta=800\ {\rm GeV}, BR)$  ~& ~~ 0.99 ~~ & ~~0.005 ~~& ~~0.003 ~~& ~~$0.17\cdot 10^{-3}$ ~~& ~~$0.55\cdot 10^{-3}$ ~~\\ \hline \hline
\end{tabular}
\label{tab:vHc}
\end{table}%

\subsection{$H^0$ and $A^0$ decays}

From Eq.~(\ref{eq:Yukawa}), the neutral Higgs triplet scalars do not directly couple to the charged leptons. Thus, without the scalar mixings, the CP-even $H^0$ decays  to the final states, such as $\nu \nu$, $hh$, $W^+W^-$, and $ZZ$, whereas the CP-odd $A^0$ can only has the invisible $A^0\to \nu \nu$ decay. Including the mixings with the SM neutral Goldstone boson $G^0$ and with the SM Higgs, it can be found that $H^0$ can further decay to $t\bar{t}$ and that $A^{0}$ can decay to $t\bar{t}$ and $hZ$. Therefore, according to the introduced Yukawa and gauge couplings, the partial decay rates of the fermionic $H^0/A^0$ decays can be expressed as:
\begin{align}
\Gamma(S \to \nu \nu) &= \frac{m_S}{8\pi} \sum_j \left( {\bf h}^{\ell\dagger} {\bf h}^\ell \right)_{jj}\nonumber \\
\Gamma(H^0 \to t \bar t) &= \frac{m_{H^0}}{8\pi}\frac{m^2_t}{v^2_h} s^2_{\phi^0} \left(1-\frac{4m^2_t}{m^2_{H^0}} \right)^{3/2}\,, \nonumber \\
\Gamma(A^0 \to t \bar t )& =  \frac{m_{A^0}}{8\pi}\frac{m^2_t}{v^2_h} s^2_{\phi^0} \left(1-\frac{4m^2_t}{m^2_{A^0}} \right)^{1/2}\,,
\end{align}
whereas the $H^0/A^0$ diboson decays are given as:
\begin{align}
\Gamma(H^0 \to h h) &= \frac{m_{H^0}}{32\pi}  \left[ (\lambda_1 + \lambda_4) \frac{2 v_\Delta - v_h s_\alpha}{2 m_{H^0}} -\sqrt{2} \frac{\mu_\Delta}{m_{H^0}}\right]^2 \left( 1 - \frac{4 m^2_h}{m^2_{H^0}}\right)^{1/2}\,,\nonumber \\
\Gamma(H^0\to W^+W^- )&=\frac{g^2 m_{H^0}}{16\pi}  \left( g\frac{v_\Delta}{m_{H^0}} + \frac{m_W}{m_{H^0}} s_\alpha\right)^2  \left( 2+ \frac{(1-2z_W)^2}{4 z^2_W}  \right) \sqrt{1-4 z_W}\,,\nonumber \\
\Gamma(H^0\to ZZ) &=\frac{g^2 m_{H^0}}{32\pi c^4_W}  \left( 2 g\frac{v_\Delta}{m_{H^0}} + \frac{m_W}{m_{H^0}} s_\alpha\right)^2  \left( 2+ \frac{(1-2z_Z)^2}{4 z^2_Z}  \right) \sqrt{1-4 z_Z}\,, \nonumber \\
\Gamma(A^0 \to h Z) & = \frac{g^2 m_{A^0}}{16 \pi} \left( s_\alpha + \frac{s_{\phi^0}}{2}\right)^2 \frac{\lambda(z_Z,z_h)^{3/2}}{z_W} \,,
\end{align}
with  $z_i = m^2_i/m^2_{S}$.  When $H^0(A^0)$ is the heaviest scalar, i.e. $\lambda_4 > 0$, similar to the cases in the $H^+$ and $H^{++}$ decays,  the three-body decays $H^0(A^0) \to H^+ W^{-*}, H^- W^{+*}$ are open and the partial decay rates are written as:
\begin{equation}
\Gamma(S \to H^+ W^{-*})=\Gamma(S\to H^- W^{+*}) =  \frac{3 g^4 m_{S}}{2^9 \pi^3} J_0(z_W, z_{H^\pm})\,,~~~\lambda_4 >0\,.
\end{equation}

\begin{figure}[phtb]
\includegraphics[scale=0.55]{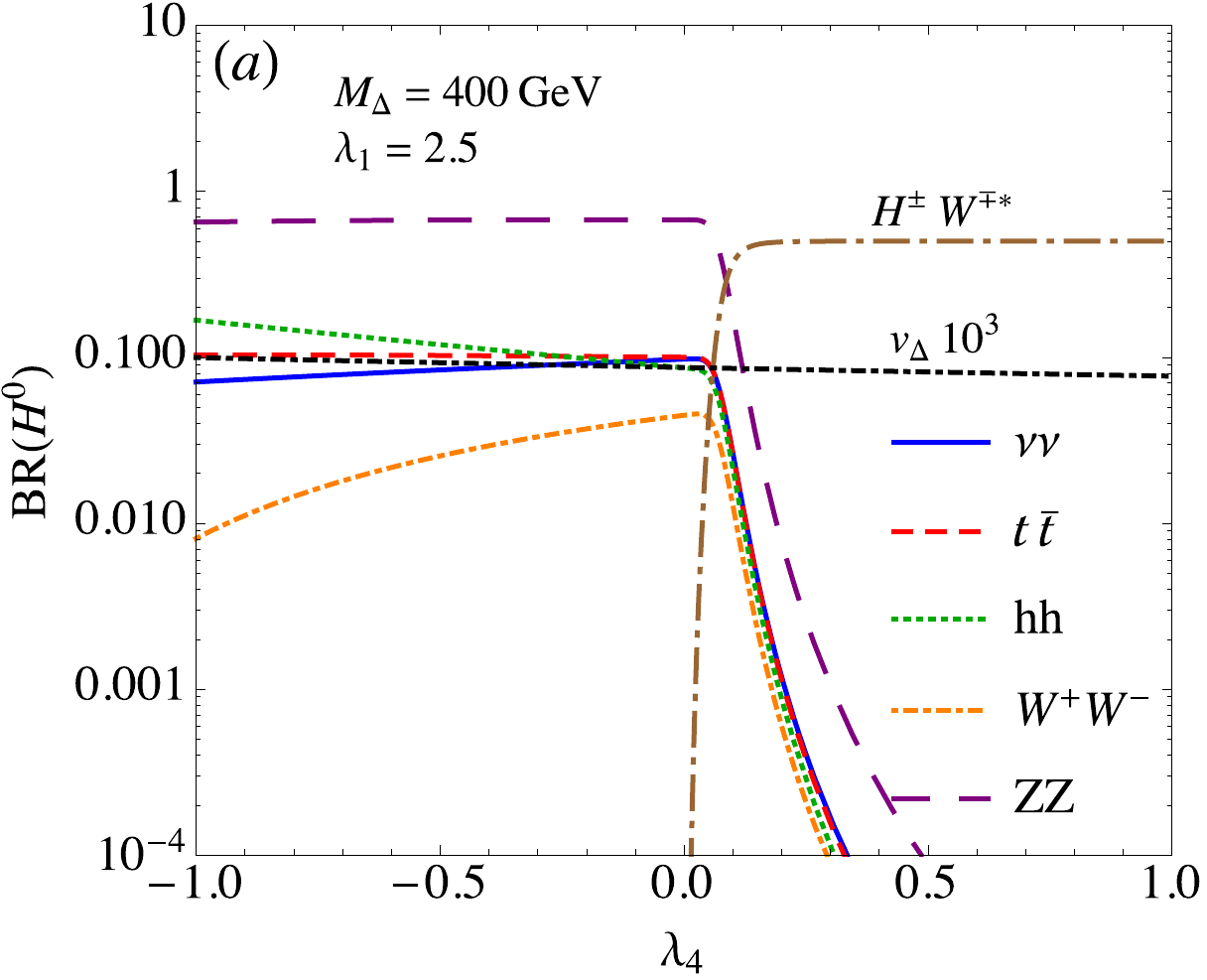}
\includegraphics[scale=0.55]{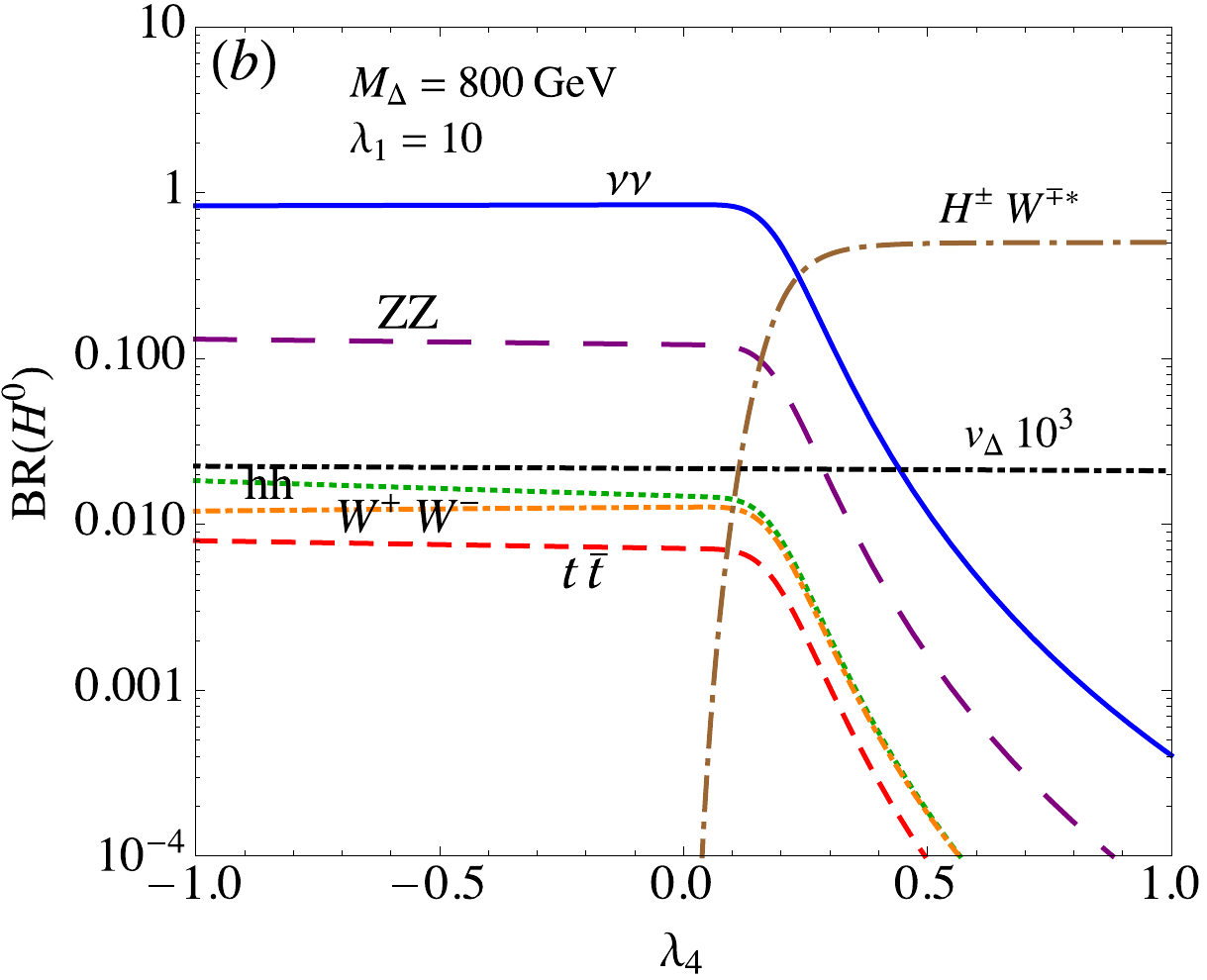}
\includegraphics[scale=0.55]{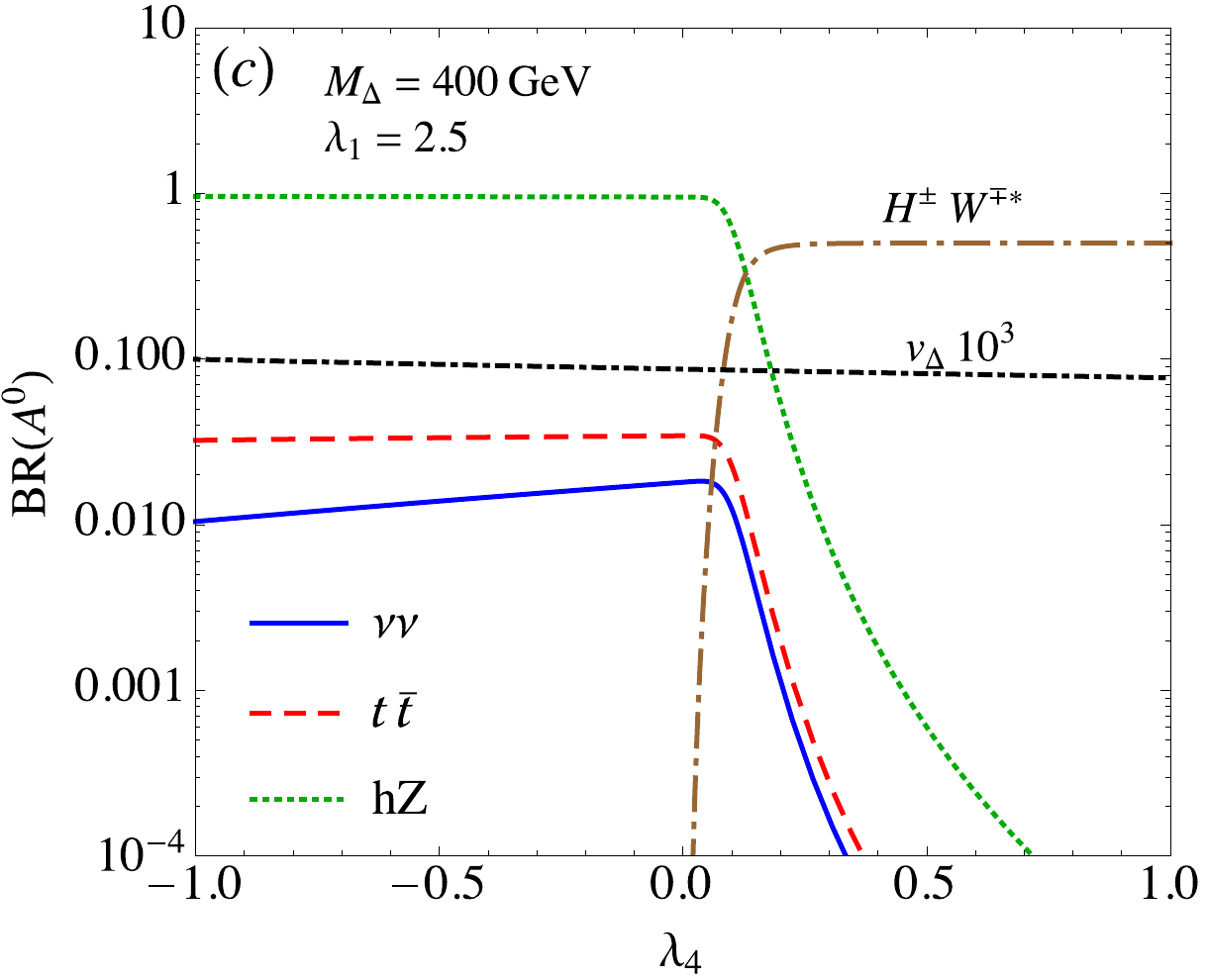}
\includegraphics[scale=0.55]{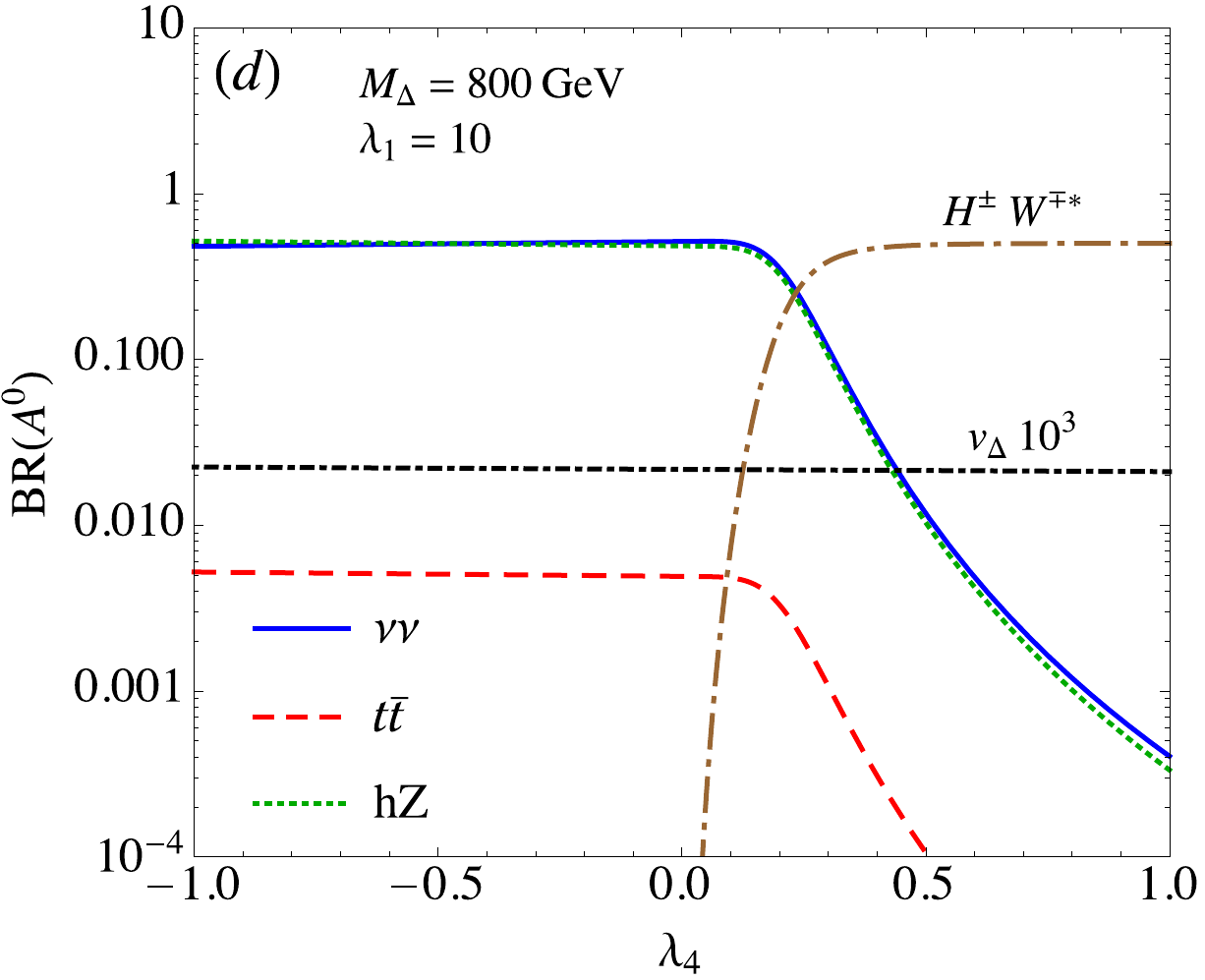}
 \caption{ This legend is the same as that shown in Fig.~\ref{fig:Hcc_decay}, where plots (a) and (b) are for $H^0$ decays, and plots (c) and (d) are for $A^0$ decays.   }
\label{fig:S_decay}
\end{figure}

\begin{figure}[phtb]
\includegraphics[scale=0.6]{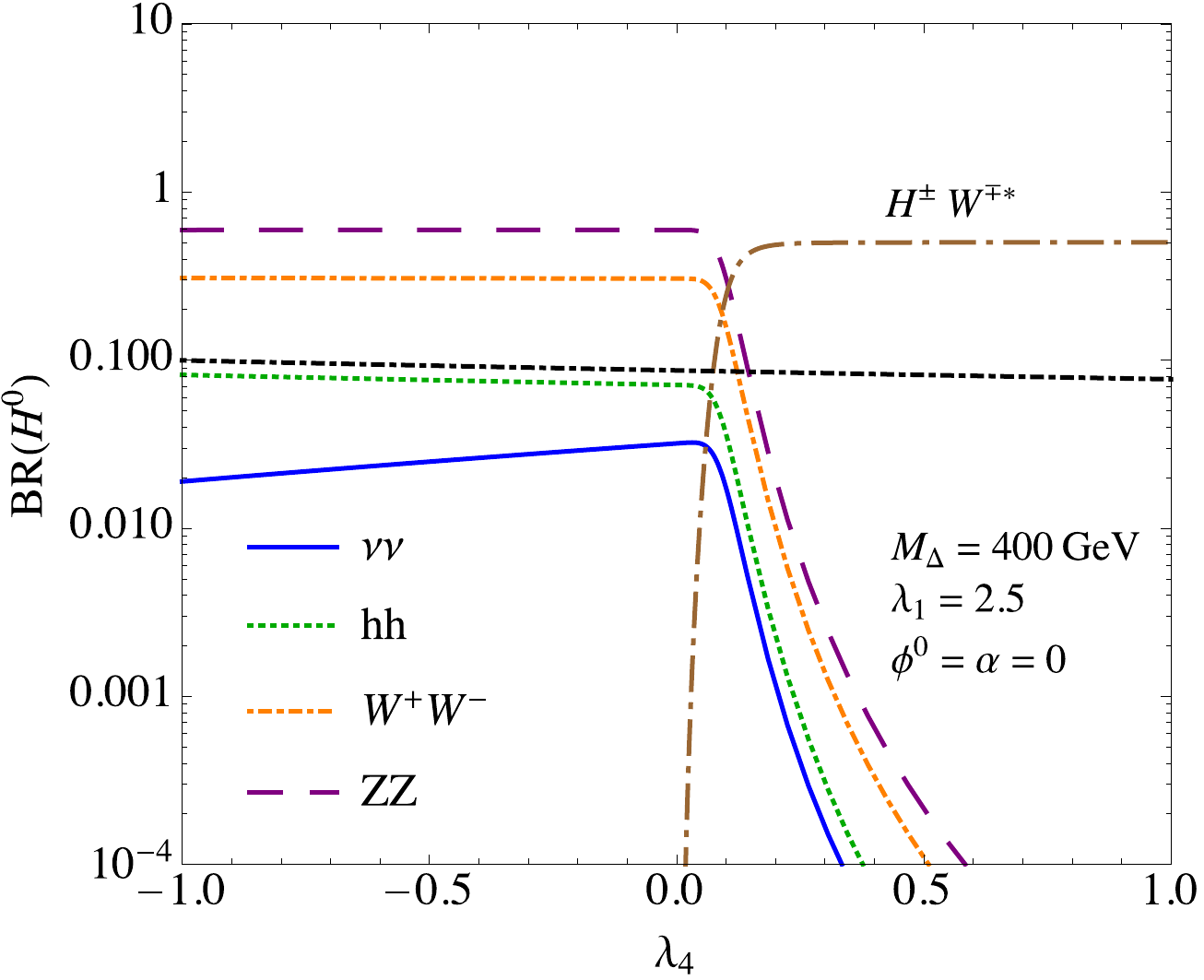}
 \caption{ BRs for $H^0$ decay into $\nu \nu$, $hh$,  $W^+W^-$, $ZZ$, and $H^{\pm}W^{\mp*}$, where $\phi^+=\alpha=0$, $M_{\Delta}=400$ GeV, and $\lambda_1=2.5$ are used. }
\label{fig:H0_no-mixing}
\end{figure}

Using the obtained partial decay rates, we show the BR for each decay channel as a function of $\lambda_4$ in Fig.~\ref{fig:S_decay}, where plots (a) and (b) denote the $H^0$ decays with $(M_\Delta=400\, {\rm GeV}, \lambda_1=2.5)$ and $(M_\Delta=800\, {\rm GeV}, \lambda_1=10)$,  and plots (c) and (d) are for the $A^0$ decays with the same parameter values taken in plots (a) and (b), respectively. From the results,  it can be seen that  the three-body decays are the dominant decay channels when $\lambda_4 \gtrsim 0.3$. However, for $\lambda_4 <0$, the $H^0(A^0)$ decay properties depend on the parameter values.  For $M_\Delta=400$ GeV and $\lambda_1=2.5$, it can be seen that the BR order in the $H^0$ two-body decays follows $BR(ZZ)\gg BR(hh) \sim BR(t \bar{t}) > BR(\nu \nu) > BR(W^+ W^-)$, and  that in the $A^0$  two-body decays is $BR(hZ)\gg BR(t\bar{t}) > BR(\nu \nu)$. For $M_\Delta=800$ GeV and $\lambda_{1}=10$, the BR order in the $H^0$ decays is $BR(\nu \bar\nu)\gg BR(ZZ) > BR(hh) > BR(W^+ W^-)> BR( t \bar t)$, and that in the $A^0$ decays is $BR(hZ)\sim  BR(\nu \bar\nu) \gg BR(t \bar{t})$. For clarity, we show the numerical values for the $H^0$  and $A^0$ decays with $\lambda_4=0$ in Table~\ref{tab:vH0A0}. In order to illustrate the $\phi^0$ and $\alpha$ mixing angle influence, we show the relative BRs as a function of $\lambda_4$  with $\phi^0=\alpha=0$ in Fig.~\ref{fig:H0_no-mixing}, where $m_\Delta=400$ GeV and $\lambda_1=2.5$ are fixed. According to the results, it can be found that  $BR(H^0\to t\bar{t})$ vanishes and that $BR(H^0\to W^+ W^-)\sim 0.3$, which is close to $BR(H^0\to ZZ)$.  Accordingly, we see that the BR of $H^0\to W^+ W^-$ obtains a destructive contribution from the $\alpha$ mixing effect.  When  $\phi^0=\alpha=0$, $A^0$ only can decay to $\nu\nu$ in the region of $\lambda_4 < 0$; therefore, we do not explicitly show the situation for the $A^0$ decay. 

\begin{table}[htp]
\caption{BRs of the $H^0$ and $A^0$ decays with $\lambda_4=0$, where $\lambda_1=2.5$ for $M_\Delta=400$ GeV and $\lambda_1=10$ for $M_\Delta=800$ GeV are used. }
\begin{tabular}{c|ccccc} \hline \hline

Mode($H^0$) ~ & $\nu \nu$  &  $t \bar{t}$  &  $hh$  &  $W^+ W^-$  & $ZZ$ \\ \hline 
$(M_\Delta=400\ {\rm GeV}, BR)$ ~ &  0.097  & 0.100  & 0.086 & 0.045 & 0.672 \\ \hline 
$(M_\Delta=800\ {\rm GeV}, BR)$  ~& ~~ $0.844$ ~~ & ~~$0.007$ ~~& ~~$0.015$ ~~& ~~$0.013$ ~~& ~~$0.121$ ~~\\ \hline \hline
Mode($A^0$)~  & $\nu \nu$  &  $t \bar{t}$  &  $hZ$  \\ \hline
$(M_\Delta=400\ {\rm GeV}, BR)$ ~ &  ~~0.018  ~~& ~~0.034 ~~ & ~~0.948~~ \\ \hline
$(M_\Delta=800\ {\rm GeV}, BR)$  ~& ~~ $0.513$ ~~ & ~~$0.005$ ~~& ~~$0.482$ \\ \hline\hline

\end{tabular}
\label{tab:vH0A0}
\end{table}%

\section{Conclusion}

Using the scotogenic approach, we studied the radiatively induced lepton-number violation dimension-3 term $\mu_\Delta H^T i \tau_2 \Delta^\dag H$ in the base of the type-II seesaw model, where the introduced dark vector-like doublet lepton $X$ and dark right-handed singlet Majorana lepton $N$ are the mediators in the loop. It was found that the dynamically induced Higgs triplet VEV is limited in the region of $10^{-5}-10^{-4}$ GeV when the relevant parameters satisfy the constraints from the DM measurements. Due to the DM direct detection constraints, only the singlet Majorana lepton can be the DM candidate in the model, and the DM mass depends on and is close to  the $m_X$ parameter. 

 In the model, the Higgs triplet VEV, $v_\Delta$,  depends not only  on the $\mu_\Delta$ and $M_\Delta$ parameters, but also on the $\lambda_{1,4}$ parameters in the scalar potential, which dictate the SM Higgs  couplings to the doubly and singly charged Higgses. Moreover, the mass ordering of the Higgs triplet scalars is dictated by the $\lambda_4$ sign.  We showed that  the Higgs diphoton decay and the oblique $T$-parameter can further bound the $\lambda_{1,4}$ parameters.  As a result, we obtain $|m_{H^{\pm\pm}}-m_{H^\pm}|\lesssim 50$ GeV. 
 
 We did not explicitly study the collider signatures in this work. Rather, we analyzed the decay channels of each Higgs triplet scalar and estimated the associated branching ratios in detail. We found that the scalar mixing effects have an important  influence on the partial decay rates of the singly charged-Higgs, CP-even scalar, and CP-odd pseudoscalar in the near degenerate masses (i.e. $\lambda_4 \ll 1$). In the non-degenerate mass region, the branching ratios of the Higgs triplet scalar decays are dominated by the three-body decays when they are kinematically allowed. \\

\appendices

\section{Scalar mass squares and mixing angles} \label{ap:mixing}

The symmetric mass-square matrices in Eqs.~(\ref{eq:G+D+}), (\ref{eq:G0ImD0}), and (\ref{eq:hReD0})  can be generally expressed as:
 \begin{equation}
 A=
\left(
\begin{array}{cc}
 a_{11}  &   a_{12}   \\
 a_{12} &  a_{22}    \\
\end{array}
\right)\,,
\end{equation}
where the $2\times 2$ symmetric matrix can be diagonalized using an orthogonal matrix $U$ through  $A^{\rm dia}= U A U^T$ with the parametrization:
 \begin{equation} 
  U=\left(
\begin{array}{cc}
 \cos\phi  &  -\sin\phi   \\
 \sin\phi &  \cos\phi    \\
\end{array}
\right)\,.
  \end{equation}
 It can be found that the two eigenvalues $A_L$ and $A_H$ and the mixing angle $\phi$ can be expressed as:
  \begin{align}
  A_{L(H)}&= \frac{a_{11}+a_{22}}{2} \mp  \frac{1}{2} \sqrt{(a_{11}-a_{22} )^2 + 4 a^2_{12}}\,, \nonumber \\
  \tan2\phi & = \frac{2 a_{12}}{a_{22} -a_{11}}\,.  \label{eq:eigen_V}
  \end{align}

Since the $(G^+, \Delta^+)$ and $(G^0, Im\Delta^0)$ states have massless Goldstone bosons, their physical mass squares  can be straightforwardly obtained by taking traces of the mass-square matrices, i.e. $m^2_{H^+}={\rm Tr} A_{G^+ \Delta^+}$ and $m^2_{A^0}={\rm Tr}A_{G^0 Im\Delta^0}$. From Eq.~(\ref{eq:eigen_V}), the corresponding mixing angles for diagonalizing $A_{G^+ \Delta^+}$ and $A_{G^0 Im\Delta^0}$ shown in Eqs.~(\ref{eq:G+D+}) and (\ref{eq:G0ImD0}) are given as:
 \begin{align}
 \tan2\phi^+ & = \frac{-2\sqrt{2} v_\Delta v_h }{ v^2_h - 2 v^2_\Delta } \approx - \frac{2\sqrt{2} v_\Delta}{v_h}\,, \nonumber \\
 \tan2\phi^0 & = \frac{-4 v_\Delta v_h}{v^2_h -4 v^2_\Delta}\approx - \frac{4 v_\Delta}{v_h}\,.
 \end{align}
 Clearly, if $v_\Delta \ll v_h$, the mixing angles are small.
In the case of the  $(Re\Phi^0, Re\Delta^0)$ states, we do not have a simple way to obtain their eigenvalues. If we use $h$ and $H^0$  to denote the light and heavy scalars, their eigenvalues $m_{h(H^0)}$ and mixing angles should follow Eq.~(\ref{eq:eigen_V}), where the associated matrix elements are:
 \begin{align}
 a_{11} & = \frac{\lambda v^2_h}{2}\,,  \nonumber \\
 a_{12} & = (\lambda_1 + \lambda_4) v_h v_\Delta -\sqrt{2} v_h \mu_\Delta \,, \nonumber \\
 a_{22} & = \frac{\mu_\Delta v^2_h }{\sqrt{2} v_\Delta } + 2v^2_\Delta \left( \lambda_2 + \lambda_3\right)\,.
 \end{align}
 As a result, the mixing between $Re\Phi^0$ and $Re\Delta^0$ can be formulated as:
  \begin{equation}
  \tan2\alpha \approx \frac{2(\lambda_1 + \lambda_4)  v_\Delta - 2\sqrt{2}  \mu_\Delta}{\mu_\Delta v_h/(\sqrt{2} v_\Delta) - \lambda v^2_h /2 }\,, \label{eq:alpha}
  \end{equation}
  where we have used  $\alpha$ instead of $\phi$, and the $v^2_\Delta$ effect in the denominator is dropped due to $v^2_\Delta \ll 1$.  In addition to $v_\Delta < \mu_\Delta$, the numerator  in Eq.~(\ref{eq:alpha}) is much smaller than the denominator; hence, the $\alpha$ angle should be of the order of $\sim \mu_\Delta v_h /M^2_\Delta$. Using $\mu_\Delta=10^{-3}$ GeV, $v_h=246$ GeV, and $M_\Delta=400$ GeV, the $\alpha$ value can be estimated to be $\alpha\sim 1.54 \times 10^{-6}$. 

\section{ Higgs triplet gauge coupling}\label{ap:gauge}

The Higgs triplet couplings to the gauge bosons can be obtained from the $\Delta$ kinetic term shown in Eq.~(\ref{eq:kin}), where the covariant derivation can be found in Eq.~(\ref{eq:cov}).  Accordingly, we can derive the triple couplings of the Higgs triplet scalars and the gauge bosons as:
 \begin{align}
 {\cal L}_{\rm kin} & = Tr[ (D_\mu \Delta)^\dag (D^\mu \Delta)] \nonumber \\
 & \supset \left\{ ig \left( H^{--} \partial_\mu H^+ - H^+ \partial_\mu H^{--} \right) W^{+\mu}  + \frac{ig}{\sqrt{2}} \left( H^0 \partial_\mu H^{-1} -H^{-1} \partial_\mu H^0 \right) W^{+\mu} \right.\nonumber \\
 & \left.- \frac{g}{\sqrt{2}}  \left( A^0 \partial_\mu H^{-1} -H^{-1} \partial_\mu A^0 \right) W^{+\mu} +H.c.\right\}
  -\frac{g}{c_W} \left( H^0\partial_\mu A^0 - A^0 \partial_\mu H^0 \right)Z^\mu \nonumber \\
  & + i \left( H^+ \partial_\mu H^{-} - H^{-} \partial_\mu H^{+}\right) \left( e A_\mu - \frac{g s^2_W}{c_W} Z_\mu\right) \nonumber \\
  & + i  \left( H^{++} \partial_\mu H^{--} - H^{--} \partial_\mu H^{++}\right) \left(2 eA_\mu + \frac{g(1-2s^2_W)}{c_W} Z_\mu \right) \nonumber \\
  & + g^2 v_\Delta H^0 W^+_\mu W^{-\mu} + \frac{1}{2} \left(\frac{2g^2 v_\Delta}{c^2_W} \right)H^0 Z_\mu Z^\mu \nonumber \\
  & - \frac{g v_\Delta}{\sqrt{2}} \left[ H^{-} W^{+\mu} \left(3eA_\mu + \frac{g}{c_W}(1-3s^2_W) Z_\mu \right) + H.c. \right]\nonumber \\
  & - \frac{1}{2} \left(\sqrt{2} g^2 v_\Delta \right) \left( H^{--} W^+_\mu W^{+\mu} + H.c. \right)\,. \label{eq:Kin_TH}
 \end{align}
We note that although Eq.~(\ref{eq:Kin_TH}) does not include the $\phi^{+,0}$ and $\alpha$ mixing effects, we have used the physical state notations for $H^+$, $H^0$, and $A^0$.

\section{Loop integral functions }\label{ap:loop}

The loop integral functions $A^h_{0,1/2,1}$ for $h\to Z\gamma$ shown in Eq.~(\ref{eq:As}) are given as:
 \begin{align}
 A^h_0(\tau_h, \tau_Z)&= I_1(\tau_h, \tau_Z)\,, A^h_{1/2} = I_1(\tau_h, \tau_Z) - I_2(\tau_h, \tau_Z)\,, \nonumber \\
 A^h_1(\tau_h, \tau_Z) & = 4(3-\tan^2\theta_W) I_2(\tau_h,\tau_Z) + \left[ \left( 1+ \frac{2}{\tau_h}\right)\tan^2_W-\left( 5+ \frac{2}{\tau_h}\right) \right] I_1(\tau_h, \tau_Z)\,,
 \end{align}
with 
 \begin{align}
 I_1(x,y) & =\frac{x y}{2(x-y)} + \frac{x^2 y^2}{2(x-y)^2} \left( f(x) - f(y) \right) + \frac{x^2 y}{(x-y)^2} \left( g(x) -g(y) \right)\,, \nonumber \\
 I_2(x,y) & = - \frac{x y}{ 2(x-y)} \left( f(x) - f(y) \right)\,,
  \end{align}
 where the function $f(\tau)$ can be found in Eq.~(\ref{eq:f_function}),  and the function $g(\tau)$ is given as: 
  \begin{align}
  g(\tau)=\left\{
\begin{array}{c}
 \sqrt{\tau-1}\sin^{-1}(1/\sqrt{\tau}) \,,  ~~~~~~~~ (\tau \geq 1)\,, \\
  \frac{\sqrt{1-\tau}}{2} \left( \ln\frac{1+\sqrt{1-\tau}}{1-\sqrt{1-\tau}} -i \pi\right)\,, ~~~~~(\tau< 1)\,.
\end{array}
\right.
  \end{align} 
  \\

\noindent{\bf Acknowledgments}

This work was partially supported by the Ministry of Science and Technology of Taiwan,  
under grants MOST-106-2112-M-006-010-MY2 (CHC).\\


\begin{thebibliography}{99}

\bibitem{Englert:1964et} 
  F.~Englert and R.~Brout,
  Phys.\ Rev.\ Lett.\  {\bf 13}, 321 (1964).

\bibitem{Higgs:1964pj} 
  P.~W.~Higgs,
  Phys.\ Rev.\ Lett.\  {\bf 13}, 508 (1964).

\bibitem{Guralnik:1964eu} 
  G.~S.~Guralnik, C.~R.~Hagen and T.~W.~B.~Kibble,
  Phys.\ Rev.\ Lett.\  {\bf 13}, 585 (1964).




\bibitem{Schechter:1980gr} 
  J.~Schechter and J.~W.~F.~Valle,
  Phys.\ Rev.\ D {\bf 22}, 2227 (1980).
  
\bibitem{Magg:1980ut} 
  M.~Magg and C.~Wetterich,
  Phys.\ Lett.\  {\bf 94B}, 61 (1980).

\bibitem{Cheng:1980qt} 
  T.~P.~Cheng and L.~F.~Li,
  Phys.\ Rev.\ D {\bf 22}, 2860 (1980).

\bibitem{Lazarides:1980nt} 
  G.~Lazarides, Q.~Shafi and C.~Wetterich,
  Nucl.\ Phys.\ B {\bf 181}, 287 (1981).

\bibitem{Mohapatra:1980yp} 
  R.~N.~Mohapatra and G.~Senjanovic,
  Phys.\ Rev.\ D {\bf 23}, 165 (1981).

\bibitem{Chun:2003ej} 
  E.~J.~Chun, K.~Y.~Lee and S.~C.~Park,
  Phys.\ Lett.\ B {\bf 566}, 142 (2003)
  [hep-ph/0304069].

\bibitem{Franceschini:2013aha} 
  R.~Franceschini and R.~N.~Mohapatra,
  Phys.\ Rev.\ D {\bf 89}, no. 5, 055013 (2014)
  [arXiv:1306.6108 [hep-ph]].

\bibitem{Cai:2017jrq} 
  Y.~Cai, J.~Herrero-Garc\'{i}a, M.~A.~Schmidt, A.~Vicente and R.~R.~Volkas,
  Front.\ in Phys.\  {\bf 5}, 63 (2017)
  [arXiv:1706.08524 [hep-ph]].

\bibitem{Ma:2006km} 
  E.~Ma,
  Phys.\ Rev.\ D {\bf 73}, 077301 (2006)
  [hep-ph/0601225].

\bibitem{Fraser:2015mhb} 
  S.~Fraser, C.~Kownacki, E.~Ma and O.~Popov,
  Phys.\ Rev.\ D {\bf 93}, no. 1, 013021 (2016)
  [arXiv:1511.06375 [hep-ph]].


\bibitem{Brdar:2013iea} 
  V.~Brdar, I.~Picek and B.~Radovcic,
  Phys.\ Lett.\ B {\bf 728}, 198 (2014)
  [arXiv:1310.3183 [hep-ph]].


\bibitem{Ma:2014cfa} 
  E.~Ma,
  Phys.\ Lett.\ B {\bf 732}, 167 (2014)
  [arXiv:1401.3284 [hep-ph]].

\bibitem{Molinaro:2014lfa} 
  E.~Molinaro, C.~E.~Yaguna and O.~Zapata,
  JCAP {\bf 1407}, 015 (2014)
  [arXiv:1405.1259 [hep-ph]].

\bibitem{Vicente:2014wga} 
  A.~Vicente and C.~E.~Yaguna,
  JHEP {\bf 1502}, 144 (2015)
  [arXiv:1412.2545 [hep-ph]].

\bibitem{Merle:2015gea} 
  A.~Merle and M.~Platscher,
  Phys.\ Rev.\ D {\bf 92}, no. 9, 095002 (2015)
  [arXiv:1502.03098 [hep-ph]].

\bibitem{Culjak:2015qja} 
  P.~Culjak, K.~Kumericki and I.~Picek,
  Phys.\ Lett.\ B {\bf 744}, 237 (2015)
  [arXiv:1502.07887 [hep-ph]].

\bibitem{Merle:2015ica} 
  A.~Merle and M.~Platscher,
  JHEP {\bf 1511}, 148 (2015)
  [arXiv:1507.06314 [hep-ph]].
  
\bibitem{Yu:2016lof} 
  J.~H.~Yu,
  Phys.\ Rev.\ D {\bf 93}, no. 11, 113007 (2016)
  [arXiv:1601.02609 [hep-ph]].

\bibitem{Ahriche:2016cio} 
  A.~Ahriche, K.~L.~McDonald and S.~Nasri,
  JHEP {\bf 1606}, 182 (2016)
  [arXiv:1604.05569 [hep-ph]].

\bibitem{Ferreira:2016sbb} 
  P.~M.~Ferreira, W.~Grimus, D.~Jurciukonis and L.~Lavoura,
  JHEP {\bf 1607}, 010 (2016)
  [arXiv:1604.07777 [hep-ph]].

\bibitem{Rocha-Moran:2016enp} 
  P.~Rocha-Moran and A.~Vicente,
  JHEP {\bf 1607}, 078 (2016)
  [arXiv:1605.01915 [hep-ph]].

\bibitem{Chowdhury:2016mtl} 
  T.~A.~Chowdhury and S.~Nasri,
  JCAP {\bf 1701}, no. 01, 041 (2017)
  [arXiv:1611.06590 [hep-ph]].

\bibitem{Hessler:2016kwm} 
  A.~G.~Hessler, A.~Ibarra, E.~Molinaro and S.~Vogl,
  JHEP {\bf 1701}, 100 (2017)
  [arXiv:1611.09540 [hep-ph]].

\bibitem{Diaz:2016udz} 
  M.~A.~D\'{i}az, N.~Rojas, S.~Urrutia-Quiroga and J.~W.~F.~Valle,
  JHEP {\bf 1708}, 017 (2017)
  [arXiv:1612.06569 [hep-ph]].

\bibitem{Borah:2017dfn} 
  D.~Borah and A.~Gupta,
  Phys.\ Rev.\ D {\bf 96}, no. 11, 115012 (2017)
  [arXiv:1706.05034 [hep-ph]].

\bibitem{Abada:2018zra} 
  A.~Abada and T.~Toma,
  JHEP {\bf 1804}, 030 (2018)
  [arXiv:1802.00007 [hep-ph]].

\bibitem{Hagedorn:2018spx} 
  C.~Hagedorn, J.~Herrero-Garc\'{i}a, E.~Molinaro and M.~A.~Schmidt,
  JHEP {\bf 1811}, 103 (2018)
  [arXiv:1804.04117 [hep-ph]].

\bibitem{Hugle:2018qbw} 
  T.~Hugle, M.~Platscher and K.~Schmitz,
  Phys.\ Rev.\ D {\bf 98}, no. 2, 023020 (2018)
  [arXiv:1804.09660 [hep-ph]].
  
\bibitem{Baumholzer:2018sfb} 
  S.~Baumholzer, V.~Brdar and P.~Schwaller,
  JHEP {\bf 1808}, 067 (2018)
  [arXiv:1806.06864 [hep-ph]].

  
\bibitem{Rojas:2018wym} 
  N.~Rojas, R.~Srivastava and J.~W.~F.~Valle,
  Phys.\ Lett.\ B {\bf 789}, 132 (2019)
  [arXiv:1807.11447 [hep-ph]].

\bibitem{Borah:2018rca} 
  D.~Borah, P.~S.~B.~Dev and A.~Kumar,
  Phys.\ Rev.\ D {\bf 99}, no. 5, 055012 (2019)
  [arXiv:1810.03645 [hep-ph]].

\bibitem{CentellesChulia:2019gic} 
  S.~Centelles Chuli\'a, R.~Cepedello, E.~Peinado and R.~Srivastava,
  arXiv:1901.06402 [hep-ph].

\bibitem{Ma:2019yfo} 
  E.~Ma,
  Phys.\ Lett.\ B {\bf 793}, 411 (2019)
  [arXiv:1901.09091 [hep-ph]].
  
\bibitem{Kang:2019sab} 
  Kang, O.~Popov, R.~Srivastava, J.~W.~F.~Valle and C.~A.~Vaquera-Araujo,
  arXiv:1902.05966 [hep-ph].
  
\bibitem{Chen:2019nud} 
  C.~H.~Chen and T.~Nomura,
  arXiv:1903.03380 [hep-ph].
  
 
\bibitem{Kanemura:2012rj} 
  S.~Kanemura and H.~Sugiyama,
  Phys.\ Rev.\ D {\bf 86}, 073006 (2012)
  [arXiv:1202.5231 [hep-ph]].
 
 
\bibitem{Nomura:2016dnf} 
  T.~Nomura, H.~Okada and Y.~Orikasa,
  Phys.\ Rev.\ D {\bf 94}, no. 11, 115018 (2016)
  [arXiv:1610.04729 [hep-ph]].
 
 
\bibitem{Nomura:2017emk} 
  T.~Nomura and H.~Okada,
  Phys.\ Lett.\ B {\bf 774}, 575 (2017)
  [arXiv:1704.08581 [hep-ph]].
 
 
   
\bibitem{Peskin:1991sw} 
  M.~E.~Peskin and T.~Takeuchi,
  Phys.\ Rev.\ D {\bf 46}, 381 (1992).

  
  
  

\bibitem{CMS:2017pet} 
  CMS Collaboration [CMS Collaboration],
  CMS-PAS-HIG-16-036.


\bibitem{Aaboud:2017qph} 
  M.~Aaboud {\it et al.} [ATLAS Collaboration],
  Eur.\ Phys.\ J.\ C {\bf 78}, no. 3, 199 (2018)
  [arXiv:1710.09748 [hep-ex]].

\bibitem{Aaboud:2018qcu} 
  M.~Aaboud {\it et al.} [ATLAS Collaboration],
  Eur.\ Phys.\ J.\ C {\bf 79}, no. 1, 58 (2019)
  [arXiv:1808.01899 [hep-ex]].
  
\bibitem{Ucchielli:2018koe} 
  G.~Ucchielli [ATLAS Collaboration],
  PoS CHARGED {\bf 2018}, 008 (2019).




\bibitem{Akeroyd:2005gt} 
  A.~G.~Akeroyd and M.~Aoki,
  Phys.\ Rev.\ D {\bf 72}, 035011 (2005)
  [hep-ph/0506176].

\bibitem{delAguila:2008cj} 
  F.~del Aguila and J.~A.~Aguilar-Saavedra,
  Nucl.\ Phys.\ B {\bf 813}, 22 (2009)
  [arXiv:0808.2468 [hep-ph]].

\bibitem{Melfo:2011nx} 
  A.~Melfo, M.~Nemevsek, F.~Nesti, G.~Senjanovic and Y.~Zhang,
  Phys.\ Rev.\ D {\bf 85}, 055018 (2012)
  [arXiv:1108.4416 [hep-ph]].

\bibitem{Aoki:2011pz} 
  M.~Aoki, S.~Kanemura and K.~Yagyu,
  Phys.\ Rev.\ D {\bf 85}, 055007 (2012)
  [arXiv:1110.4625 [hep-ph]].

\bibitem{Akeroyd:2011zza} 
  A.~G.~Akeroyd and H.~Sugiyama,
  Phys.\ Rev.\ D {\bf 84}, 035010 (2011)
  [arXiv:1105.2209 [hep-ph]].
  
\bibitem{Arhrib:2011vc} 
  A.~Arhrib, R.~Benbrik, M.~Chabab, G.~Moultaka and L.~Rahili,
  JHEP {\bf 1204}, 136 (2012)
  [arXiv:1112.5453 [hep-ph]].
  
\bibitem{Akeroyd:2012nd} 
  A.~G.~Akeroyd, S.~Moretti and H.~Sugiyama,
  Phys.\ Rev.\ D {\bf 85}, 055026 (2012)
  [arXiv:1201.5047 [hep-ph]].
  
\bibitem{Chiang:2012dk} 
  C.~W.~Chiang, T.~Nomura and K.~Tsumura,
  Phys.\ Rev.\ D {\bf 85}, 095023 (2012)
  [arXiv:1202.2014 [hep-ph]].
  
\bibitem{Chun:2012zu} 
  E.~J.~Chun and P.~Sharma,
  JHEP {\bf 1208}, 162 (2012)
  [arXiv:1206.6278 [hep-ph]].
  

  
\bibitem{Chun:2013vma} 
  E.~J.~Chun and P.~Sharma,
  Phys.\ Lett.\ B {\bf 728}, 256 (2014)
  [arXiv:1309.6888 [hep-ph]].

\bibitem{Chabab:2014ara} 
  M.~Chabab, M.~C.~Peyranere and L.~Rahili,
  Phys.\ Rev.\ D {\bf 90}, no. 3, 035026 (2014)
  [arXiv:1407.1797 [hep-ph]].


\bibitem{Han:2015hba} 
  Z.~L.~Han, R.~Ding and Y.~Liao,
  Phys.\ Rev.\ D {\bf 91}, 093006 (2015)
  [arXiv:1502.05242 [hep-ph]].

\bibitem{Guo:2016dzl} 
  S.~Y.~Guo, Z.~L.~Han and Y.~Liao,
  Phys.\ Rev.\ D {\bf 94}, no. 11, 115014 (2016)
  [arXiv:1609.01018 [hep-ph]].

\bibitem{Mitra:2016wpr} 
  M.~Mitra, S.~Niyogi and M.~Spannowsky,
  Phys.\ Rev.\ D {\bf 95}, no. 3, 035042 (2017)
  [arXiv:1611.09594 [hep-ph]].

\bibitem{Ghosh:2017pxl} 
  D.~K.~Ghosh, N.~Ghosh, I.~Saha and A.~Shaw,
  Phys.\ Rev.\ D {\bf 97}, no. 11, 115022 (2018)
  [arXiv:1711.06062 [hep-ph]].

\bibitem{Dev:2018sel} 
  P.~S.~B.~Dev, M.~J.~Ramsey-Musolf and Y.~Zhang,
  Phys.\ Rev.\ D {\bf 98}, no. 5, 055013 (2018)
  [arXiv:1806.08499 [hep-ph]].

\bibitem{Dev:2018kpa} 
  P.~S.~Bhupal Dev and Y.~Zhang,
  JHEP {\bf 1810}, 199 (2018)
  [arXiv:1808.00943 [hep-ph]].


\bibitem{Du:2018eaw} 
  Y.~Du, A.~Dunbrack, M.~J.~Ramsey-Musolf and J.~H.~Yu,
  JHEP {\bf 1901}, 101 (2019)
  [arXiv:1810.09450 [hep-ph]].

\bibitem{Antusch:2018svb} 
  S.~Antusch, O.~Fischer, A.~Hammad and C.~Scherb,
  JHEP {\bf 1902}, 157 (2019)
  [arXiv:1811.03476 [hep-ph]].

\bibitem{Bhattacharya:2018fus} 
  S.~Bhattacharya, P.~Ghosh, N.~Sahoo and N.~Sahu,
  arXiv:1812.06505 [hep-ph].

\bibitem{Barman:2019tuo} 
  B.~Barman, S.~Bhattacharya, P.~Ghosh, S.~Kadam and N.~Sahu,
  arXiv:1902.01217 [hep-ph].

\bibitem{Primulando:2019evb} 
  R.~Primulando, J.~Julio and P.~Uttayarat,
  arXiv:1903.02493 [hep-ph].


\bibitem{Aprile:2018dbl} 
  E.~Aprile {\it et al.} [XENON Collaboration],
  Phys.\ Rev.\ Lett.\  {\bf 121}, no. 11, 111302 (2018)
  [arXiv:1805.12562 [astro-ph.CO]].

\bibitem{Amole:2017dex} 
  C.~Amole {\it et al.} [PICO Collaboration],
  Phys.\ Rev.\ Lett.\  {\bf 118}, no. 25, 251301 (2017)
  [arXiv:1702.07666 [astro-ph.CO]].
  
\bibitem{Aprile:2019dbj} 
  E.~Aprile {\it et al.} [XENON Collaboration],
  arXiv:1902.03234 [astro-ph.CO].

\bibitem{Bonilla:2015eha} 
  C.~Bonilla, R.~M.~Fonseca and J.~W.~F.~Valle,
  Phys.\ Rev.\ D {\bf 92}, no. 7, 075028 (2015)
  [arXiv:1508.02323 [hep-ph]].
  
\bibitem{Arcadi:2019lka} 
  G.~Arcadi, A.~Djouadi and M.~Raidal,
  arXiv:1903.03616 [hep-ph].
  
\bibitem{Alves:2015pea} 
  A.~Alves, A.~Berlin, S.~Profumo and F.~S.~Queiroz,
  Phys.\ Rev.\ D {\bf 92}, no. 8, 083004 (2015)
  [arXiv:1501.03490 [hep-ph]].
  
\bibitem{Belanger:2008sj} 
  G.~Belanger, F.~Boudjema, A.~Pukhov and A.~Semenov,
  Comput.\ Phys.\ Commun.\  {\bf 180}, 747 (2009)
  [arXiv:0803.2360 [hep-ph]].
  

 
  
\bibitem{Lavoura:1993nq} 
  L.~Lavoura and L.~F.~Li,
  Phys.\ Rev.\ D {\bf 49}, 1409 (1994)
  [hep-ph/9309262].
  
   
  
\bibitem{Arhrib:2011uy} 
  A.~Arhrib, R.~Benbrik, M.~Chabab, G.~Moultaka, M.~C.~Peyranere, L.~Rahili and J.~Ramadan,
  Phys.\ Rev.\ D {\bf 84}, 095005 (2011)
  [arXiv:1105.1925 [hep-ph]].
  
\bibitem{Kannike:2012pe} 
  K.~Kannike,
  Eur.\ Phys.\ J.\ C {\bf 72}, 2093 (2012)
  [arXiv:1205.3781 [hep-ph]].
  
\bibitem{Bonilla:2015eha} 
  C.~Bonilla, R.~M.~Fonseca and J.~W.~F.~Valle,
  Phys.\ Rev.\ D {\bf 92}, no. 7, 075028 (2015)
  [arXiv:1508.02323 [hep-ph]].
  
    \bibitem{PDG} M. Tanabashi et al. (Particle Data Group), Phys. Rev. D {\bf 98}, 030001 (2018). 
    
\bibitem{deSalas:2017kay} 
  P.~F.~de Salas, D.~V.~Forero, C.~A.~Ternes, M.~Tortola and J.~W.~F.~Valle,
  Phys.\ Lett.\ B {\bf 782}, 633 (2018)
  [arXiv:1708.01186 [hep-ph]].
    
\bibitem{Gunion:1989we} 
  J.~F.~Gunion, H.~E.~Haber, G.~L.~Kane and S.~Dawson,
  Front.\ Phys.\  {\bf 80}, 1 (2000); Errata in hep-ph/9302272.
  
  

  
\bibitem{Denner:2011mq} 
  A.~Denner, S.~Heinemeyer, I.~Puljak, D.~Rebuzzi and M.~Spira,
  Eur.\ Phys.\ J.\ C {\bf 71}, 1753 (2011)
  [arXiv:1107.5909 [hep-ph]].
  
\bibitem{ATLAS:2019slw} 
  The ATLAS collaboration [ATLAS Collaboration],
  ATLAS-CONF-2019-005.
  
\bibitem{CMS:1900lgv} 
  CMS Collaboration [CMS Collaboration],
  CMS-PAS-HIG-18-029.

\bibitem{Chun:2012jw} 
  E.~J.~Chun, H.~M.~Lee and P.~Sharma,
  JHEP {\bf 1211}, 106 (2012)
  [arXiv:1209.1303 [hep-ph]].
  
 
\bibitem{Ade:2015xua} 
  P.~A.~R.~Ade {\it et al.} [Planck Collaboration],
  Astron.\ Astrophys.\  {\bf 594}, A13 (2016)
  [arXiv:1502.01589 [astro-ph.CO]].

\bibitem{Belanger:2008sj} 
  G.~Belanger, F.~Boudjema, A.~Pukhov and A.~Semenov,
  Comput.\ Phys.\ Commun.\  {\bf 180}, 747 (2009)
  [arXiv:0803.2360 [hep-ph]].



\bibitem{Cahn:1978nz} 
  R.~N.~Cahn, M.~S.~Chanowitz and N.~Fleishon,
  Phys.\ Lett.\  {\bf 82B}, 113 (1979).

\bibitem{Bergstrom:1985hp} 
  L.~Bergstrom and G.~Hulth,
  Nucl.\ Phys.\ B {\bf 259}, 137 (1985)
  Erratum: [Nucl.\ Phys.\ B {\bf 276}, 744 (1986)].


\bibitem{Arbabifar:2012bd} 
  F.~Arbabifar, S.~Bahrami and M.~Frank,
  Phys.\ Rev.\ D {\bf 87}, no. 1, 015020 (2013)
  [arXiv:1211.6797 [hep-ph]].




\bibitem{Dev:2013ff} 
  P.~S.~Bhupal Dev, D.~K.~Ghosh, N.~Okada and I.~Saha,
  JHEP {\bf 1303}, 150 (2013)
  Erratum: [JHEP {\bf 1305}, 049 (2013)]
  [arXiv:1301.3453 [hep-ph]].




\end{thebibliography}
\end{document}